\documentclass[11pt,a4paper]{article}
\usepackage{cite}

\usepackage{amsmath, amsthm,mathtools,setspace, hyperref, url,amssymb,upgreek,textgreek,slashed}

 \usepackage[mathscr]{eucal}
\usepackage{ifpdf}
\usepackage{siunitx}
\ifpdf
  \usepackage[pdftex]{graphicx}
  \usepackage{epstopdf}
\else
  \usepackage[dvips]{graphicx}
\fi
\parskip=\baselineskip
\numberwithin{equation}{section}

\begin{document}

\begin{titlepage}
\begin{flushright}

\end{flushright}

\vskip 1.5in
\begin{center}
  {\bf\Large{Vacuum Branching, Dark Energy, Dark Matter}}

\vskip
0.5cm  { Don Weingarten} \vskip 0.05in {\small{ \textit{donweingarten@hotmail.com}\vskip -.4cm
}
}
\end{center}
\thanks{donweingarten@hotmail.com}
\vskip 0.5in
\baselineskip 16pt
\begin{abstract}

  Beginning with the Everett-DeWitt many-worlds interpretation of quantum mechanics, there have
  been a series of proposals for how the state vector of a quantum system might split at any instant
  into orthogonal branches, each of which exhibits approximately classical behavior.
  In an earlier version of the present work, we proposed a decomposition of a state vector into branches 
  by finding the minimum of a measure of the mean squared quantum complexity of the branches in the branch 
  decomposition. In the present article, we adapt the earlier version to quantum electrodynamics of
  electrons and protons on a lattice in Minkowski space.
  The earlier version, however, here is simplified by
  replacing a definition of complexity which takes the physical vacuum as 0 complexity starting point, with
  a definition which takes the bare vacuum as starting point.
  As a consequence of this replacement, the physical vacuum itself is expected to
  branch yielding branches with
  energy densities slightly larger than that of the unbranched vacuum.
  If the vacuum energy renormalization constant is chosen as usual to give 0 energy
  density to the unbranched vacuum, in an expanding universe vacuum branches will appear to have
  a combination of dark energy and dark matter densities.
  The hypothesis that vacuum branching
  is the origin of
  the observed dark energy and dark matter densities
  leads to an estimate of $\mathcal{O}(10^{-18} \si{m}^3)$ for
  the parameter $b$ which enters the complexity measure 
  governing branch formation  and sets the
  boundary between quantum and classical behavior.
  
\end{abstract}

%\date{October, 2021}
\end{titlepage}

%\receipt{}
%\pacs{03.65.Ca, 02.30.Cj, 03.65.Ta}

\tableofcontents

\section{\label{sec:intro}Introduction}

According to the
the many-worlds interpretation of quantum mechanics \cite{Everett, DeWitt},
combined with environmentally-induced decoherence \cite{Zeh, Zurek, Zurek1, Zurek2, Wallace, Riedel},
the quantum state of the universe,
as time goes along, naturally splits into a set of orthogonal branch states each of which displays
a distinct
configuration of macroscopic reality.
In \cite{Weingarten1, Weingarten}, however, it's argued that the rules according to which these proposals are to be applied
to the real world are instrinsically uncertain and can be made precise only by the arbitrary choice
of auxiliary parameters. This uncertainty is not simply the approximate nature of the macroscopic
description of an underlying microscopic system but rather that the branching process
of the microscopic system, in each of these proposals, occurs according to uncertian rules.
In addition, missing from these proposals is a mathematical structure which allows
even the process of choosing the auxiliary parameters to be stated precisely.

To overcome these limitation, it's proposed in \cite{Weingarten1, Weingarten}
that the set of orthogonal branches into which a quatum system splits at any time
is given by the minimum of a measure of the mean squared complexity \cite{Nielsen}
of the branch decomposition. In a particular Lorentz frame, for a system beginning in
a state of low complexity, branching occurs
repeatedly over time with each branch splitting successively into further sub-branches
among which the branch followed by the real world is chosen according to the Born rule.
In an explicitly Lorentz covariant formulation, the real world is a single random draw from the set of
branches at asymptotically late time \cite{Kent, Kent1, Kent2}, which can then be restored to finite time
in a particular Lorentz frame
by sequentally retracing the set of
branching events implied by the late time choice.
In the present article, we adapt \cite{Weingarten} to quantum electrodynamics
of electrons and protons in
temporal gauge on a lattice in Minkowski space \cite{Wilson, Kogut}.

A main feature of the proposals here and in \cite{Weingarten1, Weingarten}
is that branch formation does not follow from unitary time
evolution by itself nor does it 
entail a modification of unitary time evolution.
Instead, branch formation consists of an additional layer
of the world that sits on top of unitary time evolution.

In the determination of complexity according to \cite{Weingarten},
the physical vacuum and product states based on the physical vacuum
are assigned 0 complexity. The complexity of any other state
is then given by the minimum of the amount of computational work needed
for it to be produced starting from a state of 0 complexity.
For a relativistic field theory, as a consequence of the Reeh-Schlieder theorem,
the use of the physical vacuum in the determination
of complexity leads to some mathematical baggage beyond
what is required for a non-relativistic field theory.
Here the extra baggage is avoided by
replacing the physical vacuum with the bare vacuum
as a 0 complexity state and in the determination
of complexity of any other state.
The presence of virtual processes in the physical vacuum which are missing
from the bare vacuum may
seem a potential source of trouble for a definition of complexity
based on the bare vacuum.
We will argue, however, that
this replacement has consequences arising only from 
sufficiently long distance components of virtual process.
Among these consequences is
that the physical vacuum itself is expected to branch.
The branches which result will have disrupted configurations of virtual
fermion-antifermion pairs and energy density greater than the energy density of
the unbranched vacuum as a consequence.
The bare vacuum, to be defined explicitly as part of the definition
of lattice QED,  is the state lacking all bare photons and fermions,
in particular the Dirac sea of negative energy fermion states.

While both the 
physical vacuum and the process of branching are Lorentz and translationally invariant, the fermion-antifermion components missing
from each vacuum branch will have a small nonzero mass
so that the choice of a vacuum branch will entail a spontaneous breaking of Lorentz invariance.
The entanglement removed from the vacuum
as a consequence of branching we will argue leaves behind
a set of possible branches each consisting approximately of a tensor product of
localized factors.
Any such product will fail to be translationally invariant.
Translational invariance we therefore expect to be sponaneously broken in addition.
As time evolves, in an expanding universe
the missing components of
virtual fermion-antifermion wave functions
will be progressively red shifted and thereby drift away from the distribution in the
optimal branch ensemble. The remaining virtual fermion-antifermion
pairs will then branch again according to the optimal distribution.
Thus rather than eventually being red shifted away, as would
a distribution of ordinary matter, the restored branches will persist
as would a distribution of dark energy.  
If the vacuum energy renormalization constant is chosen as
usual to give 0 energy density to the system's lowest energy state, the unbranched vacuum,
the persistent branches will mimic the effect of a dark energy term
in the Friedman equations.

The branched vacuum with disrupted long distance component of fermion-antifermion wave functions
is equivalent to an unbranched vacuum with a slight contribution of fermion-antifermion pairs with small relative momentum.
For photon and fermion propagation through the branched vacuum,
the result of these changes is
renormalized quantities such as
fermion masses and charge which differ
slightly from their values in the unbranched vacuum.
Since the renormalized quantities of the unbranched vacuum are
for the most part not directly observable, the discrepancy between these quantities
and their observable values in the branched vacuum will also be unobservable.
The one exception to this is
vacuum energy density assumed
to be 0 in the unbranched vacuum
and slightly elevated 
in the branched vacuum yielding
the appearance of dark energy.

The distribution of fermion-antifermion pairs responsible for
the appearance of dark energy, however, arises from a random
branching process and will therefore not be entirely uniform.
The gravitational effect of the inhomogeniety of this distribution will
show itself as dark matter in the formation and
evolution of galaxies.
Whether the resulting dark matter is quantitatively consistent with
what is known about galaxy formation and evolution, however,
is beyond the scope of the present work.

A feature of the branching process proposed here,
we argue in Section \ref{subsec:timeevolution}
and in Appendix \ref{app:pairsplits},
is that while branches which form in complete isolation
can recombine, branches which form
and then become entangled with the rest of the universe
with high probability will never recombine.
The improbability of recombination of entangled
branches is, in effect, a version of
environmentally induced decoherence.

While a fully satisfactory quantum theory of gravity has notoriously yet to be found,
a reasonable assumption \cite{Donoghue} is that according to any such theory
each branch in a branch decomposition of a state of the universe
will consist of the product of a state of matter and a state of the gravitational field
to which the energy density of the matter gives rise.
In particular, the expansion of the universe in each branch
will be the expansion
corresponding to the gravitational field and energy density of the matter in that branch.
If this hypothesis is correct,
the corresponding entanglement of vacuum
branches and their associated energy densities with the expansion of the universe 
will, on the one hand, make it possible to observe the value of the parameter $b$ which enters
the complexity measure governing branch formation and thereby determines the energy density
of vacuum branches and, 
on the other hand,
will result in vacuum branches which, as a consequence of this entanglement, 
with highly probablity will not recombine.

The hypothesis that vacuum branching
is the origin of
the observed dark energy and dark matter densities
leads to an estimate of $\mathcal{O}(10^{-18} \si{m}^3)$ for
$b$.
The hypothesis that the vacuum will branch could,
in principle, be tested by a numerical strategy
which, combined with the observed dark energy and dark matter 
densities,
might yield a more accurate value of $b$.
A numerical evaluation of vacuum branching
should also, in principle, yield a prediction
for the ratio of dark energy and dark matter densities.

An analogous branching process
might occur also in the ground state of a solid at 0 K.
This hypothesis could also be tested
by a numerical strategy and,
if confirmed by laboratory observation,
yield a second determination of $b$.

The absence of human experience of brain states superposing
contradictory ideas suggests an upper bound for $b$ of $ \mathcal{O} (10^{-9} \si{m}^3)$,
consistent with the astrophysical estimate.

For the sake of clarity, it is perhaps worth repeating that
the possibility of determining $b$ from dark energy and dark matter
arises not because branching acts back on the unitary time
evolution of the universe but rather because the universe
that humans experience, according to the proposal here, is
a consequence of two independent process, unitary time evolution
and branching. Branching selects out of the unitarily evolving
universe a slice experienced by humans.  What would
the resulting slice look like for a different value of $b$?
I am not sure. Once possibility is that the result
would look rather like the universe we see but with
a different trajectory of expansion and different structure
of galaxies. Another possibility, however, is that
even a small change in $b$ would result in a world
drastically different from ours.
Just as a small change in the charge of the electron
might drastically alter chemistry and make human life impossible.

As a byproduct somewhat independent of the
main proposal, we are led to the hypothesis that branching
events are the physical substance of
the phenomenon of experience, both human and otherwise.
If so, the private experience of a machine which
passes the Turning test with hardware nothing
like a human brain would be correspondingly
unlike human private experience.

A superselection rule which comes about as a result of
measuring complexity with respect to the bare vacuum
is that the total number of
protons in the world, measured with respect to the bare vacuum, must equal the total number of electrons,
measured with respect to the bare vacuum.
Restated with respect to the physical vacuum,
the total number
protons minus antiprotons must equal the total number of electrons minus positrons.

While it is unlikely that a 0 lattice spacing limit exists for
lattice QED, we will assume that
for lattice spacing sufficiently small 
the theory gives an accurate
representation of
the world on the length scale
relevant to branching.

The formulation of complexity and branching
presented here can be extended to nonabelian
gauge theories, some of which may have 0 lattice spacing limits.
For the sake of simplicity the present article is
restricted almost entirely to QED, with the sole
exception of a brief mention in 
Section \ref{subsec:branchcomplexity} of
an extension to QCD.
For the extension to QCD,
an immediate consequence is
a possible resolution of the strong CP
problem which is nearly equivalent to
a superselection rule requiring
the strong CP violating angle $\Theta$ to be
0.

We begin in Section \ref{sec:qed} with a definition of lattice QED for electrons and
protons and a corresponding definition of complexity. In Section \ref{sec:entangledstates}
we place upper and lower bound on the complexity of a family of entangled multi-fermion states.
In Section \ref{sec:branching} we define net complexity which
the optimal set of branches is defined to minimize.
In Section \ref{sec:examples} we give examples of
branches and their time evolution. Based on these examples,
in Section \ref{sec:residual} we present a conjecture
for the structure of states which result immediately following
a branching event. In Section \ref{sec:covariant} we
propose that a $t \rightarrow \infty$ limit of branching
is Lorentz covariant. In Section \ref{sec:rare} we
consider the relation between branching events and
experience, both human and otherwise.
Finally, in Section \ref{sec:vacuum}, based
on all of the foregoing, we present an argument in support of the hypothesis
that the vacuum itself undergoes branching.

Sections \ref{subsec:operatorspace}, \ref{subsec:complexitydef} and \ref{sec:entangledstates} - \ref{sec:examples} of the present article closely follow
corresponding sections of \cite{Weingarten} but
with the changes required to
adapt \cite{Weingarten} to a gauge theory and to the
replacement of the physical vacuum with the bare vacuum
in the definition of 0 complexity.  Two sections of \cite{Weingarten}
have been included as Appendices \ref{app:properties} and \ref{app:secondlaw} without change to make the
present article easier to follow.

\section{\label{sec:qed} Complexity for  Lattice QED}

We will construct a complexity measure \cite{Nielsen}
for quantum electrodynamics of electrons and protons in temporal gauge
on a 3-dimensional cubic lattice.
The Hilbert space and Hamiltonian for development in real time
are taken from the transfer matrix of the Euclidean theory in the limit of 0 lattice
spacing in the time direction \cite{Wilson, Kogut}.
The result is a $U(1)$ operator version
of the path integral $SU(3)$ construction of \cite{Wilson}
for a single time slice but with time development adapted from
\cite{Kogut}.

We use the Schroedinger representation, thus 
time independent field operators and time dependent states.
A summary of various properties of complexity
appears in Appendix \ref{app:properties}
taken from \cite{Weingarten} without change.
Also essentially unchanged from \cite{Weingarten},
Appendix \ref{app:secondlaw} consists of
a derivation from the conjectured second law of quantum complexity
of \cite{Brown} of an estimate of the change in complexity over time of a system
evolving according to the Hamiltonian $h$ of Eq. (\ref{hamiltonian}).

\subsection{\label{subsec:hilbertspace} Lattice QED}

Let $L$ be a cubic lattice with coordinates $a \hat{x}^1, a \hat{x}^2, a \hat{x}^3$,
integer $\hat{x}^i$, lattice spacing $a$, spanning the region $a B \leq a \hat{x}^i < a B$. 
Let $\Psi^{v \dagger}_s( x)$ and $\Psi^v_s( x)$ be fermion creation and annihilation
operators for lattice site $x$, spin index $0 \le s \le 3$, flavor index $v = 0$ for
protons and $ v = 1$ for electrons.
These operators have the anticommutation relation
\begin{subequations}
\begin{eqnarray}
  \label{psianti}
  \{ \Psi^v_s(x), \Psi^{v'\dagger}_{s'}( x') \} &=& \delta_{xx'} \delta_{ss'} \delta_{vv'}, \\
  \label{psianti1}
  \{ \Psi^v_s(x), \Psi^{v'}_{s'}( x') \} &=& 0, \\
  \label{psianti2}
  \{ \Psi^{v\dagger}_s(x), \Psi^{v'\dagger}_{s'}( x') \} &=& 0, 
\end{eqnarray}
\end{subequations}
Let $\mathcal{H}_x$ be the Hilbert space spanned by all polynomials in
the  $\Psi^{v\dagger}_s(x)$ for any $s, v,$ acting on the bare
vacuum $|\Omega\rangle_x  $. The bare vacuum $|\Omega\rangle_x  $
is annihilated by all $\Psi^v_s( x)$.

For each nearest neighbor pair $\{x,y\}$, let $U( x, y)$
be the corresponding link field. For use in
the perturbation expansion to be considered in Section \ref{sec:vacuum},
define also the Hermitian photon field $A( x, y)$
related to $U( x, y)$ by
\begin{equation}
  \label{ua}
  U(x, y) = \exp [ i e A( x, y)],
\end{equation}
where $e$ is the electric charge.
For $\{x,y\}$ in reverse order we adopt the convention
\begin{subequations}
\begin{eqnarray}
    \label{areversed}
    A( y, x) &=& -A( x, y), \\
    \label{udagger}
    U( y, x) &=& U^\dagger( x, y).
\end{eqnarray}
  \end{subequations}
All $U( x, y), U^\dagger( x, y), $ commute with each other and with all $\Psi^v_s( x), \Psi^{v\dagger}_s( x)$.
Let $|\Omega\rangle_{x y} $ be
the corresponding bare vacuum and $\mathcal{H}_{x y}$ the Hilbert space with orthonormal basis
\begin{equation}
  \label{ubasis}
  \{ U( x, y)^m|\Omega \rangle_{xy}, 0 \le m < n_U \} \cup  \{ U^\dagger( x, y)^m|\Omega \rangle_{xy}, 1 \le m < n_U \},
\end{equation}
for some large $n_U$.
Finite $n_U$ makes $\mathcal{H}_{xy}$ finite dimensional which
is convenient for construction of the operator Hilbert space
in Section \ref{subsec:operatorspace}
and for the argument for the persistence of pair split in Appendix
\ref{app:pairsplits}. In both cases, we will eventually
take $n_U \rightarrow \infty$.

Let $E(x,y)$ be the electric
field for link $\{x, y\}$ given by
\begin{equation}
  \label{defE}
  E( x, y) = \frac{-i\partial}{\partial A(x,y)},
  \end{equation}
and therefore
\begin{subequations}
  \begin{eqnarray}
        \label{exyyx}
    E( x, y)  &=&  -E( y, x), \\
   \label{eu}
         [ E( x, y), U( x, y)] & = & e U( x, y), \\
         \label{evac}
         E( x, y) |\Omega \rangle_{xy} &=& 0.
 \end{eqnarray}
\end{subequations}

The photon fields and electric fields commute with all fermion operators
\begin{subequations}
  \begin{eqnarray}
    \label{apsi}
    [ A( x, y), \Psi^v_s( z)] & = & 0, \\
    \label{apsi1}
    [ A( x, y), \Psi^{v\dagger}_s( z)] & = & 0, \\
    \label{epsi}
    [ E( x, y), \Psi^v_s( z)] & = & 0, \\
    \label{epsi1}
          [ E( x, y), \Psi^{v\dagger}_s( z)] & = & 0.
  \end{eqnarray}
\end{subequations}

Define the space $\mathcal{H}$ to be
\begin{equation}\label{tensorproduct}
\mathcal{H} = \bigotimes_x \mathcal{H}_x \bigotimes_{yz} \mathcal{H}_{yz},
\end{equation}
and the bare vacuum $|\Omega \rangle $ to be
\begin{equation}\label{tensorproductvacuum}
|\Omega \rangle  = \bigotimes_x |\Omega \rangle _x \bigotimes_{yz} |\Omega \rangle _{yz}.
\end{equation}
For any site $x$ define the generator of gauge
transformations $g(x)$ to be
\begin{equation}
  \label{gaugegen}
  g(x) = \sum_y E( x, y) + e \Psi^{0 \dagger}(x) \Psi^{0}(x) - e \Psi^{1 \dagger}(x) \Psi^{1}(x),
\end{equation}
where sums over spin indices are assumed in the terms $\Psi^{0 \dagger}(x) \Psi^{0}(x)$ and
$\Psi^{1 \dagger}(x) \Psi^{1}(x)$ and in similar products in the remainder of the paper.
Realizable physical states
are required to be in the subspace of  $\mathcal{H}$ invariant under all
gauge transformations, thereby enforcing Gauss's law.

From the field operators, we construct a Hamiltonian. 
Let the antihermitian matrices $\gamma^1, \gamma^2, \gamma^3,$ and
Hermitian $\gamma^0$ satisfy
\begin{equation}
  \label{gamma}
  \{ \gamma^\mu, \gamma^\nu \} = 2 g^{\mu\nu}.
\end{equation}
For a positive nearest neighbor step $\hat{j}$ in direction $1 \le j < 4$, let
\begin{subequations}
\begin{eqnarray}
  \label{gamma1}
  \gamma( x + \hat{j}, x) & = & -1 + i\gamma^j, \\
  \label{gamma2}
  \gamma(x, x + \hat{j}) & = & -1 - i\gamma^j.
\end{eqnarray}
\end{subequations}
For $\{ x, y \}$ not nearest neighbors, let
\begin{equation}
  \label{gamma3}
  \gamma( x, y) = 0.
  \end{equation}
Define the field operator $\bar{\Psi}^v(x)$ to be
\begin{equation}
  \label{defbarpsi}
  \bar{\Psi}^v(x) = \Psi^{v \dagger} \gamma^0.
\end{equation}

For any plaquette $p$ consisting of a closed loop of 4 nearest neighbor lattice sites
$\{u, v, x, y\}$ define the plaquette field $U( p)$
\begin{equation}
  \label{plaquette}
  U( p) = U( u, v) U( v, x) U( x, y) U( y, u).
\end{equation}
Plaquettes differing by a cyclic permutation will be treated as identical.
Plaquettes with order reversed will be treated as distinct.

The lattice Hamiltonian operator $h$ is then
\begin{multline}
  \label{hamiltonian}
  h = \frac{1}{2} \sum_{xy} \bar{\Psi}^0( x) \gamma( x, y) U( x, y) \Psi^0( y) + \frac{1}{2} \sum_{xy}\bar{\Psi}^1( x) \gamma( x, y) U^\dagger( x, y) \Psi^1( y) + \\
   \sum_{x v} (3 + m^v) \bar{\Psi}^v(x) \Psi^v(x) +  
 \frac{1}{4} \sum_{xy} E^2( x, y) + \frac{1}{2e^2} \sum_p[ 1 - U( p)] - \xi_\Omega, 
\end{multline}
where the sums are variously over all nearest neighbor pairs $\{x,y\}$, sites $x$, flavors  $v$, and plaquettes $p$ and the masses $m^v$ are unitless.
Here we avoid the problem of species doubling \cite{NielsenNinomiya} by use of Wilson fermions \cite{Wilson} rather than the staggered fermions of \cite{Kogut}.
Although we have defined $\mathcal{H}$ starting from the bare vacuum $|\Omega \rangle$,
$\mathcal{H}$ contains also the usual physical vacuum, namely the energy eigenstate
of $h$ with lowest eigenvalue.
The constant $\xi_\Omega$ is determined by the requirement that the physical
vacuum have 0 energy.
If we had defined the theory by a Euclidean path integral, the lowest energy eigenstate
would automatically be assigned 0 energy.
The determination of $\xi_\Omega  $ will be considered 
in the discussion of vacuum branching.

We define in $\mathcal{H}$ a set of gauge invariant product states.
For a complex-valued wave function $q_{ss'}(x)$ over postion and spins,
define the gauge invariant
operator $d^\dagger( q)$ to be
\begin{equation}
\label{extended}
d^\dagger( q) = \sum_{x s s'} q_{ss'}(x)\Psi^{0\dagger}_s( x) \Psi^{1\dagger}_{s'}( x).
\end{equation}
From a set of $n$ wave functions $q^i_{ss'}( x)$
define the gauge invariant product state 
\begin{equation}
  \label{productstate}
d^\dagger( q^{n-1}) ... d^\dagger( q^0) |\Omega \rangle .
\end{equation}
Let $\mathcal{P}$ be the set of all product states.
It will turn out that no product state factors are needed composed purely of
gauge fields.

Although this definition of product state differs from the notion of
product state in many-body physics, it is the natural
definition here and
serves best technically.
Since physical states all need to be gauge invariant,
the most general way to accomplish this without introducing gauge
fields requires products of 
$\Psi^{0\dagger}_s( x)$ and  $\Psi^{1\dagger}_{s'}( x)$ at a single point.
Gauge invariant states which include both fermions and gauge fields
can all be built from product states by the set of unitary transformations
defined below in the definition of complexity.

For the limiting case $n_U \rightarrow \infty$ we introduce a larger set of product states.
For any pair sites $w, z,$ and pair of spins $s, s',$ let $T_{ss'}(w, z)$ be a product of $U( x, y)$
along a path from $z$ to $w$ multiplied by a function of the plaquette variables $U(p)$ such that
the product $ T_{ss'}(w , z) \Psi^{0\dagger}_s( w) \Psi^{1\dagger}_{s'}( z)$ is gauge invariant.
Define $d^\dagger(T)$ to be
\begin{equation}
\label{extended1}
d^\dagger( T) = \sum_{wszs'} T_{ss'}( w, z) \Psi^{0\dagger}_s( w) \Psi^{1\dagger}_{s'}( z) .
\end{equation}
From a set of $n$ $T^i_{ss'}( w, z)$,
define the gauge invariant product state 
\begin{equation}
\label{productstate2}
d^\dagger( T^{n-1}) ... d^\dagger( T^0 ) |\Omega \rangle .
\end{equation}
Again let $\mathcal{P}$ be the set of all product states.

\subsection{\label{subsec:operatorspace} Hermitian Operator Hilbert Space}

Complexity according to the formulation of \cite{Nielsen} is determined by a family
of unitary operators. Which in turn are generated by Hermitian operators
taken from a Hilbert space over the reals of operators on $\mathcal{H}$.
The Hermitian operators we now define are taken to be nearest neighbor local
so that the corresponding unitary operators and complexity will become a measure
of the distance over which entanglement in any state is spread.
These operators, in addition, will be assumed gauge invariant and fermion number
preserving so that the
unitary operators defining complexity move states only within the physical Hilbert space
of gauge invariant states and act only by rearranging fermion positions without adding or
removing any.  For convenience in defining the inner product on the
operator Hilbert space, the operators will be decomposed into those which
act on a pair of nearest neighbor sites and those which act purely on single
sites, with both classes required to have vanishing partial trace on
each of the individual sites on which they act.
The effect of the vanishing trace conditions will eventually be compensated by allowing
an overall arbitrary phase transformation on top of the unitary transformations
these operators generate.

For each site $x$ and flavor $v$, let $N^v(x)$ be the fermion number operator on $\mathcal{H}_x$,
for nearest neighbor $\{x,y\}$
let $N^v( x, y)$ be $N^v(x) + N^v( y)$ and 
let $N^v$ be the total of $N^v(x)$ over all $x$. 
The Hamiltonian $h$ of Eq. (\ref{hamiltonian}) conserves $N^v$ for $v = 0, 1$.

For each site $x$, let $\mathcal{F}_x$ be the vector space over the reals of Hermitian operators $f_x$
on $\mathcal{H}_x$ which
conserve $N^v(x)$ and have
\begin{subequations}
\begin{eqnarray}
  \label{trx}
  \mathrm{Tr}_x f_x  &=& 0, \\
  \label{prx}
  f_x P_x &=& f_x, 
\end{eqnarray}
\end{subequations}
where $\mathrm{Tr}_x$ is the trace on $\mathcal{H}_x$ and $P_x$ is the projection onto the
gauge invariant subspace of $\mathcal{H}_x$.

For any pair of nearest neighbor sites $\{x,y\}$, 
every  Hermitian operator $f$ on
$\mathcal{H}_x \otimes \mathcal{H}_{xy} \otimes \mathcal{H}_y$
which conserves $N^v(x, y)$ and has
\begin{subequations}
\begin{eqnarray}
  \label{trxy}
  \mathrm{Tr}_{xxyy}  f  &=& 0, \\
  \label{prxy}
  f P_{xxyy} &=& f, 
\end{eqnarray}
\end{subequations}
can be split into 
\begin{subequations}
  \begin{eqnarray}
  \label{fxy}
  f &=& \left[ f_{xy} + I_x \otimes I_{xy} \otimes f_y + f_x \otimes I_{xy} \otimes I_y \right] P_{xxyy}  , \\
  \label{fxy1}
  f_x &\in & \mathcal{F}_x \\
  \label{fxy2}
  f_y & \in & \mathcal{F}_y,
  \end{eqnarray}
\end{subequations}
where $f_{xy}$ has 
\begin{subequations}
\begin{eqnarray}
  \label{ptrfx}
  \mathrm{Tr}_x f_{xy} &=& 0,\\
  \label{ptrfy}
  \mathrm{Tr}_y f_{xy} &=& 0, \\
  \label{prxy2}
  f_{xy} P_{xxyy} &=& f_{xy}.
\end{eqnarray}
\end{subequations}
In these equations $\mathrm{Tr}_{xxyy}$ is the trace on $\mathcal{H}_x \otimes \mathcal{H}_{xy} \otimes \mathcal{H}_y$, $P_{xxyy}$ the projection onto the gauge invariant subspace of $\mathcal{H}_x \otimes \mathcal{H}_{xy} \otimes \mathcal{H}_y$
and $I_x$, $I_y$ and $I_{xy}$ are the identify operators on $\mathcal{H}_x$, $\mathcal{H}_y$, and
$\mathcal{H}_{xy}$, respectively.
Eqs. (\ref{fxy}) - (\ref{prxy2}) hold as a result of Eqs. (\ref{trxy}), (\ref{prxy}) requiring $f$ to be traceless and act only on gauge invariant states. As a consequence of gauge invariance, the gauge components of $f$ on the link $\{ x, y\}$ are determined entirely by its fermion components on sites $x$ and $y$. Let $\mathcal{F}_{xy}$ be the vector space over the reals of all such $f_{xy}$.

For any plaquette $p = \{u, v, x, y\}$, define
\begin{equation}
  \label{defplaquette1}
  \mathcal{H}_p = \mathcal{H}_{ u v} \otimes \mathcal{H}_{ v x} \otimes \mathcal{H}_{ x y} \otimes\mathcal{H}_{ y u}.
  \end{equation}
Let $\mathcal{F}_p$ be the vector space over the reals of Hermitian operators $f_p$ 
on $\mathcal{H}_p$ for which
\begin{subequations}
\begin{eqnarray}
  \label{trst}
  \mathrm{Tr}_{st}  f_p &=& 0, \\
    \label{pp}
   f_p P_p &=& f_p,
\end{eqnarray}
\end{subequations}
where $P_p$ is the projection onto the gauge invariant subspace of $\mathcal{H}_p$,
$\{s,t\}$ is any one of the 4 links of $p$ and $\mathrm{Tr}_{st}$ is the trace
on $\mathcal{H}_{st}$.

Any $f_x$ in some $\mathcal{F}_x$ can be made into an
operator $\hat{ f}_x$ on $\mathcal{H}$ by
\begin{equation}
\label{defhfx}
\hat{ f}_x =  f_x \bigotimes_{z \ne x} I_z \bigotimes_{uv} I_{uv}, 
\end{equation}
any $f_{xy}$ in some $\mathcal{F}_{xy}$ can be made into an
operator $\hat{ f}_{xy}$ on $\mathcal{H}$ by
\begin{equation}
\label{defhfxy}
\hat{ f}_{xy} =  f_{xy} \bigotimes_{z \ne x,y} I_z \bigotimes_{uv \ne xy} I_{uv},
\end{equation}
and any $f_p$ in some $\mathcal{F}_p$ can be made into an operator $\hat{f}_p$ on $\mathcal{H}$ by
\begin{equation}
\label{defhfp}
\hat{ f}_p =  f_p \bigotimes_x I_x \bigotimes_{uv \notin p} I_{uv}.
\end{equation}
where $\{u,v\} \notin p$ for any plaquette $\{w, x, y, z\}$
means $\{u,v\} \ne \{w,x\}, \{x,y\}, \{y, z\}, \{z, w\}$.
We now overload notation and drop the hats on $\hat{f}_x$, $\hat{f}_{xy}$ and
$\hat{f}_p$.

Let $K$ be the vector space over the reals of Hermitian linear
operators $k$ on $\mathcal{H}$
given by sums of the form
\begin{equation}
\label{defk}
k = \sum_x f_x + \sum_{x y} f_{x y} + \sum_p f_p, 
\end{equation}
for any $f_x \in \mathcal{F}_x$ for a set of sites $x$, 
$f_{x y} \in \mathcal{F}_{x y}$ for a set of nearest neighbor pairs $\{x,y\}$,
and $f_p \in \mathcal{F}_p$ for a set of plaquettes $p$.

Define the inner product and norm on $K$ to be 
\begin{subequations}
  \begin{eqnarray}
  \label{defkkprime}
  \langle k, k' \rangle &=& \frac{ \mathrm{Tr} ( k k')}{ \mathrm{Tr}( I)}, \\
  \label{defnorm}
  \parallel k \parallel^2 & = & \langle k, k \rangle,
  \end{eqnarray}
  \end{subequations}
where  $\mathrm{Tr}$ is the trace and $I$ the identity on Hilbert space $\mathcal{H}$.
Eq. (\ref{defkkprime})
is the inner product of \cite{Nielsen} reformulated for $K$.

The trace conditions Eqs. (\ref{trx}), (\ref{ptrfx}), (\ref{ptrfy}) and (\ref{trst}) then imply
for any $f_x$, $f_{yz}$ and $f_p$
\begin{subequations}
  \begin{eqnarray}
    \label{orthog0}
    \langle f_x, f_{yz} \rangle & = & 0, \\
    \label{orthog1}
    \langle f_x, f_p \rangle & = & 0, \\
    \label{orthog2}
    \langle f_p, f_{yz} \rangle & = & 0.
  \end{eqnarray}
\end{subequations}
In addition for any distinct pair of sites $x, x'$, any nearest neighbor
pairs $\{x,y\}$, $\{x',y'\}$ which differ on at least one site,
and any pair of plaquetts $p, p'$ which differ on at least one link
\begin{subequations}
  \begin{eqnarray}
    \label{orthog3}
    \langle f_x, f_{x'} \rangle & = & 0, \\
    \label{orthog4}
    \langle f_{xy}, f_{x'y'} \rangle & = & 0, \\
    \label{orthog5}
    \langle f_p, f_{p'} \rangle & = & 0.
  \end{eqnarray}
\end{subequations}

It then follows that for any $k, k' \in K$ given by Eq. (\ref{defk}),
the inner product of Eq. (\ref{defkkprime}) becomes
\begin{equation}
\label{defkkprime3}
 \langle  k, k' \rangle   = \sum_x \langle  f_x, f'_x \rangle + \sum_{xy}  \langle  f_{xy}, f'_{xy} \rangle + \sum_p \langle f_p, f'_p \rangle.
\end{equation}
But partial traces over most $\mathcal{H}_x$ and
$\mathcal{H}_{xy}$ contribute identical factors to the numerator and denominator of Eq. (\ref{defkkprime})
so that for any $x, \{ x, y \}$ and $p$ 
\begin{subequations}
  \begin{eqnarray}
    \label{fxnorm}
    \langle f_x, f'_x \rangle & = & \frac{ \mathrm{Tr}_x (f_x f'_x)}{ \mathrm{Tr}_x (I_x)}, \\
    \label{fxynorm}
    \langle f_{xy}, f'_{xy} \rangle & = & \frac{ \mathrm{Tr}_{xxyy} (f_{xy} f'_{xy})}{ \mathrm{Tr}_{xxyy} (I_{xxyy})}, \\
    \label{fpnorm}
    \langle f_p, f'_p \rangle & = & \frac{ \mathrm{Tr}_p (f_p f'_p)}{ \mathrm{Tr}_p (I_p)}.
  \end{eqnarray}
\end{subequations}

We now rescale $ \langle  k, k' \rangle$  by a factor of $\mathrm{Tr}_{xxyy} (I_{xxyy})$,
which is the same for all nearest neighbor $\{ x, y\}$. We obtain
\begin{subequations}
  \begin{eqnarray}
    \label{fxnorm1}
    \langle f_x, f'_x \rangle & = & \frac{\mathrm{Tr}_{xxyy} (I_{xxyy}) }{ \mathrm{Tr}_x (I_x)} \mathrm{Tr}_x (f_x f'_x), \\
    \label{fxynorm1}
    \langle f_{xy}, f'_{xy} \rangle & = &  \mathrm{Tr}_{xxyy} (f_{xy} f'_{xy}), \\
    \label{fpnorm1}
    \langle f_p, f'_p \rangle & = & \frac{\mathrm{Tr}_{xxyy} (I_{xxyy}) }{ \mathrm{Tr}_p (I_p)} \mathrm{Tr}_p (f_p f'_p).
  \end{eqnarray}
\end{subequations}
In the limit $n_U \rightarrow \infty$, all $\langle f_x, f'_x \rangle$ will diverge and all
$\langle f_p, f'_p \rangle$ will go to 0.

\subsection{\label{subsec:complexitydef} Complexity from Operator Trajectories}

From this machinery, for any pair of gauge invariant states $| \omega \rangle , |\psi \rangle  \in \mathcal{H}$ with equal
norms and $N^v$ we define 
the complexity $C(|\psi \rangle , |\omega \rangle )$ of $|\psi \rangle $ with
respect to $|\omega \rangle $. This definition remains unchanged from \cite{Weingarten}.
For $0 \leq \nu \leq 1$, let $k( \nu) \in K$ be a piecewise continuous trajectory of operators.
Let $|\omega_k( \nu) \rangle \in \mathcal{H}$ be the solution to the differential
equation and boundary condition
\begin{subequations}
\begin{eqnarray}
\label{udot}
\frac{d|\omega_k(\nu) \rangle}{d \nu} & = &-i k( \nu) |\omega_k( \nu) \rangle, \\
\label{uboundary0}
|\omega_k ( 0)\rangle & = & |\omega \rangle.
\end{eqnarray}
\end{subequations}

For any pair of gauge invariant $|\psi \rangle , |\omega \rangle  \in \mathcal{H}$ with equal
fermion numbers, let $S( |\psi \rangle , |\omega \rangle)$ be the
set of trajectories $k_j(\nu)$ and phases $\xi_j$ such that for the corresponding
$|\omega_{k_j}(1) \rangle$ we have
\begin{equation}
\label{sequenceki}
\lim_{j \rightarrow \infty} \xi_j |\omega_{k_j}(1) \rangle  = |\psi \rangle .
\end{equation}
The complexity $C(|\psi \rangle , |\omega \rangle )$ is defined to be the minimum 
over all such sequences of $k_j(\nu)$ of the
limit of the integral
\begin{equation}
\label{complexity}
C(| \psi \rangle , |\omega \rangle ) = \min_S \lim_{j \rightarrow \infty} \int_0^1 d \nu \parallel k_j( \nu) \parallel. 
\end{equation}

Finally, any product state in $\mathcal{P}$ we assign 0 complexity. 
The complexity $C( |\psi \rangle )$ of any state $|\psi \rangle $ not in $\mathcal{P}$
is defined to be the distance to the nearest product state
\begin{equation}
\label{cpsi1}
C( |\psi \rangle ) = \min_{|\omega \rangle  \in \mathcal{P}} C(| \psi \rangle , |\omega \rangle ).
\end{equation}
Since every product state in $\mathcal{P}$ is an eigenvector of $N^v$,
and since all operators in $K$ preserve $N^v$,  $|\psi \rangle $ will be reachable by
a sequence of trajectories in Eq. (\ref{sequenceki}) from a product
state $|\omega \rangle $ only if $|\psi \rangle $ itself is an eigenvector of $N^v$.
For states $|\psi \rangle $ which are not eigenvectors of $N^v$,  the minimum
in Eq. (\ref{cpsi1}) and thus the value of $C(|\psi \rangle )$ is, in effect, $\infty$.
Also, since every product state has equal
eigenvalues of $N^0$ and $N^1$ and
since all operators in $K$
preserve $N^0$ and $N^1$, states with finite complexity
will have equal eigenvalues of $N^0$ and $N^1$.

By an extension of Appendix 4 of \cite{Weingarten}, it can be shown that for all gauge invariant $|\psi \rangle , |\omega \rangle  \in \mathcal{H}$ with equal
fermion numbers, the set $S( |\psi \rangle , |\omega \rangle)$ is nonvanishing. We omit the details.
For $n_u \rightarrow \infty$,
proof that $S( |\psi \rangle , |\omega \rangle)$ is nonvanishing follows directly from Appendix 4
of \cite{Weingarten} combined with  Eq. (\ref{mcpsi1}) to be proved below.
It is not shown in \cite{Weingarten} that the limit in Eq. (
\ref{complexity}) is necessarily finite. Thus in principle there could be states with infinite
complexity. In the process of branching to be defined in Section \ref{sec:branching}, however,
states are split into pieces each with finite complexity. Which is what we would
expect to happen to states which begin with finite complexity
but then, in the $n_U \rightarrow \infty$ limit, acquire large complexity as a result of time
evolution.

\subsection{\label{subsec:nuinfty} Complexity for $n_U = \infty$}

In the limit $n_U \rightarrow \infty$, all $\parallel f_x \parallel$ diverge and all
$\parallel f_p \parallel$ go to 0. As a result the gauge field configuration
of any state makes no contribution to its complexity and single site operators
$f_x$ disappear from trajectories $k( \nu)$ with finite norm. Complexity is determined entirely
by trajectories of nearest neighbor fermion operators $f_{xy}$.
For a clean formulation of these simplications, we define an auxiliary Hilbert space
Hilbert space $\mathcal{H}^A$, composed purely
of fermions, corresponding operator Hilbert space $K^A$, and a linear operator $M$ taking
$\mathcal{H}$ into $\mathcal{H}^A$ and $K$ into $K^A$.

Define $\mathcal{H}^A$ to be
\begin{equation}
  \label{defhf}
  \mathcal{H}^A = \bigotimes_x \mathcal{H}_x.
\end{equation}
For each pair of nearest neighbor sites $\{x, y\}$, let $\mathcal{F}^A_{xy}$ be the vector space over the reals
of Hermitian $f_{xy}$ on $\mathcal{H}_x \otimes \mathcal{H}_y$ which, as before, conserve $N^v( x, y)$ and have
vanishing partial traces
\begin{subequations}
\begin{eqnarray}
  \label{ptrfx1}
  \mathrm{Tr}_x f_{xy} &=& 0,\\
  \label{ptrfy1}
  \mathrm{Tr}_y f_{xy} &=& 0.
\end{eqnarray}
\end{subequations}
Define an inner product and norm for $f_{xy} \in \mathcal{F}^A_{xy}$
by
\begin{subequations}
\begin{eqnarray}
    \label{innerf}
    \langle f_{xy}, f_{xy}' \rangle &=& \mathrm{Tr}_{xy} ( f_{xy} f_{xy}'), \\
    \label{normf}
    \parallel f_{xy} \parallel ^2 &=& \langle f_{xy}, f_{xy} \rangle_{xy}.
\end{eqnarray}
\end{subequations}
Make each $f_{xy} \in \mathcal{F}^A_{xy}$ into an operator $\hat{f}_{xy}$ on $\mathcal{H}^A$ by
\begin{equation}
  \label{hatf}
  \hat{ f}_{xy} = f_{xy} \bigotimes_{z \ne x, y} I_z,
\end{equation}
and again drop the hat on $\hat{f}_{xy}$. Let $K^A$ be the vector space over the reals of sums of the
form
\begin{equation}
  \label{defkf}
  k = \sum_{xy} f_{xy},
\end{equation}
for any set of nearest neighbor $\{ x, y \}$. Define the inner product and norm on $K^A$ to be
\begin{subequations}
\begin{eqnarray}
  \label{innerf1}
  \langle k, k' \rangle &= & \sum_{x y} \langle f_{xy}, f'_{xy} \rangle, \\
  \label{normf1}
  \parallel k \parallel ^2 & = & \langle k, k \rangle.
\end{eqnarray}
\end{subequations}

Define the linear operator $M$ taking gauge invariant eigenvectors of $N^v$ in $\mathcal{H}$
to eigenvectors of $N^v$ in $\mathcal{H}^A$
\begin{equation}
  \label{defm}
  M |\psi \rangle = |\psi^A \rangle,
\end{equation}
by replacing all occurances of any $U(x, y)$ in $|\psi \rangle$ with 1.
Define
the corresponding $M$ taking $k \in K$ to $k^A \in K^A$ by
\begin{equation}
  \label{defmk}
  M k |\psi \rangle = k^A M |\psi \rangle.
\end{equation}
We then have
\begin{equation}
  \label{mnorm}
  \parallel k \parallel = \parallel k^A \parallel.
\end{equation}

Finally, consider a set of trajectories  $k_j( \nu) \in K, 0 \le \nu \le 1$. Let $|\omega_{k_j} ( \nu) \rangle$ be the
corresponding solution to Eqs. (\ref{udot}), (\ref{uboundary0}) connecting 
$|\psi \rangle , |\omega \rangle  \in \mathcal{H}$ according to Eq. (\ref{sequenceki}).
For any short interval from $\nu$ to $\nu + \delta$ we have
\begin{equation}
  \label{update}
  |\omega_j( \nu + \delta) \rangle = [ 1 - i k_j( \nu)] |\omega_j( \nu) \rangle.
\end{equation}
By Eq. (\ref{defmk}) we then have
\begin{equation}
  \label{mupdate}
  M   |\omega_j( \nu + \delta) \rangle = [ 1 - i k^A_j( \nu)] M |\omega_j( \nu) \rangle.
\end{equation}
Eq. ( \ref{mnorm}), however, implies
\begin{equation}
  \label{mint} 
  \int_0^1 d \nu \parallel k_j( \nu) \parallel = \int_0^1 d \nu \parallel k^A_j( \nu) \parallel.
\end{equation}
Eq. (\ref{complexity}) then implies
\begin{equation}
  \label{mcomplexity}
  C( |\psi \rangle, |\omega \rangle) = C( M |\psi \rangle, M |\omega \rangle) 
\end{equation}
and therefore by Eq. ( \ref{cpsi1})
\begin{equation}
  \label{mcpsi1}
  C( |\psi \rangle) = C( M |\psi \rangle).
\end{equation}

\section{\label{sec:entangledstates} Complexity of Entangled Multi-Fermion States}

For gauge field dimension $n_U = \infty$, we introduce 
a family of
entangled multi-fermion states,
then in Appendices \ref{app:lowerbound} and \ref{app:upperbound}
prove lower and upper bounds, respectively, on their complexity.
For simplicity, the states will be built from fermions with wave functions which
are constant across cubic regions. The complexity bounds
will depend on the size of the
entangled regions. 
At the cost of additional detail, the results could
be extended to more general entangled states.
With $n_U = \infty$, Eq. (\ref{mcpsi1}) holds. We therefore construct states
directly in $\mathcal{H}^A$.

For indices $0 \le v < 2, 0 \leq i < m , 0 \leq j < n $, let $\{ D^v_{ij} \}$ be a set of 
disjoint cubic regions each with
volume $V$ in lattice units
and  $\{s^v_{ij}\}$  be a set of spins with $0 \le s^v_{ij} < 4$.
From the $\{ D^v_{ij} \} $ and $\{s^v_{ij}\}$,  define a set of fermion operators
\begin{equation}
\label{pstates}
p_i  =  
V^{-n}\prod_{vj} \left[\sum_{x \in D^v_{ij}} \Psi^{v\dagger}_{s^v_{ij}}( x) \right] .
\end{equation}
The entangled states we consider are then
\begin{equation}
  \label{entangledstate}
  |\psi \rangle  = m^{-\frac{1}{2}}\sum_i \zeta_i p_i  |\Omega \rangle
\end{equation}
for complex $\zeta_i$ with $| \zeta_i| = 1$.

The state $|\psi \rangle$ can be viewed as a version of an $n$-site $m$-way GHZ state. 

For multi-fermion entangled states of the form in Eq. (\ref{entangledstate})
with $m \ge 4$, $n \ge 2$,
we prove in Appendix \ref{app:lowerbound} a lower bound on complexity
\begin{equation}
\label{lowerb}
C( |\psi \rangle ) \geq c_0 \sqrt{m V} 
\end{equation}
with $c_0$ independent of $m, n$ and $V$.

In Appendix \ref{app:upperbound} we prove in addition
\begin{equation}
\label{upperb}
C( |\psi \rangle ) \leq c_1 \sqrt{m n V} + c_2 m n + c_3\sqrt{mn} r,
\end{equation}
where $c_1, c_2$ and $c_3$ are  independent of $ m, n$ and $V$.
The distance $r$ is given by
\begin{equation}
  \label{defsbar}
  r = \min_{x_{00}} \max_{vij} r^v_{ij}
\end{equation}
where $r^v_{ij}$ is the number of nearest
neighbor steps in the
shortest path between
lattice points $x_{ij}$ and $y^v_{ij}$
such that no pair of paths for distinct
$\{ i, j\}$ intersect,
$y^v_{ij}$ is the center point of $D^v_{ij}$
and $x_{ij}$ and is an $m \times n$ rectangular grid
of nearest neighbors in the positive $x^1$ and $x^2$ directions
with base point $x_{00}$.

Eqs. (\ref{lowerb}) and (\ref{upperb}) together suggest that asymptotially for large $V$
\begin{equation}
  \label{largev}
  C(|\psi \rangle)^2 \rightarrow c_4 V,
\end{equation}
for some constant $c_4$ independent of $V$. Eq. (\ref{largev}), in turn, is plausibly a
result of the relation
\begin{equation}
  \label{lineark}
  \parallel k_0 + k_1 \parallel^2 = \parallel k_0 \parallel^2 + \parallel k_1 \parallel^2,
\end{equation}
for operators $k_0, k_1 \in K$ localized on disjoint regions. Eq. (\ref{lineark}) can
be viewed as evidence disjoint regions tend to make additive contribution to $C(|\psi \rangle)^2$
leading to a result linear in size.

Eq. (\ref{largev}) constrains
the behavior of a possible
continuum limit of the lattice definition
of complexity.  
Assume a limit as $a \rightarrow 0$  exists for
a multiplicately 
renormalized version of $C( |\psi \rangle )$
evaluated on a $|\psi \rangle$ which is held fixed
in scaled units.
One such state is the $|\psi \rangle$ of Eq. (\ref{entangledstate})
with the regions $D^v_{ij}$ kept fixed in scaled units
and therefore $V$ of the form
\begin{equation}
  \label{rescaledv}
  V = a^{-3} \hat{V},
\end{equation}
for $\hat{V}$ fixed as $a \rightarrow 0$.
Eq. (\ref{largev}) implies
$C( |\psi \rangle )$ will have to be related to
renormalized complexity
$\hat{C}( |\psi \rangle )$ by
\begin{equation}
  \label{rescaledc}
  C( |\psi \rangle ) = a^{-\frac{3}{2}} \hat{C}( |\psi \rangle ).
\end{equation}
For renormalized complexity,
in the limit $a \rightarrow 0$, the terms in Eq. (\ref{upperb}) proportional to $c_2$ and $c_3$
will vanish.

\section{\label{sec:branching}Branching}

Using the complexity measure of Sections \ref{subsec:complexitydef} and \ref{subsec:nuinfty}
we now define a
decomposition of a state vector into a set of branches
which miminizes a measure of the aggregate complexity of
the branch decomposition.
The state vector of the real world, we will propose, follows through time
a single continuously evolving branch in the optimal decomposition.
Then at various instants the branch followed in the optimal decomposition
will split into two sub-branches. Each time a split occurs,
the real world we assume
randomly chooses one of these sub-branches according to
the Born rule.

\subsection{\label{subsec:branchcomplexity} Net Complexity of a Branch Decomposition}

For any $|\psi \rangle  \in \mathcal{H}$ let 
 $ |\psi \rangle  = \sum_i |\psi_i \rangle $
be a candidate orthogonal decomposition into branches.
We define the net complexity $Q( \{|\psi_i \rangle \})$ of this decomposition to be
\begin{equation}\label{defQ} 
Q( \{|\psi_i \rangle \})  =  \sum_i \langle \psi_i | \psi_i \rangle  [C( |\psi_i \rangle )]^2 - 
 b \sum_i \langle \psi_i | \psi_i \rangle  \ln( \langle \psi_i |\psi_i \rangle ),
\end{equation} 
with branching threshold $b > 0$. For any choice of $b$, the branch
decomposition of $|\psi \rangle $ is defined to be the $\{|\psi_i \rangle  \}$ which minimizes
$Q(\{|\psi_i \rangle  \})$. The first term in Eq. (\ref{defQ}) is the mean squared complexity
of the branches split off from $|\psi \rangle $. But each branch can also be thought
of as specifying, approximately, some macroscopic classical configuration of the
world. The second term represents the entropy of this random ensemble
of classical configurations.

Since the complexity of any state which is not an eigenvector of $N^v$
is $\infty$, each branch in a decomposition $\{|\psi_i \rangle \}$ which minimizes
$Q( \{|\psi_i \rangle \})$ will be an eigenvector of $N^v$.
The requirement that each branch be an eigenvector of $N^v$
becomes a superselection rule. Gauge invariance then
requires equal eigenvalues of $N^0$ and $N^1$.
The requirement that physical states have equal numbers
of protons and electrons thus becomes a superselection rule.
Desribed with respect to the physical vacuum,
the superselection rule is that the total number of protons minus antiprotons
in the world
must equal to total number of electrons minus positrons.

For a version of branching and complexity applied to lattice QCD,
the product states corresponding to the definition at the
end of Section \ref{subsec:hilbertspace} automatically
all have $SU(3)$ gauge field winding number of 0. It follows that
any state of finite complexity would then also
have winding number of 0.
The result is that the sequence of branches corresponding
to the physical universe will also all have
$SU(3)$ winding number of 0.
An $SU(3)$ winding number fixed at 0
implies that a CP violating term added to the
underlying Lagrangian will have no effect
and is thus nearly equivalent to
a superselection rule setting
the CP violating $\Theta$ to 0.

The quantity $Q( \{|\psi_i \rangle \})$ is nonnegative and, with
nonzero lattice spacing, there is at least
one choice of orthogonal decomposition for which
$Q( \{|\psi_i \rangle \})$ is finite.
Any gauge invariant $|\psi \rangle $  with $N^0$ and $N^1$ eigenvalues of $n$ can be expressed as a linear
combination of a finite set of product states of the form
\begin{equation}
\label{particlesatpoints}
|\{x^0_i, s^0_i\}, \{x^1_j, s^1_j \}, T \rangle  =  \prod_{i, j < n} \Psi_{s^0_i}^{0\dagger}( x^0_i) \Psi_{s^1_j}^{1\dagger}( x^1_j) T_{s_i^0s_j^1}( x^0_i, x^1_j) |\Omega \rangle ,
\end{equation}
where $T_{s^0_is^1_j}(x^0_i, x^1_j)$ is a product of $U( x, y)$
along a path from $x^0_i$ to $x^1_j$ multiplied by some function of the plaquette variables $U(p)$ such that
the product $ T_{s^0_is^1_j}(x^0_i, x^1_j) \Psi^{0\dagger}_s( w) \Psi^{1\dagger}_{s'}( z)$ is gauge invariant.

For this decomposition all $C( |\psi_i \rangle )$ are 0
and the second term in Eq. (\ref{defQ})
\begin{equation}
\label{pointstates1}
-b \sum_{ij} [ \rho_{ij}  \ln ( \rho_{ij})],
\end{equation}
is finite, where
\begin{equation}
\label{pointstates2}
\rho_{ij}  =  | \langle\{x^0_i, s^0_i\}, \{x^1_j, s^1_j \}, T | \psi \rangle |^2.
\end{equation}
Since $Q( \{|\psi_i \rangle \})$ is nonnegative and finite
for at least one $\{ |\psi_i \rangle \}$, it follows that $Q( \{|\psi_i \rangle \})$ has 
a finite minimum.
Without evidence to the contrary, we will assume without proof that
this minimum is unique and realized by some single branch
decomposition.
In Section \ref{sec:vacuum}, however,
we will argue that for branching of the physical vacuum the
minimum of $Q( \{|\psi_i \rangle \})$ is not unique
and that a kind of spontaneous symmetry breaking occurs as a result.

For $b$ either 
extremely small or extremely large, the branches which follow from Eq. (\ref{defQ}) 
will
look nothing like the macro reality we see.  For small enough $b$,
the minimum of $Q( \{|\psi_i \rangle \})$ will be dominated by the complexity term.
It follows from the discussion of Section \ref{subsec:complexitydef}
that the minimum of the complexity term will occur for a set of branches each of which is nearly
a pure, unentangled multi-particle product state. Thus bound states
will be sliced up into unrecognizable fragments. On the
other hand, for very large $b$, the minimum of $Q( \{|\psi_i \rangle \})$
will be dominated by the entropy term, leading to only 
a single branch consisting of the entire coherent quantum state. 
Again, unlike the world we see.

The result of all of which is that for the branches given by minimizing $Q( \{|\psi_i \rangle \})$
of Eq. (\ref{defQ}) to 
have any chance of matching the macro world, $b$ has to be some finite
number. The determination of $b$
will be discussed in more detail in
Section \ref{sec:vacuum}

In Section \ref{sec:entangledstates} we argued that
the results of Appendices \ref{app:lowerbound} and \ref{app:upperbound} imply
that if a continuum limit exists for the lattice definition of complexity,
the multiplicatively renomalized continuum complexity  $\hat{C}( |\psi \rangle )$
will be related to lattice complexity $C( |\psi \rangle )$
by Eq. (\ref{rescaledc}). 
For net complexity of Eq. (\ref{defQ}) to have a renormalized continuum
version,  $b$ will therefore have to be given by
\begin{equation}
  \label{rescaledb}
  b = a^{-3} \hat{b},
\end{equation}
for renomalized continuum $\hat{b}$, which will then have units of volume.

\subsection{\label{subsec:remote} Net Complexity of a Tensor Product}

The choice of $[C( |\psi_i \rangle )]^2$ in Eq. (\ref{defQ}) defining
$Q(\{|\psi_i \rangle \})$ rather than some other power of
$C( |\psi_i \rangle )$ is dictated by the plausible requirement
that branching occur independently for remote,
unentangled factors of a tensor product state.

Consider a state $|\psi \rangle $ given by the tensor product
$|\chi \rangle  \otimes |\phi \rangle $
of a pair of states
localized on regions $R_{\chi}$ and $R_{\phi}$ sufficiently distant from each other.
A candidate branch decomposition then becomes
\begin{equation}
\label{productbranches}
|\psi \rangle  = \sum_{ij} |\chi_i \rangle  \otimes |\phi_j \rangle .
\end{equation}
Eqs. (\ref{ctensor}) and (\ref{defQ}) then imply
\begin{equation}
\label{productQ}
Q( \{|\chi_i \rangle  \otimes |\phi_j\}) = 
Q( \{|\chi_i \rangle  \otimes |\Omega_\phi \rangle \}) + Q( \{|\Omega_\chi \rangle  \otimes |\phi_j \rangle \}).
\end{equation}
Thus branching of each of the remote states will occur independently unaffected
by branching of the other.
For powers of $[C( |\psi_i \rangle )]$ in Eq. (\ref{defQ} other than 2, this proof of Eq. (\ref{productQ})
would fail.

\subsection{\label{subsec:timeevolution} Time Evolution of the Optimal Branch Decomposition}

Suppose $Q(\{|\psi_i \rangle \})$ is minimized at each $t$ for
some evolving $|\psi(t) \rangle $.
The set of possible 
branch decompositions over which $Q(\{|\psi_i \rangle \})$ is
minimized can be viewed as a topological space with
topology given by the product of
the Hilbert space topology on each $|\psi_i \rangle $.
At any time $t$,
the net complexity function $Q(\{|\psi_i \rangle \})$ will then have
some set of local minima, each an absolute minimum on
a corresponding open set of branch decompositions.
The optimal
decomposition will be the global minimum over this
set of local minima. 
According to Section \ref{app:properties} for time evolution by the Hamiltonian $h$ of Eq. (\ref{hamiltonian}) with any finite $n_U$,
the complexity $C[|\psi(t) \rangle ]$ of $|\psi(t) \rangle $ and the complexity $C[|\psi_i(t) \rangle ]$ of
any branch $|\psi_i(t) \rangle $ will be continuous functions of time.
Thus the local minima of $Q(\{|\psi_i \rangle \})$ will
themselves track continuously in time.
But at a
set of isolated points in time, which of the competing
local minima is the overall global minimum can potentially change.
At such instants, the optimal decomposition
will jump discontinuously. Thus the optimal decomposition
is a piecewise continuous function of $t$.

Continuous Hamiltonian time evolution of each branch
leaves the classical entropy term in Eq. (\ref{defQ}) unchanged,
while the quantum complexity term in Eq. (\ref{defQ}) potentially changes 
during Hamiltonian time evolution, thereby causing a continuous drift in
the optimal branch configuration.
For a sufficiently large $b$, however, the classical
entropy term in Eq. (\ref{defQ}) can be made arbitrarily more important than
the quantum term.
Thus for large enough $b$, the continuous part of time evolution
will consist almost entirely of Hamiltonian time evolution of each branch.

For the discontinuous part of branch evolution,
the requirement that the $\{ |\psi_i \rangle  \}$ be an orthogonal
decomposition of $|\psi(t) \rangle $ implies that a single $|\psi_i \rangle $
cannot jump by itself.

The simplest possibile discontinuity allowed by the requirement that the
$\{ |\psi_i \rangle  \}$ be orthogonal is for some single branch $|\phi \rangle $ to split
into two pieces
\begin{equation}\label{splitphi}
|\phi \rangle  = |\phi_0 \rangle  + |\phi_1 \rangle .
\end{equation}
The terms in $Q( \{|\psi_i \rangle \})$ arising from $|\phi \rangle $ before
the split are
\begin{equation}\label{beforesplit}
 \langle \phi|\phi \rangle \{[C( |\phi \rangle )]^2 - b \ln(  \langle  \phi | \phi \rangle \}.
\end{equation}
The terms from $|\phi_0 \rangle $, $|\phi_1 \rangle $ after the split are
\begin{equation}\label{aftersplit}
 \langle \phi|\phi \rangle \{ \rho [C( |\phi_0 \rangle )]^2 + ( 1 - \rho) [C( |\phi_1 \rangle )]^2 - 
b \rho \ln( \rho) - b ( 1 - \rho) \ln( 1 - \rho)  - b \ln(  \langle  \phi| \phi \rangle ]\},
\end{equation}
for $\rho$ defined by
\begin{equation}
    \label{defofrho}
     \langle  \phi_0 | \phi_0 \rangle  = \rho  \langle  \phi | \phi \rangle .
\end{equation}
Thus a split will occur if
\begin{equation}\label{splitcondition}
[C( |\phi \rangle )]^2 - \rho [C( |\phi_0 \rangle )]^2 - ( 1 - \rho) [C( |\phi_1 \rangle )]^2 > 
-b \rho \ln( \rho) - b ( 1 - \rho) \ln( 1 - \rho).
\end{equation}
The condition for a split is a saving in average squared complexity
by an amount linear in $b$. 
Splitting occurs as soon as the required threshold 
saving in average squared complexity is crossed.

A split could also reverse itself if as a result of time evolution
the complexity of $|\phi \rangle $, $|\phi_0 \rangle $ or $|\phi_1 \rangle $ changes sufficiently to
reverse the inequality in Eq. (\ref{splitcondition}).
In general, we believe both reversible and irrevesible splits will occur.
For a system evolving through a sequence of states each with
much less than the system's maximum possible complexity,
we believe most splitting events never reverse.
In Appendix \ref{app:pairsplits} we
present an argument for the hypothesis
that for a system evolving through a sequence
of states each with much less than the system's
maximum possible complexity, branches which wander off independently into a high dimensional space almost never
recombine.
Under corresponding circumstances, other possible events merging two branches
into a single result 
we will argue are similarly improbable.
Still more complicated rearrangments at a single instant
of a set of $n$ branches and a distinct set of $n'$ with $n + n' > 3$ into a new configuration
we believe occur with zero probability.

For an example of branches which do recombine, consider
an experiment with a particle with a widely spread wave function striking a screen with a pair of holes sufficiently
far apart to insure that
separate branches arise at the two separated holes.
Interference between the two
branches might then still be observed if the experiment is constructed in such a way that
two branches are allowed to recombine.
Which in turn would be possible only
if the separated branches are not allowed to wander off separately
and become intangled with the rest of the universe.
The world experienced by an experimenter conducting such an
experiment would consist only of a single one of the two  branches
produced at the holes, but then consist of the recombined
state produced to permit interference.
Which of the branches produced at the holes
the experiment's world consists of
will be undeterminable as long as no further entanglement with
the rest of the universe is permitted.

The state vector of the real world
we propose follows through the tree of branching events a single sequence of 
branches and sub-branches, with the sub-branch at each branching
event chosen randomly according to the Born rule.
For rare cases in which branches recombine, no
selection rule is needed and the real world
simply follows the recombined branches.

Branching softens the discontinuity of $C( |\psi \rangle)$ with
respect to the inner product topology on $\mathcal{H}$. A small piece of $|\psi \rangle$
which is responsible for a sufficiently large increment in $C( |\psi \rangle)$ will split
off as a separate branch. This feature of branching makes possible the
definition of $C(|\psi \rangle , |\omega \rangle )$
from trajectories  $k( \nu)$ in Eq. (\ref{sequenceki})
which exactly span between $|\psi \rangle$ and $|\omega \rangle$
in place of the approximate version adopted in \cite{Nielsen}.

\section{\label{sec:examples} Examples of Branching}

We present two examples of branching and then, based
on the second of these,
a conjecture for the structure of states left as the
result of branching events.

\subsection{\label{subsec:scattering} Scattering Experiment}

We will apply the branching proposal of Section \ref{sec:branching} to a
model of an experiment in which a microscopic system scatters and produces a final
state recorded by a macroscopic measuring device. We assume $n_U \rightarrow \infty$.

Let $\mathcal{H}$ be the product
\begin{equation}
\label{macromicro}
\mathcal{H} = \mathcal{Q} \otimes \mathcal{R},
\end{equation}
where $\mathcal{Q}$ is the space of states of the macroscopic measuring
device and $\mathcal{R}$ is the space of states of the microscopic
system which undergoes scattering.

Assume an unentangled initial state
\begin{equation}
\label{initialstate}
|\psi^{in} \rangle  = |\psi^0_{\mathcal{Q}} \rangle  \otimes |\psi^0_{\mathcal{R}} \rangle ,
\end{equation}
for which the complexity measure $Q( |\psi^{in} \rangle )$ is already at
a minimum and cannot be reduced by splitting $|\psi^{in} \rangle $ into
orthogonal parts. For the microscopic system, this
can be accomplished by a microsopic state $|\psi^0_{\mathcal{R}} \rangle $
with probability concentrated on a
scale smaller than the branching volume $b$. The macroscopic
state $|\psi^0_{\mathcal{Q}} \rangle $ we assume spread on a scale much
larger than $b$ but without entanglement on a scale larger than $b$.
We assume the macroscopic state consists of
$n_{\mathcal{Q}}$ flavor 0 fermions and $n_{\mathcal{Q}}$ flavor 1 fermions,
the microscopic state consists of
$n_{\mathcal{R}}$ flavor 0 fermions and $n_{\mathcal{R}}$ flavor 1 fermions,
with $n_{\mathcal{Q}}$ much greater than $n_{\mathcal{R}}$

A  macroscopic state satisfying these assumptions is 
the product state of Section \ref{sec:entangledstates},
\begin{equation}
  \label{macroproductstate}
|\psi^0_{\mathcal{Q}} \rangle  = V^{-n_{\mathcal{Q}}}\prod_{vj} \left[\sum_{x \in D^v_{0j}} \Psi^{v\dagger}_{s^v_{0j}}( x) \right] |\Omega \rangle].
\end{equation}

The microscopic system then undergoes scattering which produces
a final state which is a superposition of $|\psi^i_{\mathcal{R}} \rangle $, we assume for simplicity
equally weighted, which is then detected by
the macroscopic device eventually leading to the entangled result
\begin{equation}
\label{finalstate}
|\psi^{out} \rangle  = m^{-\frac{1}{2}} \sum_{0 \le i < m} |\psi^i_{\mathcal{Q}} \rangle  \otimes |\psi^i_{\mathcal{R}} \rangle .
\end{equation}
As was the case for the initial state, the macroscopic factor $|\psi^i_{\mathcal{Q}} \rangle $ in each
of these terms we assume spread on a scale $V$ large with respect to $b$, but without
additional entanglement beyond the entanglement explicit in Eq. (\ref{finalstate}),
and the microscopic factor $|\psi^i_{\mathcal{R}} \rangle $ we assume 
spread on a scale small with respect to $b$.
We assume also the macroscopic factors for distinct $i$ are orthogonal. 
Macroscopic final states which accomplish this 
are the rest of the product states of
Section \ref{sec:entangledstates},
\begin{equation}
  \label{macroproductstate1}
|\psi^i_{\mathcal{Q}} \rangle  = V^{-n_{\mathcal{Q}}}\prod_{vj} \left[\sum_{x \in D^v_{ij}} \Psi^{v\dagger}_{s^v_{ij}}( x) \right] |\Omega \rangle.
\end{equation}

Appendix \ref{app:lowerbound} can then be adapted with almost no changes to provide a lower bound on
$C( |\psi^{out} \rangle )$ from the flavor 0 parts of the $|\psi^i_{\mathcal{Q}} \rangle$
of Eq. (\ref{macroproductstate1}). Define the sets $D^e_{ij}$ as in Appendix \ref{app:lowerbound}
and from the $D^e_{ij}$ define the sets $E_\ell$ with $2n$ replaced by $2n_{\mathcal{Q}}$.
The trajectory $k(\nu)$ determining the complexity of $|\psi^{out} \rangle$ now begins from a product state $|\omega \rangle $ with
a total of
$n_{\mathcal{Q}} + n_{\mathcal{R}}$ fermions of flavor 0 and $n_{\mathcal{Q}} + n_{\mathcal{R}}$ fermions of flavor 1. As a result, the bound on Schmidt vector
rotation angles of Eq. (\ref{thetabound}) becomes
\begin{equation}
\label{thetaboundscattering}
\int_0^1 | \theta_{\ell}(\nu)| d \nu \ge \arcsin( \sqrt{\frac{n_{\mathcal{Q}} - n_{\mathcal{R}}}{mV}}).
\end{equation}

Similarly, as a consequence of the fermion number of  $|\psi^{out} \rangle$,
the bound in Eq. (\ref{psiprojectionbound}) becomes 
\begin{equation}
\label{psiprojectionboundscattering}
\sum_{x \in D^e, y \in D^o}  \langle \omega(\nu)| [I - p(x,y)]|\omega(\nu) \rangle   \le 6(n_{\mathcal{Q}} + n_{\mathcal{R}}).
\end{equation}

The net result of these two changes is that the bound of Eq. (\ref{lowerb}) becomes
\begin{equation}
\label{cbound1}
C( |\psi^{out} \rangle ) \ge c_0 \sqrt{ \frac{mV(n_{\mathcal{Q}} - n_{\mathcal{R}})}{n_{\mathcal{Q}} + n_{\mathcal{R}}}}.
\end{equation}
For $n_{\mathcal{Q}}$ large with respect to $n_{\mathcal{R}}$, the lower bound on 
$C( |\psi^{out} \rangle )$ is unchanged from Eq. (\ref{lowerb}).

For the net complexity of $|\psi^{out} \rangle $ as a single branch we obtain
\begin{equation}
\label{psioutQ}
Q( |\psi^{out} \rangle ) \ge (c_0)^2 m V.
\end{equation}
On the other hand a decomposition of $|\psi^{out} \rangle $ 
taking each of the $m$ terms in the sum in Eq. (\ref{macroproductstate1})
as a branch and assuming
low complexity for each of the microscopic $|\psi^i_{\mathcal{R}} \rangle $ gives
\begin{equation}
\label{psioutQ1}
Q( \{(m)^{-\frac{1}{2}}|\psi^i_{\mathcal{Q}} \rangle  \otimes |\psi^i_{\mathcal{R}} \rangle  \}) = b \ln( 2 m),
\end{equation}
which will be smaller than $Q( |\psi^{out} \rangle )$ since $V$ is assumed
much larger than $b$.
Thus the final state will split into $m$ separate branches, one of which, chosen randomly
according to the Born rule, becomes the real world. 
For a more detailed model filling in the evolution from $|\psi^{in} \rangle $ to $|\psi^{out} \rangle $
the branching process would occur not in a single step
but sequentially over some short time as the entangled form of Eq. (\ref{finalstate}) emerges.

\subsection{\label{subsec:nfermionsv} Multi-Fermion System with Large Volume}

Combining the lower and upper bounds on the entangled multi-fermion states of Section \ref{sec:entangledstates}, we
will show that branching occurs if the volume occupied by the entangled
states exceed a threshold proportional to $b$.

According to Eq. (\ref{upperb}), if $|\psi \rangle $ of Eq. (\ref{entangledstate}) is split
into $r$ branches $|\psi_i \rangle $ each the sum of $\frac{m}{r}$ distinct
$ \zeta_i p_i |\Omega \rangle $,  the net complexity $Q( \{|\psi_i \rangle \})$ will be
bounded by
\begin{equation}
  \label{upperbranch}
  Q( \{|\psi_i \rangle \}) \le \frac{ c_1^2 mnV}{r} + b \ln r,
\end{equation}
where, for simplicty, we assume $mV$ sufficiently large that the $c_2$ and $c_3$
terms in Eq. (\ref{upperb}) can be dropped. The minimum of the bound in
Eq. (\ref{upperbranch}) occurs at
\begin{equation}
  \label{optimalp}
  r = \frac{c_1^2 m nV}{b},
\end{equation}
for which value Eq. (\ref{upperbranch}) becomes
\begin{equation}
  \label{optimalbound}
  Q( \{|\psi_i \rangle \}) \le b + b \ln(\frac{c_1^2 m n V}{b}).
\end{equation}

On the other hand, according to Eq. (\ref{lowerb}), if $|\psi \rangle $ is not 
split into branches
\begin{equation}
  \label{lowerbranch}
  Q( \{ |\psi \rangle  \}) \ge c_0^2 m V.
\end{equation}
Eqs. (\ref{optimalbound}) and (\ref{lowerbranch}) imply
the branch configuration $\{ |\psi_i \rangle  \}$ for
$r$ of Eq. (\ref{optimalp}) will have lower
net complexity than $|\psi \rangle $ left unsplit if
\begin{equation}
  \label{splitcondition8}
  m V  \ge s b,
\end{equation}
where $s$ is the solution to
\begin{equation}
  \label{eqfork}
  c_0^2 s = 1 + \ln ( c_1^2 n s).
\end{equation}
There may be some set of branches with
complexity still lower than $\{ |\psi_i \rangle \}$, 
but $|\psi \rangle $ left unsplit will not be the minimum.

\section{\label{sec:residual} Residual Entanglement}

 Let $|\psi \rangle $ be a branch left after a branching event and not itself
immediately subject to futher branching, and define $|\hat{\psi} \rangle$ to be
\begin{equation}
  \label{defhatpsi}
  |\hat{\psi} \rangle = M |\psi \rangle,
\end{equation}
where  $M$ is the operator of Section \ref{subsec:nuinfty}
replacing all occurance of each link field $U(x,y)$ by 1.
The example of Sections \ref{subsec:nfermionsv}
supports the hypothesis 
that entanglement in $|\hat{\psi} \rangle$ will extend only over
a finite range.

A more precise version of this hypothesis is as follows.
Assume that the limit $2 B \rightarrow \infty$ has been
taken of the number of lattice steps in the edge
of the cubic lattice $L$ of Section \ref{subsec:hilbertspace}, or
alternatively, that $2 B$ is much larger than any of the
lengths, in lattice units, that occur in the following.

Let $q$ and $r$ be disjoint convex regions of $L$
separated by a distance $d$. Let $s$ be the complement of $q \cup r$.
Define the spaces $\mathcal{Q}, \mathcal{R}$ and $\mathcal{S}$
to be
\begin{subequations}
  \begin{eqnarray}
    \label{defq3}
    \mathcal{Q} & = & \bigotimes_{x \in q} \mathcal{H}_x \bigotimes_{x,y \in q} \mathcal{H}_{xy},\\
    \label{defr3}
    \mathcal{R} & = & \bigotimes_{x \in r} \mathcal{H}_x \bigotimes_{x,y \in r} \mathcal{H}_{xy}, \\
    \label{defs3}
    \mathcal{S} & = & \bigotimes_{x \in s} \mathcal{H}_x \bigotimes_{(x \in s) \vee (y \in s)} \mathcal{H}_{xy},
  \end{eqnarray}
\end{subequations}
so that the full Hilbert space $\mathcal{H}$ is then
\begin{equation}
  \label{hyetagain}
  \mathcal{H} = \mathcal{Q} \otimes \mathcal{R} \otimes \mathcal{S}.
\end{equation}

Define the corresponding reduced versions of $|\hat{\psi} \rangle $ to be
\begin{subequations}
  \begin{eqnarray}
    \label{reducedq}
    \rho_{\mathcal{Q}}  & = & \mathrm{Tr}_{\mathcal{S} \otimes \mathcal{R}} |\hat{\psi} \rangle  \otimes \langle \hat{\psi} | \\
    \label{reducedr}
    \rho_{\mathcal{R}}  & = & \mathrm{Tr}_{\mathcal{S} \otimes \mathcal{Q}} |\hat{\psi} \rangle  \otimes \langle \hat{\psi} |, \\
    \label{reduceds}
    \rho_{\mathcal{Q} \, \mathcal{R}}  & = & \mathrm{Tr}_{\mathcal{S}} |\hat{\psi} \rangle  \otimes \langle \hat{\psi} |.
  \end{eqnarray}
  \end{subequations}
The hypothesis we propose is that if $d$ is made large, the
entanglement in $|\hat{\psi} \rangle $ across region $
\mathcal{S}$
becomes small and $\rho_{\mathcal{Q} \, \mathcal{R}}$ becomes the tensor product
\begin{equation}
    \label{schmidt01}
    \rho_{\mathcal{Q} \, \mathcal{R}}   \approx  \rho_{\mathcal{Q}}  \otimes \rho_{\mathcal{R}}. 
\end{equation}

\section{\label{sec:covariant} Lorentz Covariant Branching}

The definitions of complexity and branching 
in Sections \ref{subsec:complexitydef} and \ref{sec:branching}
have a lattice approximation to translational and rotational covariance
but are not covariant under Lorentz boosts.
We now propose a way of extracting from these definitions
a formulation of branching  with full Lorentz covariance.

An immediate problem with potential Lorentz covariance of the
branches found by minimizing $Q(\{|\psi_i \rangle \})$ in Eq. (\ref{defQ})
is that the underlying definition of complexity is based
on hyperplanes of fixed $t$, which are themselves not Lorentz
invariant.
To address this problem, in \cite{Weingarten} branching
on constant $t$ hyperplanes is reformulated for
boost invariant hyperboloids of constant
proper time $\tau$.
Hyperboloids of constant $\tau$, however, are
not translationally invariant.  

The loss of translational invariance
shows itself as
a variant of the problem exposed by the EPR experiment.
This difficulty is a general problem for any formulation 
of branches as the substance of reality \cite{Zeh, Zurek, Zurek1, Zurek2,  Wallace, Riedel}.

Consider some branch
viewed in two different frames related by a translation.
For some period of proper time assume the branch's
representation 
in each frame remain related by a translation.
But then in a pair of disjoint regions with
spacelike separation, suppose processes occur each of which,
by itself, is sufficient to cause splitting of the branch
the two processes share. Assume in addition, that in one
frame one of these events occurs at smaller $\tau$ but 
in the other frame, as a consequence of the 
of the regions' spacelike separation, the other event occurs at smaller $\tau$.
The result will be that in the proper time interval between the
events the branch structure seen by the two different 
frames will be different. But our goal is to be able
to interpret branch state vectors as the underlying 
substance of reality. That interpretation
fails if branch structure is different
according to different reference frames.

For any pair of distinct frames, however,
for any pair of spacelike separated events 
each capable of causing a branch to split,
there is some proper time sufficiently late
that splitting will have been completed in both frames.
Correspondingly,
an argument is presented in \cite{Weingarten}
in support of the hypothesis that 
the branches found 
on a hyperboloid of fixed $\tau$
approach translational covariance as 
$\tau \rightarrow \infty$.
For progressively larger $\tau$, a hyperboloid with fixed $\tau$
becomes nearly flat over progressively larger distances.
But according to the conjecture of Section \ref{sec:residual},
entanglement and therefore branching has a fixed range
determined by $b$. The result is that for 
$\tau \rightarrow \infty$ branching defined
on a hyperbeloid of fixed $\tau$ is expected to recover
translational covariance.
We refer the reader to \cite{Weingarten} for the details.
A related proposal, in a somewhat different setting,
is considered in \cite{Kent, Kent1, Kent2}.

A corrolary of the result in \cite{Weingarten},
which we now borrow,
is that
the $t \rightarrow \infty$ limit of branching on
a hypersurface of fixed $t$ yields the same
Lorentz and translationally covariant result found from
the $\tau \rightarrow \infty$
limit of branching on a hyperbeloid of fixed $\tau$.

Macroscopic reality we then assume is a single random draw
from the set of branches at asymptotically late $t$.
For this purpose it is convenient to simplify slightly the
discussion of Section \ref{subsec:timeevolution},
by assuming the
the evolving state vector of the real world is
chosen only from the tree of persistent branching events.
These we 
defined to be
those branching events in which the resulting
pair of branches do not recombine and in which none of the descendents
of one of the branches ever recombines with a descendent of the other branch.
If the branches which make up macroscopic reality
are persistent once formed, a random
choice among the accumulated set of branches
at late $t$ is equivalent to the sequence
of choices of Section \ref{sec:branching}
but with the bookkeeping for the choice process performed all 
at once rather than in sequential steps.
The real world at finite $t$ in any particular frame
can be recovered from the asymptotic late $t$ choice
by transforming to that frame then tracing back
in time through the branching tree.

The indirect relation between branches found
as $t \rightarrow \infty$ and branches
found at finite time
is qualitatively similar to the indirect relation
between a final out scattering state and
a Schroedinger representation state
at some finite time.

Construction of the limit $t \rightarrow \infty$
and recovery of the real world at finite $t$
taken almost without change from \cite{Weingarten}
appear in Appendix \ref{app:lorentz}

\section{\label{sec:rare} Experience}

Central to our proposal is that the macroscopic reality experienced by each person is
a branch of the world's state vector. From which follow several consequences and
an additional hypothesis.

In a more complete version of
the experiment of Section \ref{subsec:scattering}
extended to include a human observer, each branch of the final state would
include a configuration of the observer's brain. 
The brain configuration in each branch
would then give rise to the experience
of observing a particular experimental result.
Since there is no human experience corresponding
to a state superposing two macroscopically distinct
experimental results, a constraint on the value
of the branching paramenter $b$ is that
the scattering experiment's final state
should yield no branch
which is a superposition of two macroscopically distinct experimental results.

Similarly, for a thought process disconnected from
observation of the world leading to a decision between
a pair of alternatives, there would be a distinct brain
configuration associated with each of the alternatives.
The occurance of one or the other of these configurations
at the end of the thought process
is experienced as having made a decision.
But there is no experience corresponding
to a state superposing at a single time
a pair of distinct results of a decision. Thus
a further constraint on $b$ is that
brain states corresponding to distinct
decision results should never occur
superoposed in a single branch.

To prevent the occurance of branches superposing distinct decision
results, a plausible guess is that $b$ should be no larger
than the smallest brain volume corresponding to the result
of a decision. 
The cerebral cortex is $\mathcal{O} (10^{-3} \si{m})$ thick
so an upper bound for $b$, which has volume units, might be
$ \mathcal{O} (10^{-9} \si{m}^3)$.

Left unanswered in all of this is what exactly
is the experience component of the
proposal that macroscopic reality experienced
by each person is a branch of the world's state vector.
What does human experience consist of?
Or for that matter experience
as presumably had by many, possibly all, other living things?
Lucid discussions of the difficulty of accounting for the occurance of
experience, human or otherwise, are presented in \cite{Nagel}
and \cite{Chalmers}.
Given the relation of experience to branches, however,
a simple hypothesis here is that each branching event is itself
an experience. Experience occurs not as an additional
layer on top of a branching event but instead
is exactly the branching event itself.
A person's experience at any instant
is an aggregate of the most recent branching events
in that person's neurological circuitry.
Human free will is the random choice process
of subsequent branch at each branching event.

A corollary to all of which would be
that all branching events in the universe,
both in living things and not in living things,
also constitute experiences.
The hypothesis that animals have experiences
is sufficiently similar to the cetainty each person has
about the occurance of their own experience to
seem at least plausible.
But the possibility
that branching in things that are not alive
also constitute experience
seems pretty much impossible to imagine.
Experience however is irreducably private.
There is no reason to assume that
experience outside living things
should in any way be sufficiently similar
to human experience to make it imaginable.
In particular, the private experience of a machine which
passes the Turning test with hardware nothing
like a human brain would be correspondingly
unlike human private experience.

The critically peculiar feature of this proposal, perhaps worth underlining, is the
notion that experience
is itself a physical object.

In any case, 
there is no way
either to confirm or to refute the hypothesis
that all branching events outside living things also constitute experiences.
It is supported only by its natural
role in the framework
supporting the hypothesis that branching
is the origin of macrosopic reality.

The account of the time evolution of branching for a system
with less than maximal complexity
presented in Section \ref{subsec:timeevolution}
leaves open the possibility that in some rare set of cases distinct branches
might still recombine. 
Macroscopic reality we propose is formed from
branches which persist.
But if all branching events constitute an experience,
what is the experience associated with events yielding branches that do not persist?
Once again, presumably something entirely
unlike any normal human experience
and correspondingly impossible to imagine.

\section{\label{sec:vacuum} Vacuum Branching}

The physical vacuum, the state of lowest energy, we now argue is itself expected to undergo branching.
The branches which result have disrupted configurations of virtual
fermion-antifermion pairs and energy density greater than the energy density of the unbranched vacuum as a consequence.
While the
physical vacuum is Lorentz and translationally invariant, the states it branches into, as a consequence of their nonzero
energy density, cannot be Lorentz invariant. Lorentz invariance will instead be spontaneously broken.
Based on the conjecture of Section \ref{sec:residual}, 
we expect the entanglement removed from the vacuum
as a consequence of branching to leave behind
a set of possible branches each consisting approximately of a tensor product of
localized factors.
But any such product of localized factors will fail to be translationally invariant.
Translational invariance we therefore expect to be sponaneously broken in addition.
In an expanding universe,
vacuum branching 
leads to the appearance of both dark energy and dark matter terms in the evolution of the universe.
The hypothesis that vacuum branching
is the origin of the observed dark energy and dark matter densities
leads to an estimate of $\mathcal{O}(10^{-18} \si{m}^3)$ for the parameter $b$, consistent
with the upper bound  $ \mathcal{O} (10^{-9} \si{m}^3)$ based on the proposed relation between
branching and experience.

Lorentz invariance of the branching process, according to the discussion of Secton \ref{sec:covariant}, is realized in the limit of $t \rightarrow \infty$. For which at the price of additional technical baggage, the present discussion could be reformulated. Once Lorentz invariance has been spontaneously broken, however, the remaining process we proposed now to
approximate on a slice with some large fixed $t$. 

For the present discussion, it is convenient to replace the free boundary conditions assumed
in the definition of QED in Section \ref{subsec:hilbertspace} with periodic boundary condition.
Following \cite{Creutz}, the Hamiltonian $h$
of Eq. (\ref{hamiltonian}), for its action on gauge invariant states, will then be reexpressed in a form
with no appearence of the longitudinal pieces of $A(x,y)$ and $E( x, y)$.
The reexpression will depend, in part, on the definition of a new
set of fermion fields with the longitudinal pieces of $A(x,y)$ built in.
With $h$ rexpressed without the longitudinal pieces of $A(x,y)$ and $E( x, y)$,
the physical vacuum and an estimate of the states it branches into
can then be found by
a perturbation expansion in the electric charge $e$.
We assume the lattice spacing is large enough for this expansion
to be well-behaved but smaller than the length scale relevant to branching.
What might happen in the limit of zero lattice spacing is beyond the
scope of the present discussion.

While a fully satisfactory quantum theory of gravity has notoriously yet to be found,
a reasonable assumption \cite{Donoghue} is that any such theory
will imply a superposed state of matter with
different energy density tensors in each component
will becomes entangled with a corresponding superpostion of
states of the universe each with a different state
of curvature and time evolution in each component
and therefore a different record in each
component of any device detecting expansion of the universe.
If this hypothesis is correct,
the corresponding entanglement of vacuum
branches with the rest of the universe and thus
any device measuring the evolution of the universe,
will result in vacuum branches which 
with highly probablity will not recombine.

\subsection{\label{subsec:htrans} Transverse Hamiltonian}

The Fourier transform $\tilde{A}_j( k)$ of $A( x, y)$, related to $U( x, y)$ by Eq. (\ref{ua}),
we define by
\begin{equation}
  \label{tildea}
  A( x + \frac{1}{2} \hat{j}, x - \frac{1}{2} \hat{j}) = \frac{1}{\sqrt{8 B^3}} \sum_k \exp( i k \cdot x) \tilde{A}_j( k),
\end{equation}
where the sum is over momenta $k$ determined by the lattice periodicity $2B$ in each direction and
$\hat{j}$ is a unit lattice vector in the $j$ direction. Divide $\tilde{A}_j( k)$ into longitudinal
and transverse components by
\begin{subequations}
  \begin{eqnarray}
    \label{avariant}
        \tilde{A}( k) & = & \frac{ \sum_i [\sin( \frac{1}{2} k_i) \tilde{A}_i( k)]} { \sum_j [\sin( \frac{1}{2} k_j)]^2}, \\
    \label{along}
    \tilde{A}^L_j( k) & = & \sin( \frac{1}{2} k_j) \tilde{A}( k), \\
        \label{atrans}
        \tilde{A}^T_j( k) & = & \tilde{A}_j( k) - \tilde{A}^L_j( k).
  \end{eqnarray}
\end{subequations}
Since $A( x, y)$ is Hermitian, the Fourier transforms fulfill
\begin{subequations}
  \begin{eqnarray}
    \label{astar}
    \tilde{A} (k) & = & -\tilde{A}^\dagger( -k), \\
    \label{alstar}
    \tilde{A}_j^L (k) & = & \tilde{A}_j^{L\dagger}( -k), \\
    \label{atstar}
    \tilde{A}_j^T (k) & = & \tilde{A}_j^{T\dagger}( -k).
  \end{eqnarray}
\end{subequations}
If we recover from $\tilde{A}( k)$ and $ \tilde{A}^L_j( k)$ the fields
$A( x)$ and $A^L( x, y)$
\begin{subequations}
  \begin{eqnarray}
  \label{ALxy}
  A^L( x + \frac{1}{2} \hat{j}, x - \frac{1}{2} \hat{j}) &=& \frac{1}{\sqrt{8 B^3}} \sum_k \exp(i k \cdot x) \tilde{A}^L_j( k), \\
  A( x) &=& \frac{1}{\sqrt{8 B^3}} \sum_k \exp(i k \cdot x) \tilde{A}( k),
  \end{eqnarray}
\end{subequations}
we then have
\begin{equation}
  \label{dela}
 A( x + \frac{1}{2} \hat{j}) - A( x - \frac{1}{2} \hat{j}) = 2 i   A^L( x + \frac{1}{2} \hat{j}, x - \frac{1}{2} \hat{j}).
\end{equation}

Define the gauge invariant $A^T( x, y), U^T( x, y), \Upsilon_s^0(x) $ and $\Upsilon_s^1(x)$ by
\begin{subequations}
  \begin{eqnarray}
  \label{ATxy}
  A^T( x + \frac{1}{2} \hat{j}, x - \frac{1}{2} \hat{j}) &=& \frac{1}{\sqrt{8 B^3}} \sum_k \exp(i k \cdot x) \tilde{A}^T_j( k), \\
  \label{UTxy}
  U^T( x + \frac{1}{2} \hat{j}, x - \frac{1}{2} \hat{j}) &=& \exp[ i e A^T( x + \frac{1}{2} \hat{j}, x - \frac{1}{2} \hat{j})], \\
  \label{chi0x}
  \Upsilon_s^0(x) & = & \exp[ -\frac{1}{2} A( x )] \Psi_s^0(x), \\
  \label{chi1x}
  \Upsilon_s^1(x) & = & \exp[ \frac{1}{2} A( x )] \Psi_s^1(x).
  \end{eqnarray}
\end{subequations}
The fields $ \Upsilon_s^0(x) $ and $\Upsilon_s^1(x)$ are operators for protons and
electrons, respectively, dressed with their Coulomb fields built out of
longitudinal photons.

By Eqs. (\ref{chi0x})) and (\ref{chi1x}), the first three terms in $h$ of Eq. (\ref{hamiltonian}  can be rewritten
\begin{multline}
    \label{hamiltoniant}
  h_\Upsilon = \frac{1}{2} \sum_{xy} \bar{\Upsilon}^0( x) \exp[ -\frac{1}{2} A( x )] \gamma( x, y) U( x, y)\exp[\frac{1}{2} A( y )] \Upsilon^0( y) + \\ \frac{1}{2} \sum_{xy}\bar{\Upsilon}^1( x)\exp[ \frac{1}{2} A( x )] \gamma( x, y) U^{\dagger}( x, y)\exp[-\frac{1}{2} A( y )] \Upsilon^1( y) + \\
  \sum_x (3 + m^v) \bar{\Upsilon}^0(x) \exp[ -\frac{1}{2} A( x )] \exp[ \frac{1}{2} A( x )] \Upsilon^0(x) + \\
  \sum_x (3 + m^v) \bar{\Upsilon}^1(x)\exp[ \frac{1}{2} A( x )] \exp[ -\frac{1}{2} A( x )] \Upsilon^1(x)
\end{multline}
and therefore by Eqs. (\ref{atrans}), (\ref{dela}) and (\ref{UTxy}
\begin{multline}
  \label{hamiltoniant1}
  h_\Upsilon = \frac{1}{2} \sum_{xy} \bar{\Upsilon}^0( x) \gamma( x, y) U^T( x, y) \Upsilon^0( y) + \frac{1}{2} \sum_{xy}\bar{\Upsilon}^1( x) \gamma( x, y) U^{T\dagger}( x, y) \Upsilon^1( y) + \\
   \sum_{x v} (3 + m^v) \bar{\Upsilon}^v(x) \Upsilon^v(x). 
\end{multline}

Define the fields $\tilde{E}_j ( k), \tilde{E}(k), \tilde{E}^L_j(k)$ and $\tilde{E}^T_j(k)$ by
Eqs. (\ref{tildea}) - (\ref{atrans}) with $E( x, y)$ in place of $A( x, y)$.
All physical states are in the gauge invariant subspace of $\mathcal{H}$ defined
by the requirement
\begin{equation}
  \label{gaugegen1}
  \sum_y E( x, y) + e \Psi^{0 \dagger}(x) \Psi^{0}(x) - e \Psi^{1 \dagger}(x) \Psi^{1}(x) = 0
\end{equation}
for all $x$. The Fourier transform of Eq. (\ref{gaugegen1}) then gives
\begin{equation}
  \label{gaugegen2}
  \tilde{E}^L_i(k) =  \frac{ ie \sin( \frac{1}{2} k_i) } {2\sqrt{ 8 B^3} \sum_j [\sin( \frac{1}{2} k_j)]^2}
  \sum_x  \exp ( -i k \cdot x) [ \Psi^{0 \dagger}(x) \Psi^{0}(x) - e \Psi^{1 \dagger}(x) \Psi^{1}(x)].
\end{equation}
The electric field term in $h$ with $E^L( x, y)$ reexpressed using Eq. (\ref{gaugegen2}) becomes
\begin{subequations}
  \begin{eqnarray}
    \label{helectric}
    h_E & = & h_E^c + \frac{1}{2} \sum_{xy} [E^T( x, y)]^2, \\
    \label{helectricc}
    h^c_E & = & \frac{e^2}{2} \sum_{xy} c( x - y) [ \Psi^{0 \dagger}(x) \Psi^{0}(x) - e \Psi^{1 \dagger}(x) \Psi^{1}(x)] [ \Psi^{0 \dagger}(y) \Psi^{0}(y) - e \Psi^{1 \dagger}(y) \Psi^{1}(y)], \; \; \;
  \end{eqnarray}
\end{subequations}
where $c( x - y)$ is the Coulomb potential
\begin{equation}
  \label{coulomb}
  c( z) = \sum_k  \frac{\exp( i k \cdot z) } {4 \sum_j [\sin( \frac{1}{2} k_j)]^2}.
\end{equation}

For any plaquette $\{u, v, x, y\}$ of 4 nearest neighbors, define the transverse plaquette field $U^T( p)$ as
\begin{equation}
  \label{plaquettet}
  U^T( p) = U^T( u, v) U^T( v, x) U^T( x, y) U^T( y, u).
\end{equation}
Eq. (\ref{dela}) then implies the plaquette field $U(p)$ of Eq. (\ref{plaquette}) and
the transverse plaquette field $U^T(p)$ are equal.
The plaquette term in $h$ becomes
\begin{equation}
  \label{hplaquette}
  h_U =  \frac{1}{2e^2}\sum_p[1- U^T( p)].
\end{equation}

\subsection{\label{subsec:perturbation} Perturbation Expansion $\mathcal{O}(e^0)$ }

In the limit $e \rightarrow 0$ the  Hamiltonian $h$ becomes
\begin{equation}
  \label{h0}
  h^0 = h^0_\Upsilon + h^0_E + h^0_U + \xi^0_\Omega,
\end{equation}
where $h^0_\Psi, h^0_E, h^0_U$ are
\begin{subequations}
  \begin{eqnarray}
  \label{hamiltoniant0}
  h^0_\Upsilon &=& \frac{1}{2}\sum_{xy} \bar{\Upsilon}^0( x) \gamma( x, y) \Upsilon^0( y) +\frac{1}{2}\sum_{xy}\bar{\Upsilon}^1( x) \gamma( x, y) \Upsilon^1( y) +
    \sum_{x v} (3 + m^v) \bar{\Upsilon}^v(x) \Upsilon^v(x), \;\;\;\;\;\;\;\; \\
  \label{helectric0}
  h^0_E &=& \frac{1}{2}\sum_{xy} [E^T( x, y)]^2, \\
  \label{hplaquette0}
  h^0_U &=& \frac{1}{2} \sum_p [A^T( p)]^2,
  \end{eqnarray}
  \end{subequations}
for the plaquette field $A^T( p)$ defined by
\begin{equation}
  \label{aplaquette}
  U^T( p) = \exp[ i e A^T( p)].
\end{equation}

Expanded in annihilation operators for eigenvectors of $h^0$, the field operators
$\Upsilon^v_s( x)$ become
\begin{equation}
  \label{freepsi}
  \Upsilon^v_s( x) = \frac{1}{\sqrt{8B^3}}\sum_{k j} a^{vj}( k) u^{vj}_s( k) \exp( i k \cdot x),
\end{equation}
where the sum is over  momentum components $k^i$ determined by
the lattice period of $2B$ in each direction and spin states $0 \le j < 4$. The annihilation and creation operators, $a^{vj}, a^{vj\dagger}$,
are normalized to 
\begin{equation}
  \label{annihilation}
  \{ a^{vj}( k), a^{v'j'\dagger}( k') \} = \delta_{vv'} \delta_{jj'} \delta_{kk'},
\end{equation}
and the spinors complex number valued $u^{vj}_s( k)$ satisfy the lattice Dirac equation
\begin{equation}
  \label{diraceq}
  \Bigl[\gamma^0 k^{0vj} - \sum_i \sin( k^i) \gamma^i   -  m^v - 2 \sum_i \sin^2( \frac{k^i}{2}) \Bigr] u^{vj}(k) = 0,
\end{equation}
where $k^{0vj}$ is fermion energy, negative for spin states
$j = 0, 1,$ and positive for $j = 2, 3,$ obtained from
\begin{equation}
  \label{k0}
  [k^{0vj}]^2 - \sum_i \sin^2(k^i) - [ m^v + \sum_i 2 \sin^2( \frac{ k^i}{2})]^2 = 0.
\end{equation}
The $u^{vj}_s$ are normalized to
\begin{equation}
  \label{unorm}
  \sum_s u^{vj*}_s(k) u^{vj'}_s(k) = \delta_{jj'},
\end{equation}
thereby insuring Eq. (\ref{annihilation}).

Expanded in annihilation and creation operators
for eigenvectors of $h^0$, the transverse field $\tilde{A}^T_j( k)$ of Eq. (\ref{atrans}) and
the corresponding transverse $\tilde{E}^T_j( k)$ are given by
\begin{subequations}
\begin{eqnarray}
  \label{freeatrans}
  \tilde{A}^T_j( k) &=& \frac{1}{2 [\sum_m \sin^2( \frac{1}{2} k^m)]^{\frac{1}{4}}} \sum_{\ell = 0, 1} [\epsilon^\ell_j( k) b^\ell( k) + \epsilon^\ell_j( -k) b^{\ell \dagger}( -k)],\\
  \label{freeatrans1}
  \tilde{E}^T_j( k) &=& -i[\sum_m \sin^2( \frac{1}{2} k^m)]^{\frac{1}{4}} \sum_{\ell = 0, 1} [\epsilon^\ell_j( k) b^\ell( k) - \epsilon^\ell_j( -k) b^{\ell \dagger}( -k)],
\end{eqnarray}
\end{subequations}
where $b^\ell( k)$ and $b^{\ell \dagger}( k)$ have
\begin{equation}
  \label{aell}
        [ b^\ell( k), b^{\ell' \dagger}( k')] = \delta_{ \ell \ell'} \delta_{ k k'},
\end{equation}
and the polarization vectors $\epsilon^\ell_j( k),\ell = 0, 1,$ satisfy
\begin{equation}
  \label{epsnorm}
  \sum_\ell \epsilon^\ell_j( k) \epsilon^\ell_{j'}( k) = \delta_{jj'} - \frac{\sin( \frac{1}{2} k^j) \sin( \frac{1}{2} k^{j'})}{\sum_i \sin^2( \frac{1}{2} k^i)}.
  \end{equation}
The $b^{\ell \dagger}( k)$ create states with energy eigenvalue $k^0$
\begin{equation}
  \label{k0a}
  k^{02} = 4 \sum_i \sin^2( \frac{1}{2} k^i).
\end{equation}

The physical vacuum, defined to be the lowest energy state, to order
$\mathcal{O}( e^0)$ is given by the action on the bare vacuum of all negative energy $a^{vj\dagger}(k)$  
\begin{equation}
  \label{physvac}
  |\Omega_{phys}^0 \rangle = \prod_{vk}\prod_{j=0,1} a^{vj\dagger}( k) |\Omega \rangle.
\end{equation}
The $a^{vj\dagger}( k)$ are gauge invariant, so $|\Omega_{phys}^0 \rangle$
is gauge invariant as required of physical states. According to Eq. (\ref{productstate2}),
$|\Omega_{phys}^0 \rangle$ is also as a product state, so has complexity 0
and will not branch.  No transverse photons are present 
in $|\Omega_{phys}^0 \rangle$ since all would contribute some positive
increment of energy. Longitudinal components of the $A(x,y)$ are
present, however, built into the $a^{vj\dagger}( k)$. The vacuum energy
renormalization constant $\xi_\Omega^0$ is chosen to cancel out
the negative energy contributions of the $a^{vj\dagger}(k)$ and photon 0-point
energies and assign
$|\Omega_{phys}^0 \rangle$ an $h^0$ eigenvalue of 0.

\subsection{\label{subsec:perturbation1} Perturbation Expansion $\mathcal{O}(e)$, $\mathcal{O}(e^2)$ }

We now expand $h$ as a power series in $e$, define $h^1$ to be the resulting $\mathcal{O}(e)$ terms,
\begin{equation}
  \label{expandh}
  h = h^0 + h^1 + \mathcal{O}(e^2),
\end{equation}
and consider the corrections to  $|\Omega_{phys}^0 \rangle$ and to $\xi_\Omega^0$ arising from $h^1$.
For $h^1$ we obtain
\begin{equation}
  \label{h1}
  h^1 = -\frac{e}{2}\sum_{x y} \bar{\Upsilon}^0(x)  A^T( x, y) \hat{\gamma}(x,y) \Upsilon^0( y) +
  \frac{e}{2}\sum_{x y} \bar{\Upsilon}^1(x) A^T( x, y) \hat{\gamma}(x,y) \Upsilon^1( y) ,
\end{equation}
where $\hat{\gamma}(x,y)$ is 
  \begin{subequations}
  \begin{eqnarray}
  \label{gamma1h}
 \hat{ \gamma}( x + \hat{j}, x) & = & \gamma^j, \\
  \label{gamma2h}
 \hat{ \gamma}(x, x + \hat{j}) & = & \gamma^j,
\end{eqnarray}
  \end{subequations}
  and for $\{x,y\}$ not nearest neighbors
  \begin{equation}
    \label{gamma3h}
    \hat{ \gamma}( x, y) = 0.
    \end{equation}

  Since $|\Omega^0_{phys} \rangle$ has all negative energy fermion states filled and all positive
  energy states empty, the corresponding $\mathcal{O}(e)$ shift in vacuum
energy is 0
\begin{equation}
  \label{xi1}
  \xi_\Omega^1 = \langle \Omega^0_{phys} | h^1 |\Omega^0_{phys} \rangle = 0.
\end{equation}

The nonvanishing terms in the $\mathcal{O}(e)$ correction to $|\Omega^0_{phys} \rangle$
are
\begin{equation}
  \label{omega1}
  |\Omega^1_{phys} \rangle = \sum_{k_1} \sum_{j_1= 2,3} \sum_{k_2} \sum_{j_2 = 0,1} \sum_{\ell v} \frac{|k_1,j_1,k_2,j_2,v, \ell \rangle
    \langle k_1,j_1,k_2,j_2,v, \ell | h^1 | \Omega^0_{phys} \rangle}{ k_2^0 - k_1^0 - k^0_3},
\end{equation}
where $k_1^0$ is the energy of a momentum $k_1$, spin state $j_1$, flavor $v$ fermion,
$k_2^0$ is the energy of a momentum $k_2$, spin state $j_2$, flavor $v$ fermion,
$k^0_3$ is the energy of a momentum $k_2 - k_1$, polarization $\ell$ photon,
and $|k_1,j_1,k_2,j_2,v, \ell \rangle$ is $|\Omega^0_{phys} \rangle$ with a $k_2, j_2$ fermion
removed and a $k_2 - k_1, \ell$ photon and $k_1, j_1$ fermion added
\begin{equation}
  \label{vacuumplus}
  |k_1,j_1,k_2,j_2,v, \ell \rangle = a^{v{j_1}\dagger}( k_1) b^{\ell \dagger}(k_2 - k_1) a^{v{j_2}}( k_2) |\Omega^0_{phys} \rangle .
\end{equation}
The denominator of the fraction in Eq. (\ref{omega1}) is strictly negative since
the fermion energy $k_2^0$ is strictly negative for spin states $j_2 = 0, 1$.

Since $h^1$ preserves fermion number and total momentum, the nonvanishing terms in Eq. (\ref{omega1})
consist of
adding some fermion state to $|\Omega^0_{phys} \rangle$, removing another fermion state, and
balancing the change in momentum by adding a photon. Since $|\Omega^0_{phys} \rangle$ already includes
all states with $j = 0, 1$ and no states with $j = 2, 3$, the fermion removed from $|\Omega^0_{phys} \rangle$
can have only $j = 0, 1$ and the fermion added can have only $j = 2, 3$. The
sum in Eq. (\ref{omega1}) is over all such terms.
In more familiar language, $|\Omega^1_{phys} \rangle$ is
a sum of virtual fermion-antifermion pairs.

The $\mathcal{O}(e)$ correction $|\Omega^1_{phys} \rangle$ to $|\Omega^0_{phys} \rangle$ then contributes
an $\mathcal{O}(e^2)$ correction $\xi^2_\Omega$ to $\xi^0_\Omega$
\begin{subequations}
  \begin{eqnarray}
    \label{xi2}
    \xi^2_\Omega & = & \langle \Omega^0_{phys}| h^1 | \Omega^1_{phys} \rangle, \\
    \label{xi3}
  & = &  \sum_{k_1} \sum_{j_1= 2,3} \sum_{k_2} \sum_{j_2 = 0,1} \sum_{\ell v} \frac{\langle \Omega^0_{phys} | h^1|k_1,j_1,k_2,j_2,v, \ell \rangle
    \langle k_1,j_1,k_2,j_2,v, \ell | h^1 | \Omega^0_{phys} \rangle}{ k_2^0 - k_1^0 - k^0_3}.
  \end{eqnarray}
\end{subequations}
The fraction in Eq. (\ref{xi3}) has a strictly positive numerator and strictly negative
denominator so that $\xi^2_\Omega$ is strictly negative. The shift $\xi^2_\Omega$ has
to be negative since $|\Omega^0_{phys} \rangle + |\Omega^1_{phys} \rangle$ is the
lowest lying energy eigenvector of $h^0 + h^1$ and therefore must have
energy lower than $|\Omega^0_{phys} \rangle$ by itself.

An additional possible $\mathcal{O}(e^2)$ correction
to vacuum energy  arising from the Coulomb term
in Eq. (\ref{helectricc}) is identically 0
\begin{equation}
  \label{xic}
  \xi^{2c}_\Omega = \langle \Omega^0_{phys} | h_E^c| \Omega^0_{phys} \rangle = 0,
\end{equation}
since, as before, all $j = 0, 1,$ states in $|\Omega^0_{phys} \rangle$ are filled and all
$j = 2, 3,$ states are empty. The one remaining $\mathcal{O}(e^2)$ correction
to vacuum energy arises from $h_U$ of Eq. (\ref{hplaquette}).
This term is negative
and, since it involves only photon operators, makes identical contributions
to the vacuum before and after branching. It will therefore
be ignored for the present discussion.

\subsection{\label{subsec:branching} Branching}

To order $\mathcal{O}(e)$, the physical vacuum becomes
\begin{equation}
  \label{physvac1}
  |\Omega_{phys} \rangle = |\Omega^0_{phys} \rangle + |\Omega^1_{phys} \rangle.
\end{equation}
Although $|\Omega^0_{phys} \rangle$ is a product state, has complexity 0 and therefore
will not branch, $|\Omega_{phys} \rangle$ of Eq. (\ref{physvac1}) is entangled. For
branching threshold $b \rightarrow \infty$, $|\Omega_{phys} \rangle$ will remain unbranched.
For $b \rightarrow 0$, $|\Omega_{phys} \rangle$ will
branch into a sum entirely of product states.
Vacuum branching for $b$ neither 0 nor $\infty$ could be found by a numerical strategy.
In the present discussion we consider what might still be plausibly guessed without
numerical work.

Suppose to begin $b$ has been made large enough that $|\Omega_{phys} \rangle$ is
not branched. If $b$ is progressively lowered, the result is a
sequence of branching events a bit similar to the sequence of
branching events discussed in Section \ref{subsec:timeevolution}
which occur as a result of time evolution.
At some
critical $b$, we expect $|\Omega_{phys} \rangle$ to branch into
a collection of sets of
orthogonal states $\{ | \phi_i \rangle \}$ 
\begin{equation}
  \label{vacuumsplit}
  |\Omega_{phys} \rangle = \sum_i | \phi_i \rangle,
\end{equation}
each of which satisfies a version of the inequality of Eq. (\ref{splitcondition})
\begin{equation}
    \label{splitcondition2}
   [C( |\Omega_{phys} \rangle )]^2 - \sum_i \langle \phi_i | \phi_i \rangle  [C( |\phi_i \rangle )]^2  > 
   -b \sum_i \langle \phi_i | \phi_i \rangle \ln( \langle \phi_i | \phi_i \rangle ).
\end{equation}
Since, however, $|\Omega_{phys} \rangle$ and condition of Eq. (\ref{splitcondition2})
are approximately Lorentz and translationally invariant, we expect the
collection of sets of branch states $\{ | \phi_i \rangle \}$ to
be approximately Lorentz and translationally invariant.
Then among this collection of sets, a particular
set will be selected by spontaneous breaking of translational invariance.
A vacuum branch will then be selected from this set
by the Born rule as usual.

If $b$ is now progressively lowered below the critical value
needed for Eq. (\ref{splitcondition2}), unlike
what happens in time evolution, the states
$\{ | \phi_i \rangle \}$ will remain unchanged.
All that will happen is that the right-hand side of Eq. (\ref{splitcondition2})
will become smaller. Thus the split will persist. Then at some still
smaller $b$, one or more of the states $\{ | \phi_i \rangle \}$ will split
yet again. And this subsequent split will also persist.
More complicated rearrangements are possible
but as discussed in Appendix \ref{subsec:norearrangements},
such rearrangements would require
cooincidental degeneracy of branches which we believe will occur with probabilty 0.
The end result of all of this is a growing
tree of branch states and sub-branch states.
All states in this tree will inherit the particular
spontaneous breaking of Lorentz and translational invariance
established at
the initial branching event as $b$ is lowered.

Since $b$ has units of volume, any finite $b$ seen from the perspective
of processes at sufficiently short distance will still appear to be extremely large and
thereby prohibit branching. For processes at sufficiently long distance, $b$ will appear to
be extremely small and branching will occur without restriction. Thus
the new branching events which occur as
$b$ is lowered will successively remove from $|\Omega_{phys} \rangle$
its components at length scale greater $\mathcal{O}(b^{\frac{1}{3}})$.
However since $|\Omega_{phys} \rangle$ is the state with lowest
energy density, the monotonic removal of its
components at progressively shorter length scale
should result in a monotonic increase in the energy density
of the remaining piece of $|\Omega_{phys} \rangle$.

The upshot of all of the preceding is that it seems
plausible that the deficit in vacuum energy $\delta \xi_\Omega$
is a continuous, monotone function of $b$
and therefore any possible value of
the observed density of dark matter
and dark energy  should be reproducable by tuning $b$.

We now present an argument in support of the hypothesis that
the residual branch  obtained by removing long
wave length modes from $|\Omega_{phys} \rangle$ of Eq. (\ref{physvac1}),
which we believe is what is left
after the branching process
we just described,
is consistent with the residual entanglement hypothesis
of Section \ref{sec:residual}.
For this purpose, we factor the lattice $L$
into short and long distance pieces.

Assume the lattice
period $2 B$ is given by $4 C D$ with $2 D \gg 2 C \gg 1/m^v, v = 0, 1,$
and divide $L$ into
a set of disjoint cubic regions $\{R_p\}$ each with edge length $2 C$.
Lattice momenta $k^i$ we then break into
\begin{subequations}
\begin{eqnarray}
  \label{ksplit0}
  k^i &=& k_c^i +k_d^1, \\
  \label{ksplit1}
  k_c^i & = & \frac{ \pi n_c^i}{C}, -C\le n_c^i < C, \\
    \label{ksplit2}
  k_d^i & = & \frac{ \pi n_d^i}{ 2C D}, -D \le n_d^i < D,
\end{eqnarray}
\end{subequations}
where $n_c^i, n_d^i$ are integers.
The annihilation operators of Eq. (\ref{freepsi}), recovered from $\Upsilon^v_s( x)$ by
\begin{equation}
  \label{defar0}
    a^{vj}( k) = \frac{1}{\sqrt{ 64 C^3 D^3}}\sum_{xs} u_s^{vj*}( k) \Upsilon^v_s( x)  \exp( -ik \cdot x ),
\end{equation}
where $u_s^{vj}( k_c)$ the solution 
to the lattice Dirac equation Eq. (\ref{diraceq}),
can then be approximated by a sum of operators corresponding to each $R_p$
\begin{subequations}
\begin{eqnarray}
  \label{defar2}
  a^{vj}( k)  & \approx & \frac{1}{\sqrt{ 8 D^3}} \sum_p a^{vj}_p( k_c) \exp( - i k_d \cdot x_p) \, \\
  \label{defar3}
    a^{vj}_p( k_c) &=& \frac{1}{\sqrt{ 8 C^3}}\sum_{x \in R_p} \sum_s u_s^{vj*}( k_c) \Upsilon^v_s( x)  \exp( -ik_c \cdot x ),
\end{eqnarray}
\end{subequations}
where $x_p$ is the center point of $R_p$. 
For $2 C \gg 1/m^v$ and $p \ne p'$, we have
\begin{equation}
  \label{approxorthog}
  \langle \Omega | a^{vj}_p( k_c) a^{v'j'\dagger}_{p'}( k_c') | \Omega \rangle \approx 0.
\end{equation}
In addition, the $a^{vj}_p( k_c)$
as before
annhilate almost purely negative energy for $j= 0,1$ and positive for $j = 2, 3$ and satisfy
\begin{equation}
  \label{aranti}
  \{ a^{vj}_p( k_c), a^{v'j'\dagger}_p( k_c') \} \approx \delta_{vv'} \delta_{jj'} \delta_{p p'} \delta_{k_c k_c'}.
\end{equation}

The $\mathcal{O}( e^0)$ physical vacuum can then be viewed as nearly given by a tensor product of local vacuua each associated
with a corresponding $R_p$
\begin{subequations}
  \begin{eqnarray}
    \label{physvac4}
    |\Omega_{phys}^0 \rangle & \approx & \bigotimes_p |\Omega_{phys \; p}^0 \rangle, \\
  \label{physvac5}
  |\Omega_{phys \; p}^0 \rangle &=& \prod_{vk_c}\prod_{j=0,1} a^{vj\dagger}_p( k_c) |\Omega_p \rangle,
  \end{eqnarray}
\end{subequations}
where $|\Omega_p \rangle$ is the bare vacuum associated with $R_p$ and the operators $a^{vj}_p( k_c)$,
of Eq. (\ref{physvac5}) acting on the space
at $R_p$, using an overloaded notation, are identified with corresponding $a^{vj}_p( k_c)$ of Eqs. (\ref{defar2}) and
(\ref{defar3}) which act on the full Hilbert space.

On each $R_p$ a corresponding photon operator $b_p^\ell(k_c)$ can be defined from the fields
$A( x, y), E(x, y)$ for $x, y \in R_p$. Versions of Eqs. Eq. (\ref{tildea}) - (\ref{atrans}), (\ref{defar2}) and (\ref{defar3})
give
\begin{subequations}
\begin{eqnarray}
  \label{tildea2}
  \tilde{A}_j( k) & \approx & \frac{1}{\sqrt{8 D^3}} \sum_p \tilde{A}_{jp}( k_c) \exp( -i k_d \cdot x_p), \\
    \label{tildea3}
  \tilde{A}_{jp}( k_c)  &=& \frac{1}{\sqrt{8 C^3}} \sum_{x \in R_p}  A( x + \frac{1}{2} \hat{j}, x - \frac{1}{2} \hat{j}) \exp( -i k_c \cdot x), \\
  \label{tildea4}
  \tilde{E}_j( k) & \approx & \frac{1}{\sqrt{8 D^3}} \sum_p \tilde{E}_{jp}( k_c) \exp( -i k_d \cdot x_p), \\
    \label{tildea41}
  \tilde{E}_{jp}( k_c)  &=& \frac{1}{\sqrt{8 C^3}} \sum_{x \in R_p}  E( x + \frac{1}{2} \hat{j}, x - \frac{1}{2} \hat{j}) \exp( -i k_c \cdot x).
\end{eqnarray}
\end{subequations}
From $\tilde{A}_{jp}( k_c), \tilde{E}_{jp}( k_c), $ extract $b_p^\ell(k_c)$ by duplicating Eqs. (\ref{avariant}) - (\ref{atrans}), (\ref{freeatrans}), (\ref{freeatrans1}) .
The result is
\begin{equation}
    \label{defb1}
    b^\ell( k) \approx \frac{1}{\sqrt{ 8 D^3}} \sum_p b^\ell_p( k_c) \exp( - i k_d \cdot x_p).
\end{equation}

Combining Eqs. (\ref{defar2}) and (\ref{defb1}) with Eq. (\ref{omega1}) for $|\Omega^1_{phys} \rangle$ and carrying
out the sums in Eq. (\ref{omega1}) over the low momentum components $k_{1d}$ and $k_{2d}$ gives
  \begin{equation}
  \label{omega11}
  |\Omega^1_{phys} \rangle  \approx  \sum_{p_1p_2p_3} |\Omega^1_{phys\,p_1p_2p_3} \rangle,
  \end{equation}
  for $|\Omega^1_{phys \, p_1p_2p_3 } \rangle$ defined by
  \begin{multline}
  \label{omega12}
  |\Omega^1_{phys \, p_1p_2p_3 } \rangle = \frac{1}{8D^3}\sum_{k_{1c}}\sum_{j_1= 2,3} \sum_{k_{2c}}\sum_{j_2 = 0,1} \sum_{\ell v}\sum_y\\
  \frac{|k_{1c},j_1,p_1,k_{2c},j_2,p_2,v, \ell, p_3  \rangle
    \langle k_{1c},j_1,p_1,k_{2c},j_2,p_2,v, \ell, p_3, y | h^1 | \Omega^0_{phys} \rangle}{ k_{2c}^0 - k_{1c}^0 - k^0(k_{2c} - k_{1c})},
\end{multline}
  where
  $|k_{1c},j_1, p_1, k_{2c},j_2, p_2,v, \ell, p_3  \rangle$ is $|\Omega^0_{phys} \rangle$ with a $k_{2c}, j_2$ fermion
removed from region $R_{p_1}$, $k_{1c}, j_1$ fermion added to region $R_{p_2}$, and a $k_{2c} - k_{1c}, \ell$ photon
added to region $R_{p_3}$. The state $|k_{1c},j_1, p_1, k_{2c},j_2, p_2,v, \ell, p_3, y  \rangle$ is
the same as  $|k_{1c},j_1, p_1, k_{2c},j_2, p_2,v, \ell, p_3,  \rangle$ but with all regions displaced
by a vector $y$.

The translational invariance of $h_1$ and of $|\Omega^0_{phys} \rangle$ allows the sum over $y$ in
Eq. (\ref{omega12}) to be carried out, the result of which is
  \begin{multline}
  \label{omega18}
  |\Omega^1_{phys \, p_1p_2p_3 } \rangle = \sum_{k_{1c}}\sum_{j_1= 2,3} \sum_{k_{2c}}\sum_{j_2 = 0,1} \sum_{\ell v}\\
  \frac{|k_{1c},j_1,p_1,k_{2c},j_2,p_2,v, \ell, p_3  \rangle
    \langle k_{1c},j_1,p_1,k_{2c},j_2,p_2,v, \ell, p_3 | h^1 | \Omega^0_{phys} \rangle}{ k_{2c}^0 - k_{1c}^0 - k^0(k_{2c} - k_{1c})}.
  \end{multline}
  Combining Eqs. (\ref{defar2}) and (\ref{defb1}) with Eq. (\ref{h1}) for $h^1$ and carrying out
  another sum over low momentum components, we obtain
  \begin{equation}
    \label{omega13}
    \langle k_{1c},j_1,p_1,k_{2c},j_2,p_2,v, \ell, p_3 | h^1 | \Omega^0_{phys} \rangle =
    \delta{p_1p_2} \delta{p_1p_3}    \langle k_{1c},j_1,p_1,k_{2c},j_2,p_1,v, \ell, p_1 | h^1 | \Omega^0_{phys} \rangle.
  \end{equation}
  
  The end result of all of this is that Eqs. (\ref{omega11}) and (\ref{omega12}) simplify to
    \begin{equation}
  \label{omega16}
  |\Omega^1_{phys} \rangle  \approx  \sum_p |\Omega^1_{phys\,p} \rangle,
  \end{equation}
  for $|\Omega^1_{phys \, p } \rangle$ defined by
  \begin{multline}
  \label{omega17}
  |\Omega^1_{phys \, p } \rangle =  \sum_{k_{1c}}\sum_{j_1= 2,3} \sum_{k_{2c}}\sum_{j_2 = 0,1} \sum_{\ell v}\sum_y\\
  \frac{|k_{1c},j_1,p,k_{2c},j_2,p,v, \ell, p  \rangle
    \langle k_{1c},j_1,p,k_{2c},j_2,p,v, \ell, p | h^1 | \Omega^0_{phys} \rangle}{ k_{2c}^0 - k_{1c}^0 - k^0(k_{2c} - k_{1c})},
\end{multline}
  For $|\Omega_{phys} \rangle$ we then have to $\mathcal{O}( e^1)$
  \begin{subequations}
    \begin{eqnarray}
      \label{omega19}
      |\Omega_{phys} \rangle & \approx & \bigotimes_p |\Omega_{phys \; p}^0 \rangle + \sum_p |\Omega^1_{phys\,p} \rangle,\\
      \label{omega111}
      & \approx & \bigotimes_p \large(|\Omega_{phys \; p}^0 \rangle + |\Omega^1_{phys\,p} \rangle \large).
    \end{eqnarray}
  \end{subequations}

The vacuum $|\Omega_{phys} \rangle$ of Eqs. (\ref{omega19}) and (\ref{omega111}) is an approximation to
$|\Omega_{phys} \rangle$ of Eq. (\ref{physvac1}) obtained by omitting from
Eq. (\ref{physvac1}) the components with a length scale greater than $2 C$.
While $|\Omega_{phys} \rangle$ of Eqs. (\ref{omega19}) and (\ref{omega111}) retains the entanglement of
Eq. (\ref{physvac1}) on the interior of each $R_p$, 
on a distance scale large with
respect to $2 C$, $|\Omega_{phys} \rangle$ of Eqs. (\ref{omega19}) and (\ref{omega111}) is a tensor product.
Thus the decomposition of $|\Omega_{phys} \rangle$ of Eqs. (\ref{omega19}) and  (\ref{omega111}),
when interpreted as a branch decomposition,
satisfies the residual entanglement hypothesis of Section \ref{sec:residual}.
The discussion accompanying Eq. (\ref{splitcondition2}) implies
\begin{equation}
  \label{chooseb0}
  2 C \approx \mathcal{O}(b^\frac{1}{3}).
\end{equation}

The translational alignment of the division of $L$ into $R_p$
breaks translational invariance.
Eqs. (\ref{omega19}) and (\ref{omega111}) thereby provide a picture of how
the spontaneous breaking
of the vacuum's translational invariance might come about.

The vacuum energy density carried by
$|\Omega_{phys} \rangle$  of Eq. (\ref{omega111}) is missing
the terms
in Eq. (\ref{xi3}) for $| k_1|, |k_2| \le \pi / C$, and therefore by Eq. (\ref{chooseb0}) for
 $| k_1|, |k_2| \le \mathcal{O}(2\pi b^{-\frac{1}{3}}) $.
The missing contribution
to $\xi_\Omega$ of Eq. (\ref{xi3}) will be approximately 
\begin{multline}
  \label{deltaxi}
   \delta \xi_\Omega \approx \\ \sum_{\frac{|k_1 + k_2|}{2} < 2 \pi b^{-\frac{1}{3}}}  \sum_{|k_1 - k_2| < 2 \pi b^{-\frac{1}{3}}} \sum_{j_1= 2,3} \sum_{j_2 = 0,1} \sum_{\ell v} \frac{\langle \Omega^0_{phys} | h^1|k_1,j_1,k_2,j_2,v, \ell \rangle
    \langle k_1,j_1,k_2,j_2,v, \ell | h^1 | \Omega^0_{phys} \rangle}{ k_2^0 - k_1^0 - k^0(k_2 - k_1)},
\end{multline}
where the detailed form of the limits on the $k_1, k_2$ sums
are somewhat arbitrary and have been chosen to simplify their evaluation. The missing piece
itself is negative so that the resulting apparent energy density has a positive sign.

Eq. (\ref{h1}) for $h^1$ along with Eq. (\ref{freepsi}) - (\ref{unorm}) for $\Upsilon^v_s(e)$ and
Eq. (\ref{freeatrans}) - (\ref{k0a}) for $\tilde{A}^T_j(k)$ imply
\begin{multline}
  \label{deltaxi2}
  \delta \xi_\Omega \approx \sum_{\frac{|k_1 + k_2|}{2} < 2 \pi b^{-\frac{1}{3}}}  \sum_{|k_1 - k_2| < 2 \pi b^{-\frac{1}{3}}} \sum_{j_1= 2,3} \sum_{j_2 = 0,1} \sum_{\ell vpq} \\
  \frac{ e^2[\bar{u}^{vj_2}( k_2)\gamma^p \epsilon_p^\ell(k_2 - k_1) u^{vj_1}( k_1)][\bar{u}^{vj_1}( k_1)\gamma^q \epsilon_q^\ell(k_2 - k_1) u^{vj_2}( k_2)]}{ 16 B^3 [k_2^0 - k_1^0 - k^0(k_2 - k_1)] k^0(k_2 - k_1)}.
\end{multline}
Then since $b^{-\frac{1}{3}} \ll m^v, v = 0, 1,$ we have
\begin{subequations}
  \begin{eqnarray}
    \label{deltaxi3}
    \sum_{j_1 = 2, 3} u^{vj_1}(k_1) \bar{u}^{vj_1}(k_1) & \approx & \frac{\gamma^0 + 1}{2}, \\
    \label{deltaxi4}
    \sum_{j_2 = 0, 1} u^{vj_2}(k_2) \bar{u}^{vj_2}(k_2) & \approx & \frac{\gamma^0 -1}{2}, \\
    \label{deltaxi5}
    k^0_2 - k^0_1 - k^0(k_2 - k_1) & \approx & -2m^v.
  \end{eqnarray}
\end{subequations}
Approximating momentum sums with momentum integrals, the increase in vacuum
energy density becomes
\begin{equation}
  \label{deltaxi1}
  -\frac{\delta \xi_\Omega}{8 B^3} \approx \frac{4 \pi e^2}{3 b^{\frac{5}{3}} m^1},
\end{equation}
A corresponding term with $m^1$, the electron mass, replaced by $m^0$, the proton mass,
is negligable by comparison.
Assuming combined dark energy and dark matter density
\begin{equation}
  \label{totaldark}
  -\frac{\delta \xi_\Omega}{8 B^3} \approx 8 \times 10^{-10} \frac{\si{J}}{\si{m}^3},
\end{equation}
then implies
\begin{equation}
  \label{finalb}
  b \approx 2.8 \times 10^{-18} \si{m}^3.
\end{equation}

\subsection{\label{subsec:tinfty} $t \rightarrow \infty$}

So far we have considered vacuum branching in a particular Lorentz frame.
We now briefly consider what might happen for covariant vacuum branching
found according to Section \ref{sec:covariant}
in the $t \rightarrow \infty$ limit.

Since the vacuum itself is Lorentz invariant, covariant branching might be expected
to yield Lorentz invariant branches. But according to the picture of vacuum branching
in Section \ref{subsec:branching} that cannot happen. 
The branches in Section \ref{subsec:branching} are missing long distance pieces of the virtual
fermion-antifermion wave functions present in the full vacuum.
The missing pieces are not Lorentz invariant and as a result the remaining branch states
are also not Lorentz invariant.

The ensemble of possible branches as a whole, however, should be Lorentz
invariant. The choice at a branching event of some particular branch from this ensemble
then entails a spontaneous breaking of Lorentz invariance. At any particular
instant the realized branch behaves like a distribution of ordinary matter.
As time evolves, however, in an expanding universe the missing
fermion-antifermion wave function components
will be progressively red shifted and thereby drift away from the distribution in the
optimal branch ensemble. The remaining fermion-antifermion wave functions will then branch again
restoring the momentum distribution of the missing fermion-antifermion wave functions components.
Thus rather than eventually being red shifted away, as would occur for
a distribution of ordinary matter, the restored branches persist
as would a distribution of dark energy.
The distribution of energy in any vacuum branch arises from a
random
branching process and will therefore not be entirely uniform.
The inhomogeniety of this distribution will appear as
a dark matter contribution in galaxy formation and evolution.
Whether the resulting dark matter is quantitatively consistent with
what is known about galaxy formation and evolution, however,
is beyond the scope of the present work.

\section{\label{sec:zerok} Branching in a Solid at 0 K}

The ground state of a solid at 0 K is an analog of the lattice QED vacuum. Correspondingly
we would expect a 0 K solid to undergo branching analogous to vacuum branching.
The resulting displacement of the solid's ground state energy density is likely
of the same order as the energy density resulting from vacuum branching.
It is probably not feasable to make a measurement of this displacement
as a result both of the difficulty of determining the expected undisplaced energy
and the difficulty of measuring an excruciatingly small energy displacement
resulting from branching. An alternative, however, might be to
detect in some way the initial occurance of  branching
which would come about as a result of progressively increasing
the size of the solid starting from something extremely small.
The threshold size at which we would expect branching
to first occur should be $\mathcal{O}( b^\frac{1}{3})$
which is $\mathcal{O}( 10^{-6}\si{m})$.

The critical problem in devising such an experiment is
to find some way to distinguish between branching which occurs
in the world outside an experimenter and branching which occurs within
the experimenter who registers branching.
For the observation of vacuum branching,
the corresponding problem is avoided.

\section{\label{sec:related} Complexity Elsewhere}

Reduced to a sentence, the present work proposes that a version of
quantum complexity which provides a quantitative
account of the relation between microscopic quantum mechanics and
macroscopic reality, without modification then provides also a
potential origin for cosmological dark energy and dark matter.
A variety of papers on subjects somewhat
related to what is considered here make use of versions
of quantum complexity without, however, making
the connection proposed in the present work.

For non-relativistic quantum mechanics,
\cite{Taylor} proposes
an approximate criterion based on a version of complexity
according to which
a pair of candidate child states
constitute independent branches of a parent.
How the candidate child states are to be determined is not discussed.
Also left unclear by this proposal is whether it concerns a system in isolation
or a system interacting with an environment.
For a system in isolation,
there is no source of error in a computational process to transform one
candidate branch state into the other. In which case the article's discussion of error
correction is unneeded. A system interacting with an environment, on the
other hand, is specified by a density matrix, not a state vector.
In which case the article's use of state vectors is not correct.

Meanwhile, with no connection to the relation between
macroscopic reality and microscopic quantum mechanics,
speculations on a possible role of complexity
in selecting a universe and corresponding
density of dark energy among the landscape of
possiblilties in string theory appears in \cite{Denef, Denef1}.
An alternative discussion of the development of
complexity in the time development of the
universe, with no explicit connection
to dark energy, appears in \cite{Bhattacharyya}.

\section{\label{sec:conclusion} Conclusion}

In \cite{Weingarten1, Weingarten} and in Section \ref{sec:intro} of the present article we argued that 
environmentally-induced decohence as formulated in \cite{Zeh, Zurek, Zurek1, Zurek2, Wallace, Riedel}
looks like it's missing something. 
The present article is a revised version of a series of conjectures in \cite{Weingarten1, Weingarten}
which attempt to fill in what's missing. 
In the determination
of complexity according to \cite{Weingarten} the physical vaccum is assigned
0 complexity which, as a result of the Reeh-Schlieder theorem,
leads to some mathematical complications. Here these complications
are eliminated by assigning 0 complexity to the bare vacuum.
A consequence of which is the prediction that the physical vacuum
itself will branch and as a result exhibit a combination of dark energy and
dark matter. Tuning the branching scale $b$ to reproduce the
observed total density of dark energy and dark matter
leads to an order of magnitude estimate for $b$ which appears to be at least plausible.

Many of the conjectures made here could, at least in principle, be tested
by numerical experiments. A numerical test should
be possible for the proposed QED vacuum branching in a particular
fixed Lorentz frame and for its asymptotic behavior as $t \rightarrow
\infty$. A consequence of which would then potentially be
a more accurate estimate for $b$ and a prediction for the
ratio between the density of dark energy and the density of dark matter. Similarly, it should
be possible to test numerically
the proposed branching of a solid at 0 K.
A laboratory experiment might then yield a second value of $b$ which
could be compared with the value obtained from the densities
of dark energy and dark matter.

\section*{Acknowledgements}

Thanks to Jess Riedel and to an anonymous reviewer for many suggested improvements incorporated here and to Eric Cavalcanti, Mackenzie Dion, Andrea Morello, Eleanor Rieffel, Antoine Tilloy and  Howard Wiseman for discussions 
at the March 2023 Wigner's Friends: Experimental Workshop. This work received no funding.

\appendix

\section{\label{app:properties} Properties of Complexity}

We repeat from \cite{Weingarten} a summary of several properties of $C(| \psi \rangle , |\omega \rangle )$ and $C(| \psi \rangle )$.

Any $| \psi \rangle, |\omega \rangle, |\phi \rangle$ with equal norms and $N^v$ satisfy the triangle inequality
\begin{equation}
  \label{triangle}
  C( | \psi \rangle, |\omega \rangle) \le C( | \psi \rangle, |\phi \rangle) + C( | \phi \rangle, |\omega \rangle).
\end{equation}
In addition $ C( | \psi \rangle, |\omega \rangle)$ is symmetric and 0 only if  $ | \psi \rangle =  |\omega \rangle$.
Therefore  $ C( | \psi \rangle, |\omega \rangle)$ defines a metric on the unit sphere in the subspace of $\mathcal{H}$
with any fixed $N^v$.

Any sequence of $| \psi \rangle$ which approaches some $|\omega \rangle$
according to the $ C( | \psi \rangle, |\omega \rangle)$ topology
approaches also according to the topology given by the inner product on $\mathcal{H}$.
In other words, the identity map from the unit sphere in $\mathcal{H}$ with fixed $N^v$ and topology given by
$C( | \psi \rangle, |\omega \rangle)$ to the unit sphere in $\mathcal{H}$ with fixed $N^v$
and topology given by the Hilbert space norm is continuous.

However, the identity map from the unit sphere in $\mathcal{H}$ with fixed $N^v$ and topology given by the Hilbert space norm
to the unit sphere with fixed $N^v$ and topology given by $C( | \psi \rangle, |\omega \rangle)$ is not continuous. For
any $|\omega \rangle$ and any positive $\delta$ and $\epsilon$, it is possible to find a $|\psi \rangle$
with the same $N^v$ as $|\omega \rangle$ and
\begin{equation}
  \label{psismall}
  \langle \psi | \psi \rangle < \delta,
\end{equation}
but in addition
\begin{equation}
  \label{psicomplex}
  C( z | \omega \rangle +  z | \psi \rangle, |\omega \rangle)  >  \epsilon,
\end{equation}
where 
\begin{equation}
  \label{psicomplexnorm}
  z \parallel | \omega \rangle + | \psi \rangle \parallel  =  1.
\end{equation}
Examples of $|\omega \rangle$ and $| \psi \rangle$ which satisfy Eqs. (\ref{psismall}) and (\ref{psicomplex})
follow from the lower bound on complexity in Section \ref{sec:entangledstates}.

The triangle inequality combined with Eq. (\ref{cpsi1}) defining $C( |\psi \rangle)$ to be
the shortest distance from $|\psi \rangle$ to a product state implies that
for any pair of states $|\psi \rangle $ and $|\phi \rangle $ with equal norms and $N^v$
\begin{subequations}
  \begin{eqnarray}
    \label{triangle1}
    C( |\psi \rangle ) &\le& C(|\phi \rangle ) + C(|\phi \rangle ,|\psi \rangle ), \\
    \label{triangle2}
    C( |\phi \rangle ) &\le& C(|\psi \rangle ) + C(|\phi \rangle ,|\psi \rangle ),
  \end{eqnarray}
\end{subequations}
and therefore 
\begin{equation}
    \label{triangle3}
    |C( |\psi \rangle ) - C(|\phi \rangle )| \le C(|\phi \rangle ,|\psi \rangle ).
\end{equation}

Although not continuous with respect to the
topology given by the norm on $\mathcal{H}$, the complexity $C[|\psi(t) \rangle ]$ of a state $|\psi(t) \rangle $
evolving in time according to $h$ of Eq. (\ref{hamiltonian})
is a continuous function of time if any finite value is chosen for $n_U$. This result follows from the triangle inequality
combined with the observation that $h$ itself is in
the operator Hilbert space $K$. A similar argument implies that if the universe begins at
time 0 in a state of finite complexity, its complexity will remain finite at all later times.

Eqs. (\ref{complexity}) and (\ref{cpsi1}) yield 
a formula for the complexity of the tensor product $|\chi \rangle  \otimes |\phi \rangle $ of a pair of states
localized on regions $R_{\chi}$ and $R_{\phi}$ sufficiently distant from each other. 
For this case we have
\begin{equation}
\label{ctensor}
C( |\chi \rangle  \otimes |\phi \rangle )^2 = 
C( |\chi \rangle  \otimes |\Omega_{\phi} \rangle  )^2 + C( |\Omega_{\chi} \rangle  \otimes |\phi \rangle )^2,
\end{equation}
where $|\Omega_{\chi} \rangle $ and $|\Omega_{\phi} \rangle $ are the vacuum states on regions $R_{\chi}$
and $R_{\phi}$, respectively. For sufficiently 
distant $R_{\chi}$ and $R_{\phi}$, the optimal trajectories 
$k_{i\chi}(\nu)$ and $k_{i\phi}(\nu)$
in Eq. (\ref{complexity})
for  $|\chi \rangle  \otimes |\Omega_{\phi} \rangle $ and $|\Omega_{\chi} \rangle  \otimes |\phi \rangle $
will commute. The optimal product state in 
Eq. (\ref{cpsi1}) for $|\chi \rangle  \otimes |\phi \rangle $ will be the product
$|\chi \rangle _0 \otimes |\phi \rangle _0$, where $|\chi \rangle _0$ and $|\phi \rangle _0$
are the optimal product states for $|\chi \rangle  \otimes |\Omega_{\phi} \rangle $ and $|\Omega_{\chi} \rangle  \otimes |\phi \rangle $
respectively, and
$k_{i\chi}(\nu) + k_{i\phi}(\nu)$ will give
an optimal trajectory in Eq. (\ref{complexity}) for $|\chi \rangle  \otimes |\phi \rangle $ if the
time parametrization of 
$k_{i\chi}(\nu)$ and $k_{i\phi}(\nu)$ are chosen to fulfill
\begin{equation}
\label{fixedratio}
\parallel k_{i\chi}(\nu)\parallel = \lambda \parallel k_{i\phi}(\nu)\parallel
\end{equation}
for some $\lambda$ independent of $t$.
Eq. (\ref{ctensor}) then follows.

\section{\label{app:secondlaw} Second Law of Quantum Complexity}

Also essentially unchanged from \cite{Weingarten}
is a derivation from
the conjectured second law of quantum complexity of \cite{Brown} of
an estimate of the change in complexity over time of a system
evolving according to the Hamiltonian $h$ of Eq. (\ref{hamiltonian}).
Let $|\phi(t) \rangle $ for $t \ge t_0$ be the trajectory in time
of a state
starting from some $|\phi(t_0) \rangle$
with much less than the system's maximum complexity.
Consider a time interval $t \ge t_0$ over which
the complexity of $|\phi(t) \rangle $ remains much less
than the system's maximum. For this discussion we assume $n_U$ finite
so that the term $\xi_m \sum_{x v} \bar{\Psi}^v(x) \Psi^v(x)$ in the defition of $h$
has finite norm in $K$ and therefore $h \in K$.
 
For a closely spaced pair of times $t, t + \delta$, since $h \in K$,
the set of trajectories $S( |\phi(t) \rangle, |\phi(t + \delta) \rangle)$
which fulfill
\begin{equation}
  \label{incrementalk}
  |\phi( t + \delta) \rangle  = \exp[  -i \delta \theta(t) -i \delta k(t)] |\phi(t) \rangle 
\end{equation}
is not empty, where $\theta(t)$ is some real-valued phase angle.
The incremental complexity $C( |\phi(t + \delta) \rangle , |\phi(t) \rangle )$ is then
given by
\begin{equation}
  \label{complexityincrement}
  C( |\phi(t + \delta) \rangle , |\phi(t) \rangle ) = \delta \parallel k(t) \parallel,
\end{equation}
for the $k(t) \in S( |\phi(t) \rangle, |\phi(t + \delta) \rangle)$ which 
minimizes $ \parallel k(t) \parallel$.
For any $t \ge t_0$, the triangle inequality then implies
\begin{equation}
  \label{complexityincrement1}
  C( |\phi(t) \rangle , |\phi(t_0) \rangle ) \le \int_{t_0}^t dt \parallel k(t) \parallel.
\end{equation}

Let $\mathcal{H}(c)$ be the region of state space $\mathcal{H}$ with complexity
bounded by $c$
\begin{equation}
  \label{defhofc}
  \mathcal{H}(c) = \{ |\phi \rangle  \in \mathcal{H} | C( |\phi \rangle ) \le c \}.
\end{equation}
According to the conjectured second law of quantum complexity,
the size of $\mathcal{H}(c)$ rises extremely rapidly
as a function of $c$, sufficiently rapidly 
that the overwhelming majority of  $|\phi \rangle  \in \mathcal{H}(c)$ have complexity $C( |\phi \rangle )$
nearly equal to $c$.
In particular, it is conjectured that 
a sequence of evolving states each
with much less than the system's maximum possible complexity,
at each time step very probably increase their complexity to
the maximum available on the region of state space accessible by
one step of Hamiltonian time evolution.
Eqs. (\ref{complexityincrement1}) then implies that
with high probability
\begin{equation}
  \label{complexityincrement2}
  C( |\phi(t) \rangle , |\phi(t_0) \rangle ) = \int_{t_0}^t dt \parallel k(t) \parallel - \epsilon
\end{equation}
for some very small $\epsilon > 0$.

\section{\label{app:lowerbound} Lower Bound on the Complexity of Entangled States}

The proof of Eq. (\ref{lowerb}) proceeds as follows.
The trajectories $k(\nu) \in K^A$ and $|\omega( \nu) \rangle $
which determine any $C( |\psi \rangle , |\omega \rangle )$ we characterize 
by a corresponding
set of trajectories of Schmidt spectrum vectors. We then
find the rotation matrices which govern the motion of these vectors
as $\nu$ changes. A 
bound on the time integral of the angles which 
occur in these matrices by a time integral of $\parallel k(\nu) \parallel$ 
yields Eq. (\ref{lowerb}). For the sake of simplicity, we
consider only the entanglement arising from fermions with flavor $v = 0$
from which, by itself, the lower bound follows.

\subsection{\label{subsec:schmidtspectra} Schmidt Spectra}

Consider some entangled $|\psi \rangle $ of form Eq. (\ref{entangledstate}).
For a trajectory $k(\nu) \in K^A$, let $|\omega(\nu)\rangle $ be the solution to 
\begin{subequations}
\begin{eqnarray}
\label{udot1}
\frac{d|\omega_k(\nu) \rangle}{d \nu} & = &-i k( \nu) |\omega_k( \nu) \rangle, \\
\label{uboundary1}
|\omega_k \rangle ( 0)\rangle & = & |\omega \rangle.
\end{eqnarray}
\end{subequations}
for some product state $|\omega \rangle $
and assume that $|\omega \rangle $ and $k(\nu)$ have been chosen
to give
\begin{equation}
\label{upsiphi1}
|\omega(1) \rangle  = \xi |\psi \rangle , 
\end{equation}
for a phase factor $\xi$.
Since all $k(\nu)$ conserve both $N^0$ and $N^1$, $|\omega \rangle $ according to
Eq. (\ref{productstate2}) must have the form
\begin{equation}
\label{productstate1}
|\omega \rangle  = 
d^\dagger( q_{n-1}, r_{n-1}) ... d^\dagger( q_0, r_0)  |\Omega \rangle ,
\end{equation}
where $d^\dagger( q, r)$ is given by Eq. (\ref{extended1}) with
$T$ set to 1.

We now divide the lattice $L$ into a collection
of disjoint regions and define a corresponding
collection of Schmidt decompositions of
the trajectory of states
which determine any $C( |\psi \rangle , |\omega \rangle )$.
Divide $L$ into subsets $L^e, L^o$, 
with, respectively, even or odd values of the sums
of components $\hat{x}_i$. The sites in each subset 
have nearest neighbors only in the other.
Let $D^e_{ij}, D^o_{ij}, D^e, D^o$ be 
\begin{subequations}
\begin{eqnarray}
\label{defdije}
D^e_{ij} & = & L^e \cap D^0_{ij} \\
D^o_{ij} & = & L^o \cap D^0_{ij} \\
D^e & = & \cup_{ij} D^e_{ij}. \\
D^o & = & \cup_{ij} D^o_{ij}.
\end{eqnarray}
\end{subequations}
Between $D^e$ and $D^o$ choose the larger,
or either if they are equal.
Assume the set chosen is $D^e$.
The corresponding collection of $D^e_{ij}$
will then include at least $\frac{nmV}{2}$ points.

From the set of $D^e_{ij}$ construct a set of subsets $E_\ell$ 
each consisting of $2n$ distinct points chosen from $2n$ distinct $D^e_{ij}$.
The total number of $E_\ell$ will then be
at least $\frac{m V}{4}$.
We will consider only the first $\frac{m V}{4}$
of these.

The Hilbert space $\mathcal{H}^A$  is given by
\begin{equation}
  \label{bigproduct}
  \mathcal{H}^A = \bigotimes_{x v} \mathcal{H}_x^v ,
\end{equation}
where $\mathcal{H}_x^v$ is the 16 dimensional Hilbert space generated
by all polynomials in $\Psi^{v\dagger}_s(x)$ acting
on the bare vacuum $|\Omega \rangle$.
For each set $E_\ell$ form the 
tensor product spaces
\begin{subequations}
\begin{eqnarray}
\label{defqell}
\mathcal{Q}_\ell &=& \bigotimes_{x \in E_\ell} \mathcal{H}_x^0, \\
\label{defrell}
\mathcal{R}_\ell &=& \bigotimes_{q \notin E_\ell} \mathcal{H}_q^0 \bigotimes_x \mathcal{H}_x^1.
\end{eqnarray}
u\end{subequations}
It follows that $\mathcal{Q}_\ell$ has dimension $16^{2n}$
and
\begin{equation}
\label{deftp}
\mathcal{H}^A = \mathcal{Q}_\ell \otimes \mathcal{R}_\ell.
\end{equation}

A Schmidt decomposition of $|\omega(\nu) \rangle $ according to
Eq. (\ref{deftp}) then becomes
\begin{equation}
\label{defomegat}
|\omega(\nu) \rangle  =  \sum_j \lambda_{j\ell}(\nu) 
|\phi_{j\ell}(\nu) \rangle |\chi_{j\ell}(\nu) \rangle ,
\end{equation}
where 
\begin{subequations}
\begin{eqnarray}
\label{defphit2}
|\phi_{j\ell}(\nu) \rangle  & \in & \mathcal{Q}_\ell \\
\label{defchit}
|\chi_{j\ell}( \nu) \rangle  & \in & \mathcal{R}_\ell,
\end{eqnarray}
\end{subequations}
for $0 \leq j < 16^{2n}$ and real non-negative $\lambda_{j\ell}( \nu)$ which
fulfill the normalization condition
\begin{equation}
\label{normalization}
\sum_j [ \lambda_{j\ell}( \nu)]^2 =  1.
\end{equation}
Each $|\phi_{j\ell}(\nu) \rangle $ consists entirely of flavor $v = 0$
fermions while
the $|\chi_{j\ell}(\nu) \rangle $ can include both flavor $v = 0$ and flavor $v = 1$.

The fermion number operators $N^0[\mathcal{Q}_\ell]$ and $N^0[\mathcal{R}_\ell]$ commute and
$|\omega(\nu) \rangle $ is an eigenvector of the sum with eigenvalue $n$. It follows that 
the decomposition of Eq. (\ref{defomegat}) can be done with $|\phi_{j\ell}( \nu) \rangle $ 
and $|\chi_{j\ell}(\nu) \rangle $
eigenvectors of $N^0[\mathcal{Q}_\ell]$ and $N^0[\mathcal{R}_\ell]$, respectively, with
eigenvalues summing to $n$. Let $|\phi_{0\ell} \rangle $ be $|\Omega_\ell \rangle $, the vacuum state
of $\mathcal{Q}_\ell$, and let 
$|\phi_{i\ell} (\nu) \rangle , 1 \le i \le 4n$, 
span the $4n$-dimensional subspace of $\mathcal{Q}_\ell$
with $N^0[\mathcal{Q}_\ell]$ of 1. 
We assume the corresponding $\lambda_{i\ell}( \nu), 1 \le i \le 4n$,
are in non-increasing order.
Consider Eq. (\ref{defomegat}) for $\nu = 1$. By Eq. (\ref{upsiphi1}), for any choice
of $\ell$ there will be a set of $2n$ nonzero orthogonal
$|\phi_{1\ell}( 1) \rangle , ... |\phi_{2n\ell}( 1) \rangle $ with 
\begin{equation}
\label{lambda1}
\lambda_{j\ell}( 1) = \sqrt{\frac{1}{mV}},
\end{equation}
for $1 \le j \le 2n$.

On the other hand, for $\nu = 0$, Eq. (\ref{defomegat}) becomes a decomposition of 
the product state $|\omega(0) \rangle $. 
Since $|\omega(0) \rangle $
is a product of 
$n$ independent single fermion states,
the space spanned by the component of each of these 
in some $\mathcal{Q}_\ell$ is at most $n$ dimensional,
and as a result at most
$n$ orthogonal $|\phi_{1\ell}(0) \rangle ,... |\phi_{n\ell}(0) \rangle $ 
can occur.
Therefore
at $\nu = 0$, there will be at most $n$ nonzero 
$\lambda_{1\ell}(0), ... \lambda_{n\ell}(0)$ . For
$n < j \le 2n$, we have
\begin{equation}
\label{lambda0}
\lambda_{j\ell}( 0) = 0.
\end{equation}
But according to Eq. (\ref{normalization}),
for each fixed value
of $\ell$
the set of components $\{\lambda_{j\ell}( \nu)\}$
indexed by $j$ is a unit vector.
Eqs. (\ref{lambda0}) and (\ref{lambda1}) then imply that
as $\nu$ goes from $0$ to $1$,
$\{\lambda_{j\ell}( \nu)\}$ 
must rotate through a total angle of at least $\arcsin(\sqrt{\frac{n}{mV}})$.

For the small interval from $\nu$ to $\nu + \delta \nu$ let
$\mu_{j\ell}(\nu)$ and $\theta_{\ell}(\nu)$ be 
\begin{subequations}
\begin{eqnarray}
\label{mudeltat}
\lambda_{j\ell}(\nu + \delta \nu) & = & \lambda_{j\ell}( \nu ) + \delta \nu \mu_{j\ell}(\nu), \\
\label{thetaoft}
\theta_{\ell}( \nu)^2 & = & \sum_j [ \mu_{j\ell}(\nu)]^2. 
\end{eqnarray}
\end{subequations}
We then have
\begin{equation}
\label{thetabound}
\int_0^1 | \theta_{\ell}(\nu)| d \nu \ge \arcsin\left(\sqrt{\frac{n}{mV}} \right).
\end{equation}

Summed over the $\frac{mV}{4}$ values of $\ell$,
Eq. (\ref{thetabound}) becomes
\begin{equation}
\label{thetaboundsum}
\sum_{\ell} \int_0^1 | \theta_{\ell}(\nu)| d \nu  \ge 
\frac{ m V}{4} \arcsin\left(\sqrt{\frac{n}{mV}} \right),
\end{equation}
and therefore
\begin{equation}
\label{thetaboundsum1}
\sum_{\ell} \int_0^1 | \theta_{\ell}(\nu)| d \nu
\ge \frac{ \sqrt{mnV}}{4}.
\end{equation}

\subsection{\label{subsec:schmidtrotation} Schmidt Rotation Matrix}

Next we find the rotation matrix acting on the $j$ indexed vector $\lambda_{j\ell}$ which
gives rise to the $\theta_\ell( \nu)$ of Eq. (\ref{thetaboundsum1}).

The rotation of $\lambda_{j\ell} (\nu)$ during the interval from $\nu$ to $\nu + \delta \nu$
will be determined by $k(\nu)$. For each $f_{xy}$ in Eq. (\ref{defk}) for $k(\nu)$ which can contribute to 
a nonzero value of $\theta_{\ell}(\nu)$, the nearest neighbor pair $\{x, y\}$ has one point, say $x$ in $E_{\ell}$.
Since $E_{\ell} \subset D^e$ and the nearest neighbors of all points in $D^e$ are in
$D^o$, $y$ cannot be in $E_\ell$. Let $g_{\ell}(\nu)$ be the sum of all such $f_{xy}$.
The effect of all other terms in Eq. (\ref{defk}) on the
Schmidt decomposition of Eq. (\ref{defomegat}) will be
a unitary transformation on $\mathcal{R}_\ell$ and identity on $\mathcal{Q}_\ell$.
All other terms will therefore leave $\lambda_{j\ell}( \nu)$ unchanged.

The effect of $g_{\ell}(\nu)$ on $\lambda_{j\ell}(\nu)$ over the
interval from $\nu$ to $\nu + \delta \nu$ can be determined from the simplification
\begin{equation}
\label{psisimp}
|\omega(\nu + \delta \nu) \rangle  = \exp[ i \delta \nu g_{\ell}(\nu)] |\omega(\nu) \rangle .
\end{equation}

From $|\omega(\nu + \delta \nu) \rangle  \langle \omega(\nu + \delta \nu)|$ of Eq. (\ref{psisimp}),
construct the density matrix $\rho(\nu + \delta \nu)$ by 
a partial trace over $\mathcal{R}_{\ell}$, 
using the basis for $\mathcal{R}_{\ell}$
from the Schmidt decomposition in Eq. (\ref{defomegat}) of $|\omega(\nu) \rangle $ at $\nu$
\begin{equation}
\label{defrho}
\rho(\nu + \delta \nu) = 
\sum_j [ \langle  \chi_{j\ell}(\nu)|\omega(\nu + \delta \nu) \rangle  \times  \langle \omega(\nu + \delta \nu)|\chi_{j\ell}(\nu) \rangle ].
\end{equation} 
An eigenvector decomposition of $\rho(\nu + \delta \nu)$ exposes
the $\lambda_{j\ell}(\nu + \delta \nu)$
\begin{equation}
\label{rhodeltat}
\rho(\nu + \delta \nu) = 
\sum_j [\lambda_{j\ell}( \nu + \delta \nu)^2 
 |\phi_{j\ell}(\nu + \delta \nu) \rangle  \langle \phi_{j\ell}( \nu + \delta \nu)|].
\end{equation}

A power series expansion through first order in 
$\delta \nu$ applied to Eqs. (\ref{psisimp}), (\ref{defrho}) and (\ref{rhodeltat})
then gives for $\mu_{j\ell}(\nu)$ of Eq. (\ref{mudeltat})
\begin{equation}
\label{ufromperturb}
\mu_{j\ell}(\nu) = \sum_k r_{jk\ell}(\nu) \lambda_{k\ell}(\nu), 
\end{equation}
for the rotation matrix $r_{jk\ell}(\nu)$
\begin{equation}
\label{rijp}
r_{jk\ell}(\nu) = 
 -\operatorname{Im}[  \langle \phi_{j\ell}(\nu)| \langle \chi_{j\ell}(\nu)| 
g_{\ell}(\nu)|\phi_{k\ell}(\nu) \rangle |\chi_{k\ell}(\nu) \rangle ].
\end{equation}

\subsection{\label{subsec:anglebounds} Rotation Angle Bounds}

Eqs. (\ref{thetaboundsum1}) and (\ref{rijp}) together yield
a lower bound on $C( |\psi \rangle)$.

Since the $f_{xy}$ contributing to $g_\ell(\nu)$
conserve total fermion number $N^0$,
$g_\ell(\nu)$ can be expanded as
\begin{subequations}
\begin{eqnarray}
\label{expandg}
g_{\ell}(\nu) &=& \sum_{xy} g_{\ell}( x, y, \nu),\\
\label{expandg1}
g_{\ell}(x,y,\nu) &=& \sum_{i = 0,1} a^i(x, y, \nu) z^i(x, y, \nu)
\end{eqnarray}
\end{subequations}
where $z^0( x, y, \nu)$ acts only on states
with $N^0( \mathcal{H}_x \otimes \mathcal{H}_y)$ of 0,
$z^1( x, y, \nu)$ acts only on states
with $N^0( \mathcal{H}_x \otimes \mathcal{H}_y)$ strictly greater than 0,
and the $z^i(x,y,\nu)$ are normalized by 
\begin{equation}
\label{normzi}
\parallel z^i(x, y, \nu) \parallel   =  1.
\end{equation}
The operator
$z^0(x, y, \nu)$ will be
\begin{equation}
\label{zprojection}
z^0(x,y,\nu) = p(x,y) \otimes h(x,y,\nu) \bigotimes_{q \ne x, y} I_q, 
\end{equation}
where $p(x,y)$ projects onto the vacuum state
of $\mathcal{H}^0_x \otimes \mathcal{H}^0_y$
and $h(x,y,\nu)$ is a normalized Hermitian
operator acting on $\mathcal{H}^1_x \otimes \mathcal{H}^1_y$.

Combining Eqs. (\ref{thetaoft}),(\ref{ufromperturb}) - (\ref{expandg1}) gives
\begin{subequations}
\begin{eqnarray}
\label{thetasum}
|\theta_\ell(\nu)| &\le& \sum_{xyi}|\theta^i_{\ell}(x,y,\nu)|\\ 
\label{defthetai}
[\theta^i_{\ell}( x,y,\nu)]^2 & = & \sum_j [ \mu^i_{j\ell}(x,y,\nu)]^2,
\end{eqnarray}
\end{subequations}
with the definition
\begin{equation}
\label{musupi}
\mu^i_{j\ell}(x,y,\nu) =  -a^i(x,y,\nu) \sum_k \operatorname{Im}\{ 
 \langle \phi_{j\ell}(\nu)| \langle \chi_{j\ell}(\nu)| 
z^i(x,y,\nu)|\phi_{k\ell}(\nu) \rangle |\chi_{k\ell}(\nu) \rangle  \lambda_{k\ell}(\nu)\}.
\end{equation}

Since the  $|\phi_{j\ell}(\nu) \rangle $ are orthonormal,
$h(x,y,\nu)$ is Hermitian
and the $\lambda_{k\ell}(\nu)$ are real,
we have
\begin{equation}
\label{isup01}
\operatorname{Im}\{
 \langle \phi_{j\ell}(\nu)|\phi_{k\ell}(\nu) \rangle \langle \chi_{j\ell}(\nu)|h(x,y,\nu)|\chi_{k\ell}(\nu) \rangle  \lambda_{k\ell}(\nu)\} = 0.
\end{equation}
Eq. (\ref{musupi}) for $i = 0$ can then be turned into
\begin{multline}
\label{musup01}
\mu^0_{j\ell}(x,y,\nu) =  a^0(x,y,\nu) \sum_k 
 \operatorname{Im}\{
 \langle \phi_{j\ell}(\nu)| \langle \chi_{j\ell}(\nu)| 
[I - p(x,y)] \\h(x,y,\nu) 
|\phi_{k\ell}(\nu) \rangle |\chi_{k\ell}(\nu) \rangle  \lambda_{k\ell}(\nu)\}.
\end{multline}

But in addition
\begin{equation}
\label{schmidt3}
|\omega( \nu) \rangle  = \sum_k |\phi_{k\ell}(\nu) \rangle |\chi_{k\ell}(\nu) \rangle  \lambda_{k\ell}(\nu).
\end{equation}
Also $I - p(x,y)$ is a projection operator so that
\begin{equation}
\label{projsq}
[I - p(x,y)]^2 = I - p(x,y).
\end{equation}
The normalization condition on $z^0(x,y,\nu)$ implies 
$[h(x,y,\nu)]^2$ has trace 1 as an operator on
$\mathcal{H}^1_x \otimes \mathcal{H}^1_y$
and therefore all eigenvalues bounded by 1.
Eqs. (\ref{zprojection}), (\ref{defthetai}), (\ref{musup01}), (\ref{schmidt3}), and
(\ref{projsq}) then give
\begin{equation}
\label{theta0bound}
[\theta^0_{\ell}(x,y,\nu)]^2 \le  [a^0(x,y,\nu)]^2  \langle  \omega(\nu)|[I - p(x,y)]|\omega(\nu) \rangle .
\end{equation}

For $\mu^1_{j\ell}(x,y,\nu)$, since $z^1(x,y,\nu)$ is nonzero only on the
subspace with $N^0(\mathcal{H}_x \otimes \mathcal{H}_y)$ nonzero, we have
\begin{equation}
\label{usup1}
\mu^1_{j\ell}(x,y,\nu) =  -a^1(x,y,\nu) \operatorname{Im}\{  \langle \phi_{j\ell}(\nu)| \langle \chi_{j\ell}(\nu)| 
z^1(x,y,\nu) [I - p(x,y)]|\omega(\nu) \rangle \}.
\end{equation}
Eqs. (\ref{defthetai}) and (\ref{usup1}) give
\begin{equation}
\label{theta1bound}
[\theta^1_{\ell}(x,y,\nu)]^2 \le [a^1(x,y,\nu)]^2  \langle  \omega(\nu)| 
[I - p(x,y)] 
[z^1(x,y,\nu)]^2[I - p(x,y)]|\omega(\nu) \rangle .
\end{equation}

But by Eq. (\ref{normzi}), $[z^1(x,y,\nu)]^2$ as an operator on 
$\mathcal{H}_x \otimes \mathcal{H}_y$, 
has trace 1 and therefore all eigenvalues
bounded by 1. Thus Eq. (\ref{theta1bound}) implies
\begin{equation}
\label{theta1bound1}
[\theta^1_{\ell}(x,y,\nu)]^2 \le  
[a^1(x,y,\nu)]^2  \langle  \omega(\nu)| [I - p(x,y)]|\omega(\nu) \rangle .
\end{equation}

By construction of $D^e$, each nearest neighbor pair $\{x,  y\}$ with $x \in D^e$
must have $y \in D^o$. Also any $x \in D^e$ is contained in at most
a single $E_\ell$.
As a result
Eqs. (\ref{thetasum}), (\ref{theta0bound}) and (\ref{theta1bound1}) imply
\begin{equation}
\label{thetafinal0}
\sum_{\ell} |\theta_{\ell}(\nu)| \le 
\sum_{x \in D^e, y \in D^o}  \{ [|a^0(x,y,\nu)| + |a^1(x,y,\nu)|] \times
 \sqrt{  \langle \omega(\nu)| [I - p(x,y)]|\omega(\nu) \rangle } \}.
\end{equation}
The Cauchy-Schwartz inequality then gives
\begin{equation}
\label{thetafinal}
[\sum_{\ell} |\theta_{\ell}(\nu)|] ^ 2 \le
\sum_{x \in D^e, y \in D^o} [|a^0(x,y,\nu)| + |a^1(x,y,\nu)|]^2 \times
 \sum_{x \in D^e, y \in D^o}  \langle \omega(\nu)| [I - p(x,y)]|\omega(\nu) \rangle .
\end{equation}

The state $|\omega(\nu) \rangle $ can be expanded as a linear combination of orthogonal states 
each with 
$n$ flavor $v = 0$ fermions each at a single position. A state with $v = 0$ fermions at $n$
positions will survive the projection
$I - p(x,y)$ only if at least one of the fermions is either at $x$ or $y$.
Each $x \in D^e$ can be the member of only a single such pair of nearest
neighbor $\{x, y\}$. A $y \in D^o$ can be in 6 $x, y$ pairs for an
$x \in D^e$. Thus a term with $n$ fermion positions in the
expansion of $|\omega(\nu) \rangle $ will pass $I - p(x,y)$ for 
at most $6n$ pairs of $x$ and $y$. Therefore 
\begin{equation}
\label{psiprojectionbound}
\sum_{x \in D^e, y \in D^o}  \langle \omega(\nu)| [I - p(x,y)]|\omega(\nu) \rangle  \le 6n.
\end{equation}

By Eq. (\ref{defkkprime}) 
\begin{equation}
\label{kfroma0}
\parallel k(\nu) \parallel ^ 2  \ge  \sum_{\ell, x \in D^e, y \in D^o} \parallel g_\ell( x, y, \nu) \parallel^2
\end{equation}
In addition, $z^0(x,y,\nu)$ is orthogonal
to $z^1(x, y, \nu)$. It follows that
\begin{equation}
\label{kfroma}
\parallel k(\nu) \parallel^2 \ge \sum_{x \in D^e, y \in D^o} [|a^0(x,y,\nu)|^2 + |a^1(x,y,\nu)|^2].
\end{equation}

Assembling Eqs. (\ref{thetafinal}), (\ref{psiprojectionbound})
and (\ref{kfroma}) gives
\begin{equation}
\label{kbound}
\parallel k(\nu) \parallel^2 \ge \frac{1}{2} \sum_{x \in D^e, y \in D^o} [|a^0(x,y,\nu)| + |a^1(x,y,\nu)|]^2 
\ge \frac{1}{12 n} [\sum_{\ell} |\theta_{\ell}(\nu)|] ^ 2
\end{equation}
Eq. (\ref{thetaboundsum1}) then implies
\begin{equation}
\label{kbound1}
\int_0^1 \parallel k(\nu) \parallel \ge \sqrt{ \frac{ mV}{192}},
\end{equation}
and therefore
\begin{equation}
\label{cbound}
C( |\psi \rangle , |\omega \rangle ) \ge \sqrt{ \frac{ mV}{192}}.
\end{equation}
Since Eq. (\ref{cbound}) holds for all product $|\omega \rangle $
we obtain
\begin{equation}
\label{cbound2}
C( |\psi \rangle ) \ge \sqrt{ \frac{ mV}{192}}.
\end{equation}

\section{\label{app:upperbound} Upper Bound on  the Complexity of Entangled States}

An upper bound on $C( |\psi \rangle )$ of the entangled state of Eq. (\ref{entangledstate}) 
is given by $C( |\psi \rangle , |\omega \rangle )$ for any product state
$|\omega \rangle $, for which in turn an upper bound is given by 
\begin{equation}
\label{cpsiomega}
C( |\psi \rangle , |\omega \rangle ) \le \int_0^1 d t \parallel k( \nu) \parallel,
\end{equation} 
for any 
trajectory $k(\nu) \in K^A$ fulfilling
\begin{subequations}
\begin{eqnarray}
\label{udot11}
\frac{d|\omega(\nu) \rangle}{d \nu} & = &-i k( \nu) |\omega( \nu) \rangle, \\
\label{uboundary11}
|\omega \rangle ( 0)\rangle & = & |\omega \rangle, \\
\label{upsiphi11}
|\omega(1) \rangle & =& \xi |\psi \rangle , 
\end{eqnarray}
\end{subequations}
for a phase factor $\xi$.
Beginning with an $|\omega \rangle $
consisting of $n$ flavor 0 and flavor 1 pairs of particles 
each at one of a corresponding set of
$n$ single points, we construct a sufficient $k(\nu)$ in three stages.
First, $|\omega \rangle $
is split into a sum of $m$ orthogonal product states, each again consisting
of $n$ particle pairs one at each of a corresponding set of $n$ single points. Then the 
position of each of the particles in the product states is moved to the center of
the wave function of one of the single particle states of Eq. (\ref{pstates}). 
Finally, by approximately $\ln( V) / \ln( 8)$ iterations of a
fan-out tree, the $2 m n$ wave functions concentrated at points are spread over the 
$2 m n$ cubes $D^v_{ij}$.

\subsection{\label{app:subsecfirst}Product State to Entangled State}

Define the set of positions $x_{ij}$ to be
\begin{subequations}
\begin{eqnarray}
\label{defxij0}
(x_{ij})^1 & = & i + (x_{00})^1 ,\\
\label{defxij1}
(x_{ij})^2 & = & j + (x_{00})^2,\\
\label{defxij2}
(x_{ij})^3 & = & (x_{00})^3,
\end{eqnarray}
\end{subequations}
for $0 \le i < m, 0 \le j < n$ and arbitrary base point $x_{00}$.
Let the set of product states $|\omega_i \rangle $ be
\begin{equation}
\label{defomega}
|\omega_i \rangle   =  \prod_{v, 0 \le j < n} \Psi_0^{v\dagger}( x_{ij}) |\Omega \rangle .
\end{equation}
The entangle state $|\chi \rangle $
\begin{equation}
\label{defchi}
|\chi \rangle  = \sqrt{\frac{1}{m}} \sum_i |\omega_i \rangle 
\end{equation}
we generate from a sequence of unitary transforms acting
on $|\omega \rangle  = |\omega_0 \rangle $.

The product state $| \omega_0 \rangle$ is transformed into the entangled state $|\chi \rangle$
in a sequence of steps, each splitting some $|\omega_i \rangle$ into
a linear combination of $|\omega_i \rangle$ and $|\omega_{i+1} \rangle$.
The task of splitting an $|\omega_i \rangle$, in turn,
proceeds by two other sequences of steps. The first of these
walks along the index $0 \le j < n$ of the states at some $x_{ij}$
and splits each state into the sum of a state with spin index 0
and a state with spin index 2. Then a second sequence walks the index  $0 \le j < n$
and turns the product of states with spin index 2 at $x_{ij}$ into
a product of states with spin index 0 at $x_{i+1j}$.
This somewhat roundabout procedure is needed to
insure that the end result
is the sum of a product over $j$ of states at  $x_{ij}$ and a product
over $j$ of states at $x_{i+1j}$ without cross terms between states at $x_{ij}$ and states at
$x_{i+1j}$.
The process of splitting off the product state with spin index 2 at each site
temporarily assigns spin index 1 to the state being updated in order
insure that at each step the corresponding update operator satisifies the partial trace conditions of
Eqs. (\ref{ptrfx1}) and (\ref{ptrfy1}).
For this purpose define
the set of spin indices $t_{ij}, 0 \le i,j < n$,
\begin{subequations}
\begin{eqnarray}
\label{defsj0}
t_{ij} & = & 2, j < i, \\
\label{defsj1}
t_{ij} & = & 1, j = i, \\
\label{defsj2}
t_{ij} & = & 0, j  >  i.
\end{eqnarray}
\end{subequations}

Let $k_{0}$ acting on $\mathcal{H}_{x_{00}} \otimes \mathcal{H}_{x_{01}}$
have matrix elements
\begin{subequations}
  \begin{eqnarray}
\label{defk01}
 \langle \beta_0| k_0  |\alpha_0 \rangle  &=& -i,\\
\label{defk10}
  \langle \alpha_0| k_0  |\beta_0 \rangle &=& i, \\
 \label{defalpha0}
 |\alpha_0 \rangle & = & \Psi_0^{0\dagger}( x_{00}) \Psi_0^{1\dagger}( x_{00})\Psi_0^{0\dagger}( x_{01}) \Psi_0^{1\dagger}( x_{01}) |\Omega \rangle, \\
 \label{defbeta0}
 |\beta_0 \rangle & = & \Psi_2^{0\dagger}( x_{00}) \Psi_2^{1\dagger}( x_{00})\Psi_1^{0\dagger}( x_{01}) \Psi_1^{1\dagger}( x_{01}) |\Omega \rangle
  \end{eqnarray}
\end{subequations}
and extend $k_0$ to $\mathcal{H}^A$ as before by appropriate factors of $I_x$.

We then have
\begin{equation}
\label{k00}
\exp( i \theta_0 k_0) |\omega_0 \rangle  = 
\sqrt{\frac{1}{m}} |\omega_0 \rangle  + 
\sqrt{\frac{m - 1}{m}} \prod_{0 \le j < n} \Psi_{t_{1j}}^{0\dagger}( x_{0j})\Psi_{t_{1j}}^{1\dagger}( x_{0j}) |\Omega \rangle ,
\end{equation}
where
\begin{equation}
\label{defarcsin}
\theta_0 = \arcsin( \sqrt{\frac{m - 1}{m}}).
\end{equation}

Now let $k_{1}$ acting on $\mathcal{H}_{x_{01}} \otimes \mathcal{H}_{x_{02}}$
have matrix elements
\begin{subequations}
  \begin{eqnarray}
\label{defk011}
 \langle \beta_1| k_1  |\alpha_1 \rangle  &=& -i,\\
\label{defk101}
  \langle \alpha_1| k_1  |\beta_1 \rangle &=& i, \\
 \label{defalpha1}
 |\alpha_1 \rangle & = & \Psi_1^{0\dagger}( x_{01}) \Psi_1^{1\dagger}( x_{01})\Psi_0^{0\dagger}( x_{02}) \Psi_0^{1\dagger}( x_{02}) |\Omega \rangle, \\
 \label{defbeta1}
 |\beta_1 \rangle & = & \Psi_2^{0\dagger}( x_{01}) \Psi_2^{1\dagger}( x_{01})\Psi_1^{0\dagger}( x_{02}) \Psi_1^{1\dagger}( x_{02}) |\Omega \rangle
  \end{eqnarray}
\end{subequations}
and extend $k_1$ to $\mathcal{H}^A$ by factors of $I_x$.
We then have
\begin{equation}
\label{k001}
\exp( i \theta_1 k_1) \exp( i \theta_0 k_0)|\omega_0 \rangle  =
\sqrt{\frac{1}{m}} |\omega_0 \rangle  +
\sqrt{\frac{m - 1}{m}} \prod_{0 \le j < n} \Psi_{t_{2j}}^{0 \dagger}( x_{0j})  \Psi_{t_{2j}}^{1 \dagger}( x_{0j}) |\Omega \rangle ,
\end{equation}
for $\theta_1$ given by $\frac{\pi}{2}$.

Continuing in analogy to Eqs. (\ref{defk011}) - (\ref{k001}),
for a sequence of operators $k_j$, $2 \le j < n-1 $, acting on
$\mathcal{H}_{x_{0j}}  \otimes \mathcal{H}_{x_{0j+1}}$, and corresponding
$\theta_j$ we obtain
\begin{equation}
\label{k0n}
\exp( i \theta_{n-2} k_{n-2}) ... \exp( i \theta_0 k_0) |\omega_0 \rangle  = 
\sqrt{\frac{1}{m}} |\omega_0 \rangle  +
\sqrt{\frac{m - 1}{m}} \prod_{0 \le j < n} \Psi_{t_{n-1j}}^{0\dagger}( x_{0j}) \Psi_{t_{n-1j}}^{1\dagger}( x_{0j})|\Omega \rangle .
\end{equation}

Let $k_{n-1}$ acting on $\mathcal{H}_{x_{00}} \otimes \mathcal{H}_{x_{10}}$
have matrix elements
\begin{subequations}
  \begin{eqnarray}
\label{defknm1}
 \langle \beta_{n-1}| k_{n-1}  |\alpha_{n-1} \rangle  &=& -i,\\
\label{defknm11}
  \langle \alpha_{n-1}| k_{n-1}  |\beta_{n-1} \rangle &=& i, \\
 \label{defalphanm1}
 |\alpha_{n-1} \rangle & = & \Psi_2^{0\dagger}( x_{00}) \Psi_2^{1\dagger}( x_{00}) |\Omega \rangle, \\
 \label{defbetanm1}
 |\beta_{n-1} \rangle & = & \Psi_0^{0\dagger}( x_{10}) \Psi_0^{1\dagger}( x_{10}) |\Omega \rangle,
  \end{eqnarray}
\end{subequations}
extend $k_{n-1}$ to $\mathcal{H}^A$ by factors of $I_x$,
let $\theta_{n-1}$ be $\frac{\pi}{2}$,
and apply $\exp(i \theta_{n-1} k_{n-1})$ to Eq. (\ref{k0n}).
For $1 \le j < n$ and $\ell = j + n - 1$, a
 similar sequence of $\exp(i \theta_\ell k_\ell)$ acting on
$\mathcal{H}_{x_{0j}} \otimes \mathcal{H}_{x_{1j}}$ gives
\begin{equation}
\label{k02n}
\exp( i \theta_{2n-2} k_{2n-2}) ... \exp( i \theta_0 k_0) |\omega_0 \rangle  = 
\sqrt{\frac{1}{m}} |\omega_0 \rangle  +
\sqrt{\frac{m - 1}{m}} |\omega_1 \rangle .
\end{equation}
Multiplying Eq. (\ref{k02n}) by $\exp(i\theta_\ell k_\ell), 0 \le j < n, \ell = j + 2n-1$, on
$\mathcal{H}_{x_{1 j}}  \otimes \mathcal{H}_{x_{1 j + 1 }}$, 
and then $\exp(i\theta_\ell k_\ell), 0 \le j < n, \ell = j + 3n -1$, on
$\mathcal{H}_{x_{1j}} \otimes \mathcal{H}_{x_{2j}}$ gives
\begin{equation}
\label{k04n}
\exp( i \theta_{4n-3} k_{4n-3}) ... \exp( i \theta_0 k_0) |\omega_0 \rangle  = 
\sqrt{\frac{1}{m}} |\omega_0 \rangle  +
\sqrt{\frac{1}{m}} |\omega_1 \rangle  + 
\sqrt{\frac{m - 2}{m}} |\omega_2 \rangle .
\end{equation}

The end result of a sequence of $2mn - m$ such steps is
$|\chi \rangle $ of Eq. (\ref{defchi})
\begin{equation}
\label{finalk}
\exp( i \theta_{2mn-m-1} k_{2mn-m-1}) ... \exp( i \theta_0 k_0) |\omega_0 \rangle  = 
\sqrt{\frac{1}{m}} \sum_i |\omega_i \rangle .
\end{equation}

The $k_i$ and $\theta_i$ of Eq. (\ref{finalk}) have
\begin{subequations}
\begin{eqnarray}
\label{normfinalk}
\parallel k_i \parallel & = & \sqrt{2},\\
\label{normfinaltheta}
| \theta_i | & \le & \frac{\pi}{2}.
\end{eqnarray}
\end{subequations}
Thus  Eq. (\ref{finalk}) implies
\begin{equation}
\label{deltac}
C( |\chi \rangle , |\omega \rangle ) \le \sqrt{2} \pi m (n - \frac{1}{2}).
\end{equation}

\subsection{\label{app:subsectionsecond}Entangled State Repositioned}

Let $y^v_{ij}$ be the center of cube $D^v_{ij}$ of Eq. (\ref{pstates}). Define the state $|\rho \rangle$
to be
\begin{equation}
\label{phinpoints}
|\rho \rangle  = \sum_{i} \zeta_i \prod_{vj} \Psi^{v\dagger}_{s^v_{ij}}( y^v_{ij}) |\Omega \rangle ,
\end{equation}
where $s^v_{ij}$ are the spin indices in Eq. (\ref{pstates}).
By a sequence of unitary operators acting on $|\chi \rangle$ of Eq. (\ref{defchi}), we now
move the flavor $v$ member of the fermion pair at each $x_{ij}$ to the corresponding $y^v_{ij}$, rotate
the spin to $s^v_{ij}$, and generate
the phase factors $\zeta_i$, thereby placing a bound
on $C( |\rho \rangle, |\chi \rangle)$.

For each $0 \le v < 2, 0 \le i < m, 0 \le j < n$, let
$z^{v0}_{ij}, z^{v1}_{ij} ... z^{vr^v_{ij}}_{ij}$ be the shortest
sequence of nearest neighbor sites
such that 
\begin{subequations}
\begin{eqnarray}
\label{z0}
z^{v0}_{ij} & = & x_{ij}, \\
\label{zlast}
z^{vr^v_{ij}}_{ij} & = & y^v_{ij},
\end{eqnarray}
\end{subequations}
for the $x_{ij}$ in Eqs. (\ref{defxij0}) - (\ref{defxij2})
and such that all $z^{v\ell}_{ij}$ for distinct $v, \ell, i, j,$ with $\ell > 0$ are
themselves distinct.
For each $0 \le v < 2, 0 \le i < m, 0 \le j < n, 0 \le \ell < r^v_{ij} - 1$,  for nearest neighbor pair $\{z^{v\ell}_{ij}, z^{v\ell+1}_{ij}\}$,
let $k^{v\ell}_{ij}$ acting on $\mathcal{H}_{z^{v\ell}_{ij}} \otimes \mathcal{H}_{z^{v\ell+1}_{ij}}$
have matrix elements
\begin{subequations}
  \begin{eqnarray}
    \label{defkellij}
    \langle \beta^{v\ell}_{ij} | k^{v\ell}_{ij} |\alpha^{v\ell}_{ij} \rangle &=& -i, \\
      \label{defkellij1}
      \langle \alpha^{v\ell}_{ij} | k^{v\ell}_{ij} |\beta^{v\ell}_{ij} \rangle &=& i, \\
        \label{defkellij2}
 | \alpha \rangle & = & \Psi_0^{v\dagger}(z^{v\ell}_{ij})|\Omega \rangle, \\
        \label{defkellij3}
 | \beta \rangle & = & \Psi_0^{v\dagger}(z^{v\ell+1}_{ij})|\Omega \rangle.
\end{eqnarray}
\end{subequations}
For the final step with $0 \le v < 2, 0 \le i < m, 0 \le j < n - 1, \ell = r^v_{ij} - 1$, and for the final step with
$0 \le v < 2, 0 \le i < m,  j = n - 1, \ell = r^v_{ij} - 1$,
let $k^{v\ell}_{ij}$ acting on $\mathcal{H}_{z^{v\ell}_{ij}} \otimes \mathcal{H}_{z^{v\ell+1}_{ij}}$
have matrix elements which both move states and rotate spins
\begin{subequations}
  \begin{eqnarray}
    \label{defkellij4}
    \langle \beta^{v\ell}_{ij} | k^{v\ell}_{ij} |\alpha^{v\ell}_{ij} \rangle &=& -i, \\
      \label{defkellij5}
      \langle \alpha^{v\ell}_{ij} | k^{v\ell}_{ij} |\beta^{v\ell}_{ij} \rangle &=& i, \\
        \label{defkellij6}
 | \alpha \rangle & = & \Psi_0^{v\dagger}(z^{v\ell}_{ij})|\Omega \rangle, \\
        \label{defkellij7}
        | \beta \rangle & = & \Psi_{s^v_{ij}}^{v\dagger}(z^{v\ell+1}_{ij}) |\Omega \rangle .
\end{eqnarray}
\end{subequations}

For the final step with $0 \le v < 2, 0 \le i < m, j = n-1, \ell = r^v_{ij} - 1$,
let $k^{v\ell}_{ij}$ acting on $\mathcal{H}_{z^{v\ell}_{ij}} \otimes \mathcal{H}_{z^{v\ell+1}_{ij}}$
have matrix elements which move states, rotate spins and generate the $\zeta_i$
\begin{subequations}
\begin{eqnarray}
  \label{defkellij8}
    \langle \beta^{v\ell}_{ij} | k^{v\ell}_{ij} |\alpha^{v\ell}_{ij} \rangle &=& -i \zeta_i, \\
      \label{defkellij9}
      \langle \alpha^{v\ell}_{ij} | k^{v\ell}_{ij} |\beta^{v\ell}_{ij} \rangle &=& i \zeta^*_i, \\
        \label{defkellij10}
 | \alpha \rangle & = & \Psi_0^{v\dagger}(z^{v\ell}_{ij})|\Omega \rangle, \\
        \label{defkellij11}
        | \beta \rangle & = & \Psi_{s^v_{ij}}^{v\dagger}(z^{v\ell+1}_{ij}) |\Omega \rangle .
\end{eqnarray}
\end{subequations}
Extend all of these $k^{v\ell}_{ij}$ to $\mathcal{H}^A$ by factors of $I_x$.

Define $r$ to be
\begin{equation}
  \label{defs}
  r = \max_{vij} r^v_{ij},
\end{equation}
and for each $v, i, j$ define
\begin{equation}
  \label{extendk}
  k ^{v\ell}_{ij} = 0, r^v_{ij} \le \ell < r.
\end{equation}
Let $k^\ell$ be
\begin{equation}
  \label{defsumk}
  k^\ell = \sum_{vij} k^{v\ell}_{ij}.
\end{equation}
We then have
\begin{equation}
\label{k0kd}
 \exp( i\frac{\pi}{2} k^{r-1}) ... \exp( i\frac{\pi}{2} k^0) |\chi \rangle  = |\rho \rangle ,
\end{equation}
for $|\chi \rangle $ of Eq. (\ref{defchi}) and $|\rho \rangle$ of Eq. (\ref{phinpoints})

The $k^\ell$ of Eqs.  (\ref{defkellij}) - (\ref{defkellij1}), (\ref{defsumk}), have
\begin{equation}
\label{normk0kd}
\parallel k^\ell_{ij} \parallel \le 2 \sqrt{ mn}.
\end{equation}
Thus  Eq. (\ref{k0kd}) implies
\begin{equation}
\label{deltac0}
C( |\rho \rangle , |\chi \rangle ) \le  \pi \sqrt{mn} r .
\end{equation}
We now minimize $r$ over the base point $x_{00}$ 
\begin{equation}
  \label{defdd}
  \hat{r} = \min_{x_{00}} r,
\end{equation}
with the result
\begin{equation}
\label{deltac1}
C( |\rho \rangle , |\chi \rangle ) \le \pi \sqrt{mn} r,
\end{equation}
where we have dropped the hat on $r$.

\subsection{\label{app:fanout}Fan-Out}

The state $|\rho \rangle $ of Eq. (\ref{phinpoints}) with particles at the centers of the cubes $D^v_{ij}$ we now fan-out
to the state $|\psi \rangle $ of Eq. (\ref{entangledstate}) with particle wave functions spread uniformly over the
cubes $D^v_{ij}$. For sufficiently small lattice spacing $a$ nearly all of the complexity of
the bound on $C(|\psi \rangle )$ is generated in this step.

Let $d$ be the length of the edge of the $D^v_{ij}$. Each edge of
$D^v_{ij}$ then consists of $d+1$ sites. The volume $V$ is then $d^3$. We begin with case
\begin{equation}
\label{rpower2}
d = 2^p,  
\end{equation}
for some integer $p$. For simpilicity we present the fan-out applied to
a prototype single particle state $|\phi_0 \rangle $ on prototype cube $G$ with edge length $d$,
and center at some point $y$
\begin{equation}
\label{defphi0}
|\phi_0 \rangle  =  \Psi^{v\dagger}_s( y) |\Omega \rangle .
\end{equation}

The first stage of the fan-out process consists of
splitting $|\phi_0 \rangle $ into a pair of components displaced from
each other in lattice direction 1.
For integer $-2^{p-2} \le i \le 2^{p-2}$ define $y(i)$ to be
$y$ incremented by $i$ nearest neighbor steps
in lattice direction 1. For $1 \le j \le 2^{p-2}$ define
$k_j$ on $\mathcal{H}_{y_{j-1}}\otimes \mathcal{H}_{y_j}$ to have
matrix elements
\begin{subequations}
\begin{eqnarray}
\label{defkofi}
 \langle \beta_j  | k_j | \alpha_j \rangle  &=& -i, \\
\label{defkofi1}
\langle  \alpha_j|  k_j| \beta_j\rangle  &=& i. \\
\label{defkofi2}
| \alpha_j \rangle & = & \Psi^{v\dagger}_s[y(j-1)]|\Omega \rangle, \\
\label{defkofi3}
| \beta_j \rangle & = & \Psi^{v\dagger}_s[y(j)]|\Omega \rangle .
\end{eqnarray}
\end{subequations}
For $-2^{p-2} \le j \le -1$ define
$k_j$ by Eqs. (\ref{defkofi}) - (\ref{defkofi3}) but with $j+1$
in place of $j-1$.
Then define $\bar{k}_j$ by
\begin{subequations}
  \begin{eqnarray}
    \label{defbark}
    \bar{k}_1 & = & \frac{1}{\sqrt{2}}( k_1 + k_{-1}), \\
   \label{defbark1}
   \bar{k}_j & = &  k_j + k_{-j}, 2 \le j \le 2^{p-2}.
  \end{eqnarray}
\end{subequations}
With these definitions it then follows that
\begin{equation}
\label{defphi1}
|\phi_1 \rangle  = 
\exp(i \frac{\pi}{2}\bar{ k}_q) ... \exp(i \frac{\pi}{2} \bar{ k}_1) |\phi_0 \rangle ,
\end{equation}
for $q = 2^{p-2}$,
is given by
\begin{equation}
\label{defphi11}
|\phi_1 \rangle  = \frac{1}{\sqrt{2}}\left\{ \Psi^{v\dagger}_s[y(-2^{p-2})] |\Omega \rangle + \Psi^{v\dagger}_s[y(2^{p-2})] |\Omega \rangle \right\}
\end{equation}

Eqs. (\ref{defbark}) and (\ref{defbark1}) imply
\begin{subequations}
  \begin{eqnarray}
       \label{defbarkn}
   \parallel \bar{k}_1 \parallel & = & \sqrt{2}, \\
   \label{defbark1n}
  \parallel \bar{k}_j\parallel & = &  2, 2 \le j \le 2^{p-2}.
  \end{eqnarray}
\end{subequations}
It then follows that
\begin{equation}
  \label{stageoneb}
  C( |\phi_1 \rangle , |\phi_0 \rangle ) < 2^{p-2} \pi,
\end{equation}
where for simplicity we have used an overestimate for $\parallel \bar{k}_1 \parallel$.

The next stage of the fan-out consists of splitting each of the 2 
components of $|\phi_1 \rangle $ but now in lattice
direction 2. For a set of $\bar{k}_j, 2^{p-2} < j \le 2^{p-1}$, defined
by adapting
of Eqs. (\ref{defkofi}) - (\ref{defbark1}),
we have
\begin{equation}
\label{defphi2}
|\phi_2 \rangle  = 
\exp(i \frac{\pi}{2}\bar{ k}_q) ... \exp(i \frac{\pi}{2} \bar{ k}_r) |\phi_1 \rangle ,
\end{equation}
with $q = 2^{p-1}, r = 2^{p-2} + 1$,
given by
\begin{equation}
\label{defphi12}
|\phi_2 \rangle  = \frac{1}{2}\sum_{i = -2^{p-2},2^{p-2}} 
\sum_{j = -2^{p-2},2^{p-2}}  \Psi^{v\dagger}_s[ y(i,j)] |\Omega \rangle ,
\end{equation}
for $y(i,j)$ defined to be $y(i)$ displaced $j$ steps in lattice direction 2.
Eqs. (\ref{defbark}) and (\ref{defbark1}) adapted to the fan-out in direction 2
give $\bar{k}_j, 2^{p-2} < j \le 2^{p-1}$ each acting on twice as
many sites as was the case for the direction 1 fan-out and therefore
\begin{subequations}
  \begin{eqnarray}
       \label{defbarkn2}
   \parallel \bar{k}_{2^{p-2} + 1} \parallel & = & 2, \\
   \label{defbark1n2}
  \parallel \bar{k}_j\parallel & = &  2\sqrt{2}, 2^{p-2} + 2 \le j \le 2^{p-1}.
  \end{eqnarray}
\end{subequations}
It then follows that
\begin{equation}
  \label{stageoneb1}
  C( |\phi_2 \rangle , |\phi_1 \rangle ) <  2^{p-2} \sqrt{2}\pi.
\end{equation}

Splitting yet again, now in lattice direction 3,
yields

\begin{equation}
\label{defphi3}
|\phi_3 \rangle  = 
\exp(i \frac{\pi}{2}\bar{ k}_q) ... \exp(i \frac{\pi}{2}\bar{ k}_r) |\phi_2 \rangle ,
\end{equation}
for $q =2^{p-1} + 2^{p-2}, r = 2^{p-1}+1$, given by
\begin{equation}
\label{defphi13}
|\phi_3 \rangle  = \frac{1}{\sqrt{8}}\sum_{i = -2^{p-2},2^{p-2}} 
\sum_{j = -2^{p-2},2^{p-2}} \sum_{\ell = -2^{p-2},2^{p-2}}  \Psi^{v\dagger}_s[ y(i,j, \ell)] |\Omega \rangle ,
\end{equation}
for $y(i,j, \ell)$ defined to be $y(i, j)$ displaced $\ell$ steps in lattice direction 3.

Eqs. (\ref{defbark}) and (\ref{defbark1}) adapted to the fan-out in direction 3
give $\bar{k}_j, 2^{p-1} < j \le 2^{p-1} + 2^{p-2}$, each acting on twice as
many sites as was the case for the direction 2 fan-out and therefore
\begin{subequations}
\begin{eqnarray}
       \label{defbarkn3}
    \parallel \bar{k}_ {2^{p-1} + 1} \parallel  &=&  2\sqrt{2}, \\
   \label{defbark1n3}
 \parallel \bar{k}_j\parallel &=&   4, 2^{p-1}+ 2 \le j \le 2^{p-1} + 2^{p-2}.
\end{eqnarray}
\end{subequations}
It then follows that
\begin{equation}
  \label{stageoneb2}
  C( |\phi_3 \rangle , |\phi_2 \rangle ) <2^{p-1} \pi.
\end{equation}

The weight originally concentrated in $|\phi_0 \rangle $ at the center point $y$ of
$G$, with edge length $d$, in $|\phi_3 \rangle $ is distributed equally over the center points
of 8 sub-cubes of $G$ each with edge length $\frac{d}{2}$. 
Combining Eqs. (\ref{stageoneb}), (\ref{stageoneb1}) and (\ref{stageoneb2}) gives
\begin{equation}
  \label{stageoneb3}
  C( |\phi_3 \rangle , |\phi_0 \rangle ) < (3+ \sqrt{2}) 2^{p-2} \pi.
\end{equation}

The fan-out process of Eqs. (\ref{defphi1}) - (\ref{stageoneb3}) we now repeat for
an additional $p-2$ iterations arriving at a state $|\phi_{3 p - 3} \rangle $ with weight
equally distributed over the center points of $2^{3 p - 3}$ cubes each with edge length
$2$.
Eqs. (\ref{stageoneb3}) rescaled for iteration $\ell$ give
\begin{equation}
\label{iterationell}
C( |\phi_{3 \ell} \rangle , |\phi_{3 \ell - 3} \rangle ) < (3+\sqrt{2}) 2^{p-\ell-1} 2^{\frac{3\ell - 3}{2}} \pi.
\end{equation}
The term $2^{p-\ell - 1}$ counts the decreasing number of lattice steps between cube centers as
the fan-out process is iterated, while the term $2^{\frac{3\ell - 3}{2}}$
counts the growing number of cubes and therefore 
of sites which each subsequent operator $\bar{k}(i)$ acts on simultaneously.

To complete the fan-out process, the weight at the center of each of the cubes with edge
length 2 needs to
be distributed to the 26 points forming its boundary.
This process can be carried out in 3 additional steps thereby defining
$|\phi_{3p -2} \rangle , |\phi_{3p-1} \rangle $ and $|\phi_{3p} \rangle $.
Each of the 26 points in the final set of boundaries, however, is shared by between 2 and
8 adjacent cubes. This complication can be handled by assigning
each of the shared points to the cube from the center of which it
can be reached solely by positive displacements in any combination
of directions 1, 2 and 3. The result is that the
weight which needs to be moved in each of the final 3 steps
is less than the weight which is moved in the generic steps so far.
The end result, 
is that the cost of this last step is strictly less than given
by Eq. (\ref{iterationell}) for $\ell = p$.
We therefore sum Eq. (\ref{iterationell}) from $\ell$ of 1 to $p$ and obtain
\begin{equation}
\label{summedfanout}
C(|\phi_{3p} \rangle , |\phi_0 \rangle ) < \frac{(3 + \sqrt{2})(1+\sqrt{2})}{4}\pi 2^{\frac{3 p}{2}}.
\end{equation}
Substituting $V$ for $2^{3p}$, we then have
\begin{equation}
\label{summedfanout1}
C(|\phi_{3p} \rangle , |\phi_0 \rangle ) < \frac{(3 + \sqrt{2})(1 + \sqrt{2})}{4}\pi \sqrt{V}.
\end{equation}

The bound of Eq. (\ref{summedfanout}) is derived assuming Eq. (\ref{rpower2}) giving the edge $d$ of cube $G$ 
as an even power of 2. Consider now the case
\begin{equation}
\label{rnotpower2}
2^{p-1} < d < 2^p.
\end{equation}

Assume again that at each iteration $\ell$ of the fan-out process, each edge length of each parent cube is
split as evenly as possible into halves to produce 8 child cubes with all edges nearly equal. Suppose
$d$ is $2^p - 1$. After iteration $\ell$ has been completed, the total number
of cubes will still be $2^{3 \ell}$. Orthogonal to each direction, the cubes can be grouped
into $2^\ell$ planes, each holding $2^{2 \ell}$ cubes. But for each direction one of these
orthogonal planes will have an edge in that direction which is one lattice unit shorter than the 
corresponding edge of the other $2^\ell$ planes. It follows that the update process in
each direction can proceed with $2^{p - \ell - 1} - 1$ steps occuring simultanously across all
cubes, and one final update skipped for the cubes with a single edge in that direction one
lattice unit shorter. The bound of Eq. (\ref{iterationell}) will hold without modification.
For $d$ given by  $2^p - 2$, after iteration $\ell$, for each direction, there will be two planes of
$2^{2 \ell}$ cubes each with the edge in that direction one lattice unit shorter. The bound of
Eq. (\ref{iterationell}) will continue to hold. Similarly for $d$ given  by $2^p - q$ for any
$q < 2^{p-1}$.

For $d$ of Eq. (\ref{rnotpower2}), when $\ell$ reaches $p - 1$ the resulting cubes (no longer exactly cubes)
will have a mix of edges of length $2$ and of length $1$. The boundary points of
these cubes can still be assigned to a single one of its adjacent cubes 
with the result that the bound of Eq. (\ref{iterationell}) remains in place for the final pass
$\ell = p$.
The bound of Eq. (\ref{summedfanout}) remains in place for the net result of the entire
fan-out process. By assumption, according to Eq. (\ref{rnotpower2}) we have
\begin{equation}
\label{rbound}
2 d > 2^p.
\end{equation}
Then since $V$ is $d^3$, Eq. (\ref{summedfanout}) gives
\begin{equation}
\label{summedfanout2}
C(|\phi_{3p-1} \rangle , |\phi_0 \rangle ) < \frac{(3 + \sqrt{2})(1 + \sqrt{2})}{\sqrt{2}} \pi  \sqrt{V},
\end{equation}
which is weaker than Eq. (\ref{summedfanout1}) and therefore holds whether or not
$d$ is an even power of 2.

The bound of Eq. (\ref{summedfanout1}) applies to a fan-out process on a 
single prototype state on cube $G$. Assume the process repeated in parallel on the
$2mn$ cubes $D^v_{ij}$, thereby generating $|\psi \rangle $ of Eq. (\ref{entangledstate}).
For $|\rho \rangle $ of Eq. (\ref{phinpoints}) we then have
\begin{equation}
\label{psiphi}
C( |\psi \rangle , |\rho \rangle ) \le  (3 + \sqrt{2})(1+\sqrt{2}) \pi \sqrt{mnV}.
\end{equation}
From Eqs. (\ref{deltac}) and (\ref{deltac1}), it follows that for a product state
$|\omega \rangle $ we have
\begin{equation}
\label{psiomega}
C(|\psi \rangle ,|\omega \rangle ) \le c_1 \sqrt{ mnV} + c_2 m n + c_3 \sqrt{mn} r, 
\end{equation}
where
\begin{subequations}
\begin{eqnarray}
\label{defc1}
c_1 & =&(3 + \sqrt{2})(1 + \sqrt{2}) \pi  , \\
\label{defc2}
c_2 & = & \sqrt{2} \pi, \\
\label{defc32}
c_3 & = & \pi, 
\end{eqnarray}
\end{subequations}
for $r$ of Eq. (\ref{deltac1}).
Eq. (\ref{upperb}) then follows.

\section{\label{app:pairsplits} Class of Pair Splits Which Persist, Other Discontinuities Absent}

In Appendix \ref{app:secondlaw}
based on the conjectured second law of quantum
complexity \cite{Brown}, we derived an estimate
for the time evolution of complexity of a system
evolving according to the Hamiltonian $h$ of Eq. (\ref{hamiltonian})
through a sequence of states
each with much less than
the system's maximum possible complexity.
Based on this estimate
we now present an argument for the hypothesis that
a pair of branches 
produced  by
a split  at some $t_0$ of a branch $|\phi \rangle $ with much less than maximal complexity,
for a system with a large number of degrees of freedom and $b$ sufficiently large,
branches which wander off independentaly into a high dimensional space, with high probability will not merge back
into $|\phi \rangle $
at $t > t_0$.

Other possible events merging two branches
into a single result 
we will argue are similarly improbable.
Still more complicated rearrangments at a single instant of $n$ branches into
a new configuration of $n'$ branches with $n + n' > 3$
we believe occur with zero probability.

The time evolution of the set of optimal branches then
yields a tree structure of branches each eventually splitting
into a pair of sub-branches.  The state vector of the real world
we propose follows through the tree a single sequence of
branches and sub-branches, with the sub-branch at each splitting
event chosen randomly according to the Born rule.

For this discussion we assume 
$n_u$ is finite. But the results hold uniformly in $n_u$, therefore apply also in the
limit $n_U \rightarrow \infty$.

\subsection{\label{subsec:after} Complexity After a Split}

The first half of the argument for the persistence
of pair splits is a bound on the change
in complexity, following a branching event, of either child branch 
by the change in complexity of the parent.
At some time $t_0$, assume a particular $|\phi \rangle $ of an optimal branch decomposition $\{|\psi_i \rangle \}$
has a pair of sub-branches $|\phi_0 \rangle $ and $|\phi_1 \rangle $
  \begin{equation}
\label{splitphi1}
|\phi \rangle  = |\phi_0 \rangle  + |\phi_1 \rangle,
  \end{equation}
  which satisfy the split condition
  \begin{equation}
\label{splitcondition1}
[C( |\phi \rangle )]^2 - \rho [C( |\phi_0 \rangle )]^2 - ( 1 - \rho) [C( |\phi_1 \rangle )]^2 > 
-b \rho \ln( \rho) - b ( 1 - \rho) \ln( 1 - \rho) ,
  \end{equation}
  where
  \begin{equation}
    \label{defofrho1}
     \langle  \phi_0 | \phi_0 \rangle  = \rho  \langle  \phi | \phi \rangle .
  \end{equation}

According to the discussion of Appendix \ref{app:secondlaw},
there is an operator $k(t)$ which 
satisfies
\begin{equation}
  \label{incrementalkx}
  |\phi( t + \delta) \rangle  = \exp[ -i \delta \theta(t) -i \delta k(t)] |\phi(t) \rangle, 
\end{equation}
with minimal $\parallel k(t) \parallel$
yielding
\begin{equation}
  \label{complexityincrementx}
  C( |\phi(t) \rangle , |\phi(t_0) \rangle ) = \int_{t_0}^t dt \parallel k(t) \parallel - \epsilon .
\end{equation}
For the branches $|\phi_0 \rangle $ and $|\phi_1 \rangle $ we can
then define $k_i(t)$ to accomplish
\begin{equation}
  \label{incrementalki}
  |\phi_i( t + \delta) \rangle  =  \exp[ -i \delta \theta_i(t) - i\delta k_i(t)] |\phi_i(t) \rangle , 
\end{equation}
with minimal $\parallel k_i(t) \parallel$.
The argument leading to Eq. (\ref{complexityincrementx})
then implies that with high probability
\begin{equation}
  \label{complexityincrement3}
  C( |\phi_i(t) \rangle , |\phi_i(t_0) \rangle ) = \int_{t_0}^t dt \parallel k_i(t) \parallel - \epsilon_i
\end{equation}
for some very small $\epsilon_i \ge 0$.

For sufficiently large $b$, for $|\phi \rangle $ the state of a system
with a large number of degrees of freedom,
we can obtain bounds on the $\parallel k_i(t) \parallel$.

For some nearest neighbor pair $\{x, y\}$, define
\begin{subequations}
  \begin{eqnarray}
    \label{defQ1}
    \mathcal{Q} & = & \mathcal{H}_x \otimes \mathcal{H}_{xy} \otimes \mathcal{H}_y, \\
    \label{defR}
    \mathcal{R} & = & \bigotimes_{z \ne x, y} \mathcal{H}_z \bigotimes_{ \{u,v\} \ne \{x,y\}} \mathcal{H}_{xy}\\
    \label{defQR}
    \mathcal{H} & = & \mathcal{Q} \otimes \mathcal{R}.
  \end{eqnarray}
\end{subequations}
Let the corresponding Schmidt decompositions of $|\phi_0 \rangle , |\phi_1 \rangle $ be
\begin{equation}
  \label{defpsixhi}
  |\phi_i \rangle  = \sum_j  |\psi_{ij} \rangle  \otimes |\chi_{ij} \rangle .
\end{equation}

For sufficiently large $b$, for a system with a large number
of degrees of freedom, the states $|\phi_0 \rangle $ and $|\phi_1 \rangle $
on reaching the branching threshold and after will have wandered off into a high dimensional space.
We therefore expect
the burden of orthogonality between $|\phi_0 \rangle $ and $|\phi_1 \rangle $
to be spread over many lattice spacings.
The reduced states produced by averaging
$|\phi_0 \rangle $ and $|\phi_1 \rangle $
over the 2 site Hilbert space $\mathcal{Q}$
should then still be orthogonal.
If so, we have
\begin{equation}
  \label{schmidtinr}
   \langle  \chi_{0j} | \chi_{1\ell} \rangle  = 0,
\end{equation}
for all $j, \ell$.

Let $h_{xy}$ be the piece of the Hamiltonian $h$ acting on $\mathcal{Q}$, $k_{xy}$ be the piece
of $k$ acting on $\mathcal{Q}$ and $k_{ixy}$ be the piece of $k_i$ acting on $\mathcal{Q}$.
We then have
\begin{subequations}
  \begin{eqnarray}
    \label{knotsplit}
    [k_{xy} + \theta(t)] \sum_{ij}   |\psi_{ij} \rangle  \otimes |\chi_{ij} \rangle & = &  h_{xy} \sum_{ij}   |\psi_{ij} \rangle  \otimes |\chi_{ij} \rangle, \\
    \label{ksplit}
          [k_{ixy} + \theta_i(t)] \sum_j  |\psi_{ij} \rangle  \otimes |\chi_{ij} \rangle & = &  h_{xy} \sum_j   |\psi_{ij} \rangle  \otimes |\chi_{ij} \rangle.
  \end{eqnarray}
\end{subequations}
Since $k_{xy}$ and $h_{xy}$  act only on $\mathcal{Q}$,  Eqs. (\ref{schmidtinr}) and (\ref{knotsplit}) then imply
  \begin{equation}
    \label{knotsplit1}
          [k_{xy} + \theta(t)] \sum_j   |\psi_{ij} \rangle  \otimes |\chi_{ij} \rangle  =  h_{xy} \sum_j   |\psi_{ij} \rangle  \otimes |\chi_{ij} \rangle.
  \end{equation}
  Thus $k_{xy}$ satisfies requirement Eq. (\ref{ksplit}) both for $k_{0xy}$ and for $k_{1xy}$. But since
  $k_{0xy}$ and $k_{1xy}$ are the operators with minimal norm which satisfy Eq. (\ref{ksplit}),
  we must have
 \begin{equation}
  \label{normki}
  \parallel k_{ixy}(t) \parallel \le \parallel k_{xy}(t) \parallel.
\end{equation}

 For any plaquette $p$ an argument identical to the derivation of Eq. (\ref{normki}) gives
  \begin{equation}
  \label{normki1}
  \parallel k_{ip}(t) \parallel \le \parallel k_p(t) \parallel,
  \end{equation}
  where $k_{ip}$ is the piece of $k_i$ acting on $\mathcal{H}_p$ and $k_p$ is the piece of
  $k$ acting on $\mathcal{H}_p$.

  Eqs. (\ref{normki}) and (\ref{normki1}) combined across
  all nearest neighbor pairs $\{x, y\}$ and all plaquettes $p$ then imply
  \begin{equation}
  \label{normki2}
  \parallel k_i(t) \parallel \le \parallel k(t) \parallel.
  \end{equation}

Combining Eq. (\ref{normki2}) with Eqs. (\ref{complexityincrementx}) and (\ref{complexityincrement3})
implies 
\begin{equation}
  \label{complexityincrementbound}
  C( |\phi_i(t) \rangle , |\phi_i(t_0) \rangle ) < C( |\phi(t) \rangle , |\phi(t_0) \rangle ) + \epsilon,
\end{equation}
We now repeat of the argument leading to Eq. (\ref{complexityincrementx}) but this time to the
trajectories of $k(\nu)$ connecting $|\phi_i(t_0) \rangle$ and $|\phi(t_0)\rangle$ to corresponding
product states in Eqs. (\ref{udot}) and (\ref{uboundary0}) entering the definition of complexity.
Eq. (\ref{complexityincrementbound}) then implies
\begin{equation}
  \label{complexityincrementbound1}
  C( |\phi_i(t) \rangle ) - C(|\phi_i(t_0) \rangle ) < 
  C( |\phi(t) \rangle ) - C(|\phi(t_0) \rangle ) + \epsilon.
\end{equation}

\subsection{\label{subsec:after1} Net Complexity After a Split}

We now show that combined with Eq. (\ref{splitcondition1}) at
$t_0$, Eq. (\ref{complexityincrementbound1}) leads
to Eq. (\ref{splitcondition1}) for all $t > t_0$.

At  $t > t_0$, the left hand side of Eq. (\ref{splitcondition1}) is given
by $p(t)$
\begin{equation}
  \label{lefthand}
  p(t) = [ D( t) + C(|\phi( t_0) \rangle ] ^2 - 
  \rho [D_0( t) + C( |\phi_0(t_0) \rangle )]^2 - 
  (1-\rho) [D_1( t) + C( |\phi_1(t_0) \rangle )]^2,
\end{equation}
with the definitions
\begin{subequations}
  \begin{eqnarray}
    \label{defD}
    D(t) & = &  C( |\phi(t) \rangle ) - C(|\phi(t_0) \rangle ) ,\\
 \label{defD0}
    D_0(t) & = &  C( |\phi_0(t) \rangle ) - C(|\phi_0(t_0) \rangle ) ,\\
 \label{defD1}
    D_1(t) & = &  C( |\phi_1(t) \rangle ) - C(|\phi_1(t_0) \rangle ).
  \end{eqnarray}
\end{subequations}
We can then rearrange $p(t)$ as a sum of three terms
\begin{subequations}
\begin{eqnarray}
  \label{alpha}
  q( t) &= &D(t)^2 - \rho D_0(t)^2 - (1 - \rho) D_1(t)^2, \\
  \label{beta}
  r( t) &=& 2 D(t)C(|\phi(t_0) \rangle ) - 2\rho D_0(t)C(|\phi_0(t_0) \rangle ) - 2 (1 - \rho) D_1(t)C(|\phi_1(t_0) \rangle ),\\
  \label{gamma4}
  s &=& C(|\phi(t_0) \rangle )^2 - \rho C(|\phi_0(t_0) \rangle )^2 - (1 - \rho)C(|\phi_1(t_0) \rangle )^2.
\end{eqnarray}
\end{subequations}

Eqs. (\ref{complexityincrementbound1}) and (\ref{defD}) - (\ref{defD1}) imply $q(t)$ is 
greater than some $-\epsilon$.
Also $s$ is the left hand side of  Eq. (\ref{splitcondition1}) 
so strictly greater than the right hand side, thus greater than 0, $D(t)$ is greater than $-\epsilon$
by a futher application
of the second law of quantum complexity, and $C(|\phi(t_0) \rangle )$ is nonnegative by the definition of
complexity. We therefore have
\begin{equation}
  \label{cauchyschwartz0}
  D( t) C(|\phi(t_0) \rangle ) \ge   \sqrt{\rho D_0(t)^2 + (1 - \rho) D_1(t)^2} \times
  \sqrt{\rho C(|\phi_0(t_0) \rangle )^2 + (1 - \rho)C(|\phi_1(t_0) \rangle )^2} - \epsilon.
\end{equation}
The Cauchy-Schwartz inequality 
\begin{multline}
  \label{cauchyschwartz}
  \sqrt{\rho D_0(t)^2 + (1 - \rho) D_1(t)^2} \times
  \sqrt{\rho C(|\phi_0(t_0) \rangle )^2 + (1 - \rho)C(|\phi_1(t_0) \rangle )^2} \ge \\  
\rho D_0(t)C(|\phi_0(t_0) \rangle ) + (1 - \rho) D_1(t)C(|\phi_1(t_0) \rangle ).
\end{multline}
then implies 
$r(t)$ is greater than some $-\epsilon$. Assembling these pieces
gives
\begin{equation}
  \label{branchingt}
  p(t) > s - \epsilon >
  -b \rho \ln( \rho) - b ( 1 - \rho) \ln( 1 - \rho) - \epsilon.
\end{equation}
Thus Eq. (\ref{splitcondition1}) is highly likely satisified for all $t > t_0$.
A split which first occurs at some time $t_0$ with
high probability persists
for all $t > t_0$.

\subsection{\label{subsec:dontmerge} Other Mergers of Pairs Similarly Improbable}

The argument supporting the hypothesis that splits persist can equally well be applied
to show that any pair of branches $|\phi_0 \rangle $ and
$|\phi_1 \rangle $ which exists at some time $t_0$, whether or
not they were born from the split
of a single shared parent branch, are highly unlikely to
merge into a single branch at $t > t_0$.

Let the sum of the branches $|\phi_0 \rangle $ and $|\phi_1 \rangle $ at $t_0$ be given
again by
\begin{equation}\label{splitphix}
|\phi \rangle  = |\phi_0 \rangle  + |\phi_1 \rangle .
\end{equation}
Then since the optimal branch decomposition
$\{ |\psi_i \rangle \}$ at $t_0$ includes $|\phi_0 \rangle $ and $|\phi_1 \rangle $, rather than
their replacement by $|\phi \rangle $, the inequality
\begin{equation}\label{splitconditionx}
[C( |\phi \rangle )]^2 - \rho [C( |\phi_0 \rangle )]^2 - ( 1 - \rho) [C( |\phi_1 \rangle )]^2 > 
-b \rho \ln( \rho) - b ( 1 - \rho) \ln( 1 - \rho).
\end{equation}
must again hold at $t_0$. The discussion of Sections \ref{subsec:after}
and \ref{subsec:after1} then supports the hypothesis that
Eq. (\ref{splitconditionx}) continues to hold for all $t > t_0$.

\subsection{\label{subsec:norearrangements} No Other Discontinuities}

The remaining class of possible discontinuities
in branch time evolution 
are events
rearranging $n$ branches
at a single instant into a new configuration of $n'$ branches with $n + n' > 3$.
Our hypothesis is that such events occur with zero probability.

Consider the case of $n$ and $n'$ both 2.
Suppose at time $t_0$ the optimal branch configuration includes
a pair of branches $|\phi_0 \rangle , |\phi_1 \rangle $.
Then at some $t > t_0$,
$|\phi_0 \rangle , |\phi_1 \rangle $
jump to
a distinct pair $|\phi_0' \rangle , |\phi_1' \rangle $ with
\begin{equation}
  \label{merger2}
  |\phi_0 \rangle  + |\phi_1 \rangle  = |\phi_0' \rangle  + |\phi_1' \rangle ,
\end{equation}
while all other branches
vary continuously with time at $t$.
Since all branches in the optimal decomposition $\{ |\psi_i \rangle \}$
aside from $|\phi_0 \rangle $ and $|\phi_1 \rangle $
vary continuously with time across $t$,
for the discontinuous jump of Eq. (\ref{merger2}) to occur at $t$, the
pair of states
$|\phi_0' \rangle , |\phi_1' \rangle, $ and the pair of states
$|\phi_0 \rangle , |\phi_1 \rangle, $
must span the same
2-dimensional subspace of $\mathcal{H}$.
Therefore, for some matrix of coefficients $z_{ij}$
the recombination event of Eq. (\ref{merger2}) can be viewed as a
simultaneous pair of branching events
\begin{subequations}
  \begin{eqnarray}
    \label{merger3}
    |\phi_0 \rangle &=&  z_{00} |\phi_0' \rangle + z_{01} |\phi_1' \rangle, \\
    \label{merger4}
      |\phi_1 \rangle &=& z_{10} |\phi_0' \rangle + z_{11} |\phi_1' \rangle.
  \end{eqnarray}
\end{subequations}
This coincidental alignment at a single instant of a pair of
otherwise independent branching events 
we believe has probability 0.

By a similar argument,
rearrangment of $n$ branches
at a single instant into a new configuration of $n'$ branches
for any other $n$ with $n + n' > 3$
we believe also occurs with zero probability.

The end result of all of which is
the hypothesis that
for a system evolving
through a sequence of states
with much less
than the system's maximum possible complexity,
the discontinuities
in branch time evolution are
highly probably only splits of a single
branch into a pair of sub-branches.

\section{\label{app:lorentz} Lorentz Covariant Branching}

\subsection{\label{subsec:latetau} $t \rightarrow \infty$}

We now assume that the limit of infinite box size $B \rightarrow \infty$
exits for the optimal branch decomposition arising from $Q(\{|\psi_i \rangle \})$,
and assume the lattice spacing $a$ has been made much smaller
than any other relevant length scales. Whether either of these
assumptions potentially leads to trouble is beyond the scope of
the present discussion.

An example of branching considered in \cite{Weingarten} suggests
that as $t \rightarrow \infty$, for a system in infinite volume,
branch splitting will continue without stop.
This possibility we will deal with
by constructing summed sets of branches.

The construction of summed sets of
branches rests on
a branch labeling scheme defined as follows.
For a system beginning in some initial state $|\psi \rangle $ with
complexity close to 0 at
time $t_0$, 
consider the set of branching events and branch states
which result
from minimizing $Q( \{|\psi_i \rangle \})$ for $t \ge t_0$.
Following Section \ref{subsec:timeevolution},
the evolving state vector of the real world we assume is
chosen from a tree of persistent branching
events which is defined to consist of those
 branching events
in which the two branches do not
eventually recombine and in which no
descendent of either branch eventually
recombines with a descendent of the other branch.
Let $E$ be the set of all persistent branching events accumulated over all $t$.

If $E$ consist entirely of events
splitting some branch permanently into a pair of sub-branches,
each branch state $|\psi( s, t) \rangle $
can be labelled with a
set of pairs  
\begin{equation}
  \label{defstring}
  s = \{ (e_0, \ell_0),  \ldots (e_{n-1}, \ell_{n-1}) \}, n > 0,
\end{equation}  
giving a corresponding history of 
splitting events $e_i \in E$ and branch indices $ \ell_i \in \{0,1 \}$.
For a splitting event $e \in E$ at time $t$
of a state $|\psi(w,t) \rangle $ with history
\begin{equation}
      \label{defw}
      w =  \{ w_0, \ldots w_{n-1}\},
\end{equation}
the resulting branch states $|\psi(u, t) \rangle , |\psi(v, t) \rangle $,
have
\begin{subequations}
  \begin{eqnarray}
    \label{defu}
    u & = & \{ w_0, \ldots w_{n-1}, (e, 0) \}, \\
    \label{defv}
    v & = & \{ w_0, \ldots w_{n-1}, (e, 1) \}.
  \end{eqnarray}
\end{subequations}

The initial state
we assign branch index 0
of an initial null branching event
$ \emptyset \in E$ at $t_0$.
Thus $|\psi \rangle $ at $t_0$ becomes
$|\psi[\{( \emptyset, 0)\}, t_0] \rangle $.
For $s$ of Eq. (\ref{defstring}),
define $|s|$ to be $n$.

For any $t \ge t_0$, let $S(t)$
be the set of $s$ corresponding to
the set of branches
which minimize $Q( \{|\psi_i \rangle \})$.
Each $S(t)$ can be viewed as a
set of maps each taking some subset of $E$ into
$\{0,1\}$.
Define $S$ to be the set of all such 
maps
\begin{equation}
  \label{defS}
  S = \cup_t S(t).
\end{equation}

For any $s \in S$, and any $t$,
define $|\chi( s, t) \rangle $ to be the sum of all the
$t$ branches with histories containing $s$
\begin{equation}
  \label{defhatchi}
  |\chi(s, t) \rangle  = \sum_{s' \in S(t), s' \supseteq s}| \psi( s', t) \rangle .
\end{equation}
For any $t$, there will be a corresponding $n_t$ such that
\begin{equation}
  \label{defnt}
  |\chi( s, t) \rangle  = 0, |s| > n_t.
\end{equation}
On the other hand, for every $s \in S$ there is a $t_s$ such that
\begin{equation}
  \label{defnt1}
  |\chi( s, t) \rangle  \ne 0, t > t_s.
\end{equation}

For any $t_0 \le t_1 \le t$,  
selecting the $s' \in S( t_1)$ which are descendants of
some $s \in S( t_0)$ yields 
\begin{equation}
  \label{defhatchi1}
  |\chi(s, t) \rangle  = \sum_{s' \in S(t_1), s' \supseteq s}| \chi( s', t) \rangle .
\end{equation}
For any $s \in S(t)$, the only $s' \in S(t)$
which satisfies $s' \supseteq s$ is $s' = s$ itself, in
which case
\begin{equation}
  \label{trivialcase}
  |\chi(s, t) \rangle  = |\psi(s, t) \rangle .
\end{equation}

Let $V(t)$ be the time development operator
\begin{equation}
  \label{defU}
  V(t) = \exp( -i t h),
\end{equation}
where $h$ is the Hamiltonian defined in Eq. (\ref{hamiltonian}).
Define $|\hat{\chi}(s, t) \rangle $ to be
the $t = 0$ representation
of $|\chi( s, t) \rangle $
  \begin{equation}
    \label{defchiu}
    |\hat{\chi}( s, t) \rangle  =  V^\dagger(t) |\chi( s, t) \rangle .
  \end{equation}
For any $t_0 \le t_1 \le t$,  
Eq. (\ref{defhatchi1}) implies
\begin{equation}
  \label{defhatchi2}
  |\hat{\chi}(s, t) \rangle  = \sum_{s' \in S(t_1), s' \supseteq s}| \hat{\chi}( s', t) \rangle .
\end{equation}
For any $t \ge t_0$
\begin{equation}
  \label{psih}
  |\hat{\chi}[ \{( \emptyset, 0)\}, t] \rangle  = V^\dagger(t_0) |\psi \rangle .
\end{equation}

The continuous
piece of the evolution of the optimal branch configuration, according to Section \ref{subsec:timeevolution},
for sufficiently large $b$
will consist almost entirely of continuous unitary
evolution with $t$ of those branches which do not split.
If the branches which do not 
split changed purely by unitary evolution in
$t$, then each $|\hat{\chi}(s, t) \rangle $ of
Eq. (\ref{defchiu}),
for any $s \in S( t')$ for any $t \ge t'$, would be
constant in $t$.
Thus the existence, for any $s \in S$, at least of the limit
\begin{equation}
  \label{chihat}
  \lim_{t \rightarrow \infty} |\hat{\chi}(s, t) \rangle  = |\hat{\chi}(s) \rangle 
\end{equation}
appears to be a plausible hypothesis.

\subsection{\label{subsec:born} Born Rule as an Invariant Measure on Branching Histories}

Consider an infinite sequence of $s_i \in S$ with
\begin{subequations}
  \begin{eqnarray}
    \label{lengthi}
    | s_i| & = & i, \\
    \label{segi}
    s_i & \subset & s_{i+1}.
  \end{eqnarray}
\end{subequations}
A version of the Born rule based on asymptotic late
time branches says the probability that a state with
history $s_i$ will at the next
branching event land in
$s_{i+1}$ is
\begin{equation}
  \label{probsi}
  P( s_{i+1} | s_i) = \frac{  \langle  \hat{\chi}(s_{i+1}) | \hat{\chi}( s_{i+1}) \rangle }{  \langle  \hat{\chi}(s_i) | \hat{\chi}( s_i) \rangle }.
\end{equation}

The Born rule we now formulate as a measure on the set of branching histories,
each extending over all time, beginning from some initial state $|\psi \rangle $.
An all-time branching history $\hat{s}$ is an infinite
set of pairs which assigns each event 
$e \in E$ 
to a corresponding
branch index $i \in \{0, 1\}$.
\begin{equation}
  \label{hatspairs}
  \hat{s} = \{ (e_0, i_0), (e_1, i_1), ... \}.
\end{equation}
Let $\hat{S}$ be the set of all such all-time histories $\hat{s}$.
For every $s \in S$, let $v( s) \subset \hat{S}$ be
the collection of $\hat{s} \in \hat{S}$
which are supersets of $s$, 
\begin{equation}
  \label{defus}
  v( s) = \{ \hat{s} \in \hat{S} | \hat{s} \supset s \}.
\end{equation}

For every such $v(s)$ define the function $\mu[v(s)]$ to be
\begin{equation}
  \label{defmu}
  \mu[ v(s)] =  \langle  \hat{ \chi}(s) |\hat{\chi}(s) \rangle .
\end{equation}
Let $\Sigma$ be the $\sigma$-algebra of sets in $\hat{S}$
generated by all $v(s)$ for $s \in S$.
The complement of any $v(s)$ is given
by the finite union
\begin{equation}
  \label{complement}
  v(s)^c = \cup_{s' \in c(s)} v(s'),
\end{equation}
where $c(s)$ is the set of $s'$ each consisting of
exactly one of
the events in $s$ but with branch index reversed
\begin{equation}
  \label{cofs}
  c[\{ (e_0, i_0), ... (e_{n-1}, i_{n-1}) \}] = 
  \Bigl\{ \{(e_0, \neg i_0)\}, ...\{ (e_{n-1},  \neg i_{n-1})\} \Bigr\}.
\end{equation}

In addition, for any $s, s' \in S$,
\begin{equation}
  \label{intersection}
  v( s) \cap v(s') = v(s \cup s').
\end{equation}
It follows that every element
of $\Sigma$ is given by a union of a countable collection
of pairwise disjoint $v(s)$.
For every countable collection of pairwise
disjoint sets $\{ v( s_i) \}$, define
\begin{equation}
  \label{defmu1}
  \mu[ \cup_i v(s_i) ] = \sum_i \mu[ v( s_i)].
\end{equation}
Eq. (\ref{defmu1}) turns $\mu$ into a probability measure on $\Sigma$.

Eq. (\ref{probsi}) follows from Eq. (\ref{defmu}). Since
the $|\hat{\chi}(s) \rangle $ are Lorentz covariant and
the algebra $\Sigma$ is frame independant, the measure
$\mu$ is Lorentz invariant.
The Born rule can then be formulated as the hypothesis that
world's history of branching events
is an $\hat{s} \in \hat{S}$ chosen randomly according to the
measure $\mu$.

\subsection{\label{subsec:framebranching} Time Dependent View of Branching History}

The Lorentz covariant set of $t \rightarrow \infty$ branches $|\hat{\chi}(s) \rangle $
and corresponding branching history $\hat{s}$ chosen according to the Born measure of
Section \ref{subsec:born} we take to be the physical objects
underlying macroscopic reality.
From these, a view of branching history
unfolding in time in any particular Lorentz frame can be
constructed.

In any particular frame, for any all-time history of events $\hat{s}$,
there is a corresponding sequence of partial branch histories
$s_n \in S, n \ge 1,$ with
\begin{subequations}
  \begin{eqnarray}
    \label{partialbranch0}
    | s_n | & = & n, \\
    \label{partialbranch1}
    s_n  & \subset & s_{n + 1}, \\
    \label{partialbranch2}
    \cup_n s_n & = & \hat{s},
  \end{eqnarray}
\end{subequations}
ordered in such a way that for every $n$  the last event in $s_n$ occurs
after the last event in $s_{n-1}$.
Let $|\hat{\chi}( s_n) \rangle $ be the corresponding sequence of states
represented at $t = 0$.
From these define $|\psi_n(t) \rangle $ to be
\begin{equation}
  \label{movedtoh}
  |\psi_n( t) \rangle  = U( t) |\hat{\chi}(s_n) \rangle.
\end{equation}
The system begins at $t_0$
in the state $|\psi \rangle $
\begin{equation}
  \label{psi1psi}
  |\psi_1( t_0) \rangle  = |\psi \rangle ,
\end{equation}
then at a sequence of times
$t_n, n \ge 1$,
successively branches from $|\psi_n( t_n) \rangle $ to $|\psi_{n+1}( t_n) \rangle $.

The $t_n$ can be found
as follows.
Define
$|\psi_n(t) \rangle $ and $\rho_n$ to be
\begin{subequations}
  \begin{eqnarray}
    \label{psi11}
    |\phi_n(t) \rangle  & = &  |\psi_n( t) \rangle  - |\psi_{n+1}(t) \rangle  \\
     \label{timeview0}
    \rho_n & = & \frac{  \langle \psi_{n+1}( t)|\psi_{n+1}( t) \rangle }{  \langle \psi_n( t)|\psi_n( t) \rangle } .
  \end{eqnarray}
\end{subequations}
From these define
\begin{multline}
  \label{defdeltan}
  \Delta_n( t) = [C( |\psi_n( t)  \rangle )]^2 - 
  \rho_n [C( |\psi_{n+1}(t) \rangle )]^2 - ( 1 - \rho_n) [C( |\phi_n \rangle )]^2 \\
+b \rho_n \ln( \rho_n) + b ( 1 - \rho_n) \ln( 1 - \rho_n).
\end{multline}
Each $t_n$ will then be the smallest $t$ such that
\begin{equation}
  \label{branchn}
  \Delta_n( t) \ge 0.
\end{equation}
By choice of the $s_n$,
the sequence of $t_n$ is guaranteed to be increasing.


\begin{thebibliography}{10}

\bibitem{Everett}
H.~Everett.
\newblock {R}elative {S}tate {F}ormulation of {Q}uantum {M}echanics.
\newblock {\em Rev. Mod. Phys.}, 29:454, 1957.

\bibitem{DeWitt}
B.~DeWitt.
\newblock Quantum mechanics and reality.
\newblock {\em Physics Today}, 23:30, 1970.

\bibitem{Zeh}
H.~D. Zeh.
\newblock On the interpretation of measurement in quantum theory.
\newblock {\em Found. Phys.}, 1:69, 1970.

\bibitem{Zurek}
W.~H. Zurek.
\newblock Pointer basis of quantum mechanics: {I}nto what mixture does the wave
  packet collapse?
\newblock {\em Phys. Rev. D}, 24:1516, 1981.

\bibitem{Zurek1}
W.~H. Zurek.
\newblock Environment-induced superselection rules.
\newblock {\em Phys. Rev. D}, 26:1862, 1982.

\bibitem{Zurek2}
W.~H. Zurek.
\newblock Decoherence, einselection and the quantum origins of the classical.
\newblock {\em Rev. Mod. Phys.}, 75:715, 2003.

\bibitem{Wallace}
D.~Wallace.
\newblock Everett and structure.
\newblock {\em Studies in the History and Philosophy of Modern Physics}, 34:87,
  2003.

\bibitem{Riedel}
C.~Jess Riedel.
\newblock Classical {B}ranch {S}tructure from {S}patial {R}edundancy in a
  {M}any-{B}ody {W}ave {F}unction.
\newblock {\em Phys. Rev. Lett.}, 118:120402, 2017.

\bibitem{Weingarten1}
D.~Weingarten.
\newblock Hidden {V}ariable {Q}uantum {M}echanics from {B}ranching from
  {Q}uantum {C}omplexity.
\newblock arXiv:1802.10136v5[quant-ph], 2018.

\bibitem{Weingarten}
D.~Weingarten.
\newblock Macroscopic {R}eality from {Q}uantum {C}omplexity.
\newblock {\em Found. Phys.}, 52:45, 2022.

\bibitem{Donoghue}
J.~F.~Donoghue.
\newblock Introduction to the {E}ffective {F}ield {T}heory {D}escription of {G}ravity.
\newblock arXiv:gr-qc/9512024, 1995.


\bibitem{Nielsen}
M.~A. Nielsen.
\newblock A geometric approach to quantum circuit lower bounds.
\newblock {\em Quantum Information and Computation}, 6:213, 2006.

\bibitem{Kent}
A.~Kent.
\newblock Real {W}orld {I}nterpretations of {Q}uantum {T}heory.
\newblock {\em Found. Phys.}, 42:421, 2012.

\bibitem{Kent1}
A.~Kent.
\newblock A {S}olution to the {L}orentzian {Q}uantum {R}eality {P}roblem.
\newblock {\em Phys. Rev. A}, 90:0121027, 2014.

\bibitem{Kent2}
A.~Kent.
\newblock Quantum {R}eality via {L}ate {T}ime {P}hotodetection.
\newblock {\em Phys. Rev. A}, 96:062121, 2017.

\bibitem{Wilson}
K.~Wilson.
\newblock Confinement of quarks.
\newblock {\em Phys. Rev. D}, 10:2445, 1974.

\bibitem{Kogut}
J.~Kogut and L.~Susskind.
\newblock Hamiltonian formulation of {W}ilson's lattice gauge theories.
\newblock {\em Phys. Rev. D}, 11:395, 1975.

\bibitem{NielsenNinomiya}
H.~B. Nielsen and M.~Ninomiya.
\newblock A no-go theorem for regularizing chiral fermions.
\newblock {\em Phys. Lett. B}, 105:219, 1981.

\bibitem{Brown}
A.~R. Brown and L.~Susskind.
\newblock The {S}econd {L}aw of {Q}uantum {C}omplexity.
\newblock {\em Phys. Rev. D}, 97:086015, 2018.

\bibitem{Nagel}
T.~Nagel.
\newblock What {I}s {I}t {L}ike to {B}e a {B}at.
\newblock {\em The Philosophical Review}, 83 (4):435, 1974.

\bibitem{Chalmers}
D.~Chalmers.
\newblock Facing up to the {P}roblem of {C}onsciousness.
\newblock {\em Journal of Consciousness Studies}, 2 (3):200, 1995.

\bibitem{Creutz}
M.~Creutz.
\newblock Quantum {E}lectrodynamics in the {T}emporal {G}auge.
\newblock {\em Annals of Physics}, 117:471, 1979.

\bibitem{Taylor}
J.~K. Taylor and I.~P. McCulloch.
\newblock Wavefunction branching: when you can't tell pure states from mixed states.
\newblock arXiv:2308.04494v2[quant-ph], 2023.

\bibitem{Denef}
  F. Denef, M.~R. Douglas, B. Greene and C. Zukovski.
  \newblock Computational complexity of the landscape I.
  \newblock {\em Annals of Physics}, 322, 1096, 2007.

\bibitem{Denef1}
  F. Denef, M.~R. Douglas, B. Greene and C. Zukovski.
  \newblock Computational complexity of the landscape II - cosmological considerations.
  \newblock {\em Annals of Physics}, 392: 93, 2018.

\bibitem{Bhattacharyya}
  A. Bhattacharyya, S. Das, S.~S. Haque and B. Underwood.
  \newblock Cosmological complexity.
  \newblock {\em Phys. Rev. D}, 101:106020, 2020.
  





\end{thebibliography}
\end{document}